\documentclass[12pt]{article}

\usepackage{amsmath, amssymb, amsfonts, bbm, mathtools}
\usepackage{amsthm}
\theoremstyle{plain}% Theorem-like structures provided by amsthm.sty
\newtheorem{theorem}{Theorem}[section]

\newtheorem{corollary}[theorem]{Corollary}

\newcommand\numberthis{\addtocounter{equation}{1}\tag{\theequation}}
\newcommand{\mt}[1]{\big[\begin{matrix}#1\end{matrix}\big]} % matrix

\theoremstyle{definition}
\newtheorem{definition}[theorem]{Definition}

\theoremstyle{remark}
\newtheorem{remark}{Remark}

\usepackage{geometry} 
\usepackage{changepage}
\usepackage{setspace}
\usepackage{fancyhdr}
\def\spacingset#1{\renewcommand{\baselinestretch}%
{#1}\small\normalsize} \spacingset{1}

\usepackage{float}
\usepackage{tikz}
\usepackage{tikz-qtree}
\usetikzlibrary{trees}
\usetikzlibrary{automata,positioning}
\usetikzlibrary{shapes,decorations,arrows,calc,arrows.meta,fit}

\usepackage{graphicx}

\newcommand{\blue}{\textcolor{blue}}
\newcommand{\E}{\mathbb{E}}

\usepackage[mathscr]{euscript}
\usepackage{soul}

\usepackage{subcaption}
\usepackage{longtable}
\usepackage{xltabular}
\usepackage{booktabs}
\usepackage{supertabular}
\usepackage{array}
\usepackage{threeparttable}
\usepackage{pifont}

\usepackage{comment}

\usepackage{enumitem}

\usepackage{url} % not crucial - just used below for the URL 
\usepackage[urlbordercolor={1 1 1}]{hyperref}
\usepackage[numbers,sort&compress]{natbib}% Citation support using natbib.sty
\bibpunct[, ]{[}{]}{,}{n}{,}{,}% Citation support using natbib.sty
\usepackage{bibunits}

\allowdisplaybreaks
\newcommand{\blind}{1}

\date{}

\fancypagestyle{firstpage}{
  \fancyhf{}
  \fancyhead[L]{\small
    \begin{tabular}{l}
      \textbf{Published version}:\\
      \textit{Statistics in Medicine}. \\ 
       Published: 27 May 2026.\\
      \textbf{DOI:} \href{https://doi.org/10.1002/sim.70606}{10.1002/sim.70606}
    \end{tabular}
  }
  \fancyfoot[C]{\thepage}
  
}

\begin{document}
\begin{bibunit}[apalike]

%%%%%%%%%%%%%%%%%%%%%%%%%%%%%%%%%%%%%%%%%%%%%%%%%%%%%%%%%%%%%%%%%%%%%%%%%%%%%%

\if1\blind
{
  \title{\bf \Large Assessing Racial Disparities in Healthcare Expenditures via Mediator Distribution Shifts \vspace{0.1cm}\\}
  \author{Xiaxian Ou, 
          Xinwei He, 
          David Benkeser, 
          Razieh Nabi\vspace{0.1cm}\\
    Department of Biostatistics and Bioinformatics, Emory University}
  \maketitle
} \fi

\thispagestyle{firstpage}

\if0\blind
{
  \bigskip
  \bigskip
  \bigskip
  \begin{center}
    {\LARGE\bf Assessing Racial Disparities in Healthcare Expenditures via Mediator Distribution Shifts}
\end{center}
  \medskip
} \fi

% \bigskip
\abstract{Racial disparities in healthcare expenditures are well-documented, yet the underlying drivers remain complex. This study develops a framework to decompose such disparities through shifts in the distributions of mediating variables, rather than treating race itself as a manipulable exposure. We define disparities as differences in covariate-adjusted outcome distributions across racial groups, and decompose the total disparity into a component attributable to differences in mediator distributions, and a residual component that remains after equalizing those distributions. Using data from the Medical Expenditures Panel Survey (MEPS), we examine the extent to which expenditure disparities would persist or be reduced if mediators such as socioeconomic status (SES), insurance access, health behaviors, or health status were equalized across racial groups. To ensure valid inference, we derive asymptotically linear estimators based on influence-function techniques and flexible machine learning, including super learners and a two-part model designed for the zero-inflated, right-skewed nature of expenditure data. 

Applying this framework to MEPS data from 2009 and 2016, substantial disparities were observed across all pairwise racial comparisons, with the largest gaps observed between non-Hispanic Whites and Hispanics in both years. Differences in SES and health status were the largest contributors to these disparities, with insurance access also playing a meaningful role, particularly for Hispanic populations, whereas health behaviors contributed minimally. Residual disparities persisted, especially in comparisons involving non-Hispanic Whites, suggesting the influence of unmeasured or structural factors.}

\vspace{0.5cm}
\noindent%

%3 to 6 keywords, that do not appear in the title
% \vfill

% \newpage
\spacingset{1.75} 
% ---  ---  ---  ---  ---  ---  ---  ---  ---  ---  ---  ---  ---  ---  ---  ---  ---  ---  ---  ---  ---  ---  ---  --- 
%######################################
\section{Introduction}
%######################################

% % Importance of studying disparities
Racial disparities in health outcomes are long-standing public health concerns \citep{dieleman_us_2021, us_department_of_health_and_human_services}, with disparities in healthcare expenditures reflecting inequities in access and utilization \citep{mahajan_trends_2021}. Using racial and ethnic categories aligned with U.S. Census definitions (non-Hispanic White, non-Hispanic Black, non-Hispanic Asian, and Hispanic), evidence from the Medical Expenditures Panel Survey (MEPS) data consistently highlights these disparities in the United States \citep{cook_comparing_2010,charron-chenier_racial_2018,simmons_racial_2019}. For example, Dieleman et al. \citep{dieleman_us_2021} estimated that in 2016, non-Hispanic White individuals comprised 61\% of the U.S. population but accounted for 72\% (95\% uncertainty interval: 71\% to 73\%) of total healthcare spending across all racial groups. Such gaps reflect differential healthcare use between advantaged and marginalized populations and are often avoidable and unjust \citep{braveman_health_2011}. Understanding how these disparities arise is essential for informing policy responses that promote more equitable healthcare systems. While aggregate comparisons can document overall gaps, they do not reveal how disparities propagate through specific social and structural mechanisms. A more informative approach decomposes these disparities into contributions from different mediating factors.

% % Common factors driving racial disparities 
Racial disparities in healthcare expenditures reflect a complex interplay of socioeconomic, structural, and behavioral factors.
\textit{Socioeconomic status} (SES) is a major driver, influencing access to resources, quality of care, and overall health outcomes \citep{adler2002socioeconomic}. Black and Hispanic populations, for instance, experience higher poverty rates and lower educational attainment than Whites, creating significant barriers to healthcare access \citep{williams_understanding_2016}.
\textit{Insurance access} further exacerbates these disparities, as uninsured or underinsured individuals are less likely to receive timely and adequate care \citep{cook_comparing_2010, mahajan_trends_2021}. Zuvekas and Taliaferro \citep{zuvekas_pathways_2003} reported that insurance explained 42\% of the Black-White and 24\% of the Hispanic-White disparity in having a usual source of care.
\textit{Health behaviors}, shaped by cultural norms and socioeconomic context, also vary by race and ethnicity \citep{barkley_factors_2008}. For instance, non-Hispanic Asians report the lowest rates of physical inactivity \citep{centers_for_disease_control_and_prevention_adult_2024}, while smoking rates are higher among non-Hispanic Whites and Blacks \citep{thomson2022association}.
\textit{Health status}, which reflects the cumulative effects of disadvantage, shows similar patterns: marginalized groups report worse self-rated health \citep{bell_role_2018} and higher chronic disease prevalence \citep{fiscella2016racial}. Despite greater medical needs, they often encounter barriers to care and lower-quality treatment \citep{charron-chenier_racial_2018}.
These differences in the distribution of mediating factors play a central role in shaping disparities in healthcare spending, and understanding their contributions is essential for designing targeted policy interventions. 

% % Measures of disparities, limitations, and alternative approaches
Empirical studies of racial disparities in healthcare expenditures often rely on regression-based methods that compare outcomes across racial groups while adjusting for mediating factors such as socioeconomic status or insurance access \citep{vanderweele_causal_2014, an_health_2015, wallace_disparities_2022}. A widely used approach in economics and epidemiology is the Oaxaca-Blinder (OB) decomposition \citep{kitagawa1955components, oaxaca1973male, blinder1973wage, zuvekas_pathways_2003, opacic2025disparity}. This method decomposes group differences in outcomes into an explained part, due to differences in observed covariates, and an unexplained part, which is often interpreted as discrimination. While informative as a point-in-time decomposition, OB approaches do not explicitly model mediating pathways and typically rely on linear model specifications. Moreover, their causal interpretation can be challenging when confounding is not explicitly addressed \citep{jackson2018decomposition, park2025identifying}. Another widely used framework for defining and measuring disparities is based on the Institute of Medicine (IOM) definition, which defines a disparity as a difference in treatment provided to members of different racial or ethnic groups that is not justified by the underlying health conditions or treatment preferences of patients \citep{nelson2002unequal, nelson2003unequal}. IOM-concordant methods operationalize this distinction through standardization or reweighting procedures. While widely used in applied work, these approaches do not directly quantify pathway-specific contributions.

While useful for estimating conditional associations, regression-based approaches often mischaracterize mediators as confounders, obscuring the pathways through which disparities arise. Causal mediation analysis has been proposed as a remedy \citep{baron_moderator-mediator_1986, jackson_interpretation_2018, beydoun_racial_2016,ko_association_2020, dickman_trends_2022}, but traditional mediation frameworks typically partition disparities into a single direct and indirect effect. This structure is often too rigid to capture the influence of multiple, interacting mediators. Moreover, standard mediation methods often rest on strong parametric assumptions, such as linearity and additivity, that may bias results when relationships are complex or nonlinear \citep{shpitser13cogsci}. 

% % Race as a cause 
These limitations are especially pronounced when studying racial disparities. As a socially constructed and deeply embedded attribute, race cannot generally be manipulated like a conventional treatment, challenging its interpretation under counterfactual mediation frameworks that rely on hypothetical interventions \citep{holland_statistics_1986, vanderweele2013causal, vanderweele2016commentary}. There are, however, important experimental paradigms in which aspects of perceived race are manipulated, such as audit and vignette studies that vary names or visual cues \citealp{schulman1999effect, bertrand2004emily}. These designs provide valuable evidence on discrimination in specific contexts, but they do not correspond to interventions on race as a lived social identity shaped by structural and historical processes. Observed differences across racial groups reflect a confluence of historical exclusion, structural disadvantage, and lived social experience, not a single treatment effect. For this reason, efforts to estimate a total or mediated ``effect of race'' are often ill-defined and difficult to interpret \citep{vanderweele_causal_2012, vanderweele_causal_2014, howe_recommendations_2022}. Recent scholarship has instead shifted toward examining how disparities might be reduced by intervening on tangible, modifiable factors, such as insurance access, education, or health behaviors, while treating race as a structural index of social position shaped by structural racism, discrimination, and related social and economic inequalities \citep{vanderweele_causal_2014, jackson2018decomposition, howe_recommendations_2022}.

% % On our approach
In this study, we adopt that perspective. We develop a nonparametric framework that decomposes racial disparities in healthcare expenditures into components attributable to differences in the distributions of specific mediators and a residual component that remains after alignment. This approach avoids assumptions about race as a treatment and does not rely on parametric models or additive decompositions. Instead, we assess how much of the observed disparity can be attributed to unequal distributions of SES, insurance access, health behaviors, and health status. Our framework uses directed acyclic graphs \citep{pearl_causality_2009} to structure assumptions, and influence-function-based estimators to enable robust estimation with flexible machine learning models  \citep{van2000asymptotic, tsiatis2007semiparametric, van_der_laan_super_2007}. In addition, our framework is complementary to IOM-concordant approaches in that it provides a formal statistical decomposition of disparities into contributions from specific mediating variables, thereby further quantifying the extent of disparities that cannot be explained by patients' underlying health conditions or treatment preferences.

By quantifying how disparities shift under hypothetical alignments of mediator distributions, our analysis identifies policy-relevant pathways through which structural inequities in healthcare spending may be reduced. For example, a large disparity component associated with insurance access suggests that aligning insurance distributions across racial groups (through policies such as Medicaid expansion or premium subsidies) could substantially reduce spending gaps \citep{davis_achievements_1976}. Similarly, if the component attributed to SES is large, interventions aimed at improving education or economic opportunity may help narrow disparities. When health behaviors or health status account for substantial variation, public health efforts and chronic disease management become key targets. 

% % Identification and estimation of causal effects 
Beyond conceptual challenges, estimation of our defined disparity components presents several methodological challenges. Relationships between race, healthcare expenditures, and mediating factors are often complex and nonlinear, making model specification a key challenge. In addition, zero-inflation and right-skewness in expenditure data introduce further complications, requiring tailored statistical techniques. Existing estimation methods---including plug-in G-computation \citep{robins1986new, young_comparative_2011}, inverse odds ratio-weighted estimators \citep{tchetgen_tchetgen_inverse_2013}, inverse treatment probability-weighted estimators \citep{lange_simple_2012}, and regression-based imputation approaches \citep{vansteelandt_imputation_2012, zhou_tracing_2023}---are widely used but often prone to model misspecification. To mitigate these issues, we employ influence function-based estimators \citep{van2000asymptotic, miles_semiparametric_2020, diaz_nonparametric_2021, zhou_semiparametric_2022}, which improve robustness against model misspecification in parametric settings. A key advantage of these estimators, however, is their ability to accommodate data-adaptive statistical machine learning techniques, even when the underlying nuisance estimates converge at rates slower than parametric. Despite this flexibility, they still retain desirable frequentist properties, such as  root-n consistency and asymptotic normality, which are crucial for constructing confidence intervals and quantifying uncertainty \citep{double17chernozhukov}. In our estimation pipeline, we employ super learners, which aggregate multiple predictive models to improve robustness and estimation accuracy while leveraging these statistical guarantees \citep{polley_super_2010}. By integrating these tools into our estimation pipeline, we improve the reliability of disparity decompositions and provide a more nuanced understanding of the mechanisms contributing to racial differences in healthcare spending.

% % Contributions 
This study makes several contributions to the literature on racial disparities in healthcare expenditures. Our goal is not to identify the full set of mechanisms underlying racial and ethnic disparities, which develop over long historical time scales, but rather to characterize disparities and their mediating pathways at specific time points and to examine how these pathways evolve over time using repeated cross-sectional data.
First, we develop a framework that decomposes disparities into components attributable to differences in mediator distributions and components that remain after alignment. This approach moves beyond traditional regression methods by offering a more detailed accounting of the pathways through which disparities arise. 
Second, we advance estimation techniques by deriving asymptotically linear estimators based on influence function theory. We integrate data-adaptive machine learning methods, such as super learners, to enhance estimation precision, improve robustness against model misspecification, and effectively handle the complex data-generating mechanisms underlying healthcare expenditures. 
Third, we apply this framework to analyze key mediators---socioeconomic status, insurance access, health behaviors, and health status---using the 2009 and 2016 MEPS data, and to compare how the contributions of these mediators differ across the two time periods. Finally, we contribute the \href{https://github.com/xxou/flexPaths}{flexPaths} \textsf{R} package, which extends the methodological toolkit for mediator-focused disparity analysis and related applications.

% % Organization  
The remainder of this paper is structured as follows. Section~\ref{sec:MEPS} describes the MEPS dataset and analytic sample. 
Section~\ref{sec:method} introduces our analytical framework for decomposing racial disparities, including intuitive motivation, formal definitions of disparity components, and estimation procedures. Section~\ref{sec:simulation} presents simulation studies evaluating the finite-sample performance and inferential properties of the proposed estimators. Section~\ref{sec:data_analysis} describes the empirical implementation, including the two-part modeling strategy, and presents the main findings along with a summary of patterns across comparisons. Section~\ref{sec:discussion} discusses the results, limitations, and policy implications, and Section~\ref{sec:conclusion} concludes. Supplementary materials contain additional implementation details and all technical proofs.

%###################################### 
\section{MEPS data and sample description}
\label{sec:MEPS}
%###################################### 

The Medical Expenditures Panel Survey (MEPS) provides individual-level data on healthcare costs, utilization, and insurance coverage. We use the 2009 and 2016 MEPS household components, focusing on self-reported race and ethnicity. Racial and ethnic categories are constructed from MEPS variables following U.S. Census-aligned definitions \citep{census_racial_ethnic_diversity}, and we focus on individuals categorized as non-Hispanic White, non-Hispanic Black, non-Hispanic Asian, and Hispanic. For brevity, we often refer to these groups as White, Black, Asian, and Hispanic. The sample sizes are 20,789 in 2009 and 19,508 in 2016. These analytic samples were obtained after restricting to adults aged 18 years or older and individuals reporting a single race/ethnicity, and excluding observations with missing data on any variables included in the analysis (complete case analysis). Detailed information on sample construction, exclusion criteria, and missing data handling is provided in Appendix Section~\ref{app:meps_description}, including Table~\ref{tab:missing_summary} and Figure~\ref{fig:exclusion}.

MEPS collects demographic, socioeconomic, and health-related data. We consider \textit{baseline characteristics} (age, sex, geographic region, and marriage); \textit{family socioeconomic status} indicator (family poverty);  \textit{socioeconomic status} (SES) indicators (income, education, and employment); \textit{insurance access}, classifying individuals as uninsured if they lacked private or public health insurance; \textit{health behaviors} (smoking status and physical activity); and \textit{health status}, including BMI, self-reported physical and mental health, functional limitations, and chronic conditions such as diabetes, hypertension, and cancer. The primary outcome is total annual healthcare expenditures, the sum of direct payments for care, including out-of-pocket spending and payments from private insurance and government programs, excluding over-the-counter drugs. 

A detailed breakdown of these datasets, including variable definitions and sample characteristics, is provided in Tables~\ref{tab:description_09_16} and \ref{tab:expenditure09_16} in Appendix Section~\ref{app:meps_description}. Table~\ref{tab:description_09_16} summarizes demographic, socioeconomic, and health-related characteristics by racial group in both years. Across both periods, Whites had the highest median healthcare expenditures, while Hispanics had the lowest. Expenditures increased across all groups from 2009 to 2016, with Whites spending a median of \$1,675 in 2009 and \$2,094 in 2016. Table~\ref{tab:expenditure09_16} further examines expenditure disparities by race and other characteristics. Older adults, females, and those with higher SES and insurance coverage had significantly higher spending. Insured individuals spent nearly \$1,400 more than the uninsured. Conversely, those who exercised regularly or reported better health status had lower expenditures. These trends were consistent across racial groups.

%######################################
\section{Disparity definition, decomposition, and estimation}
\label{sec:method}
%######################################

%  ---  ---  ---  ---  ---  ---  ---  ---  ---  ---  ---  ---  --- --
\subsection{Definition and interpretation of disparity components}
% Path-specific effects as measures of disparity}
\label{subsec:method_def}
%  ---  ---  ---  ---  ---  ---  ---  ---  ---  ---  ---  ---  --- --

Let $R$ denote racial group membership, with $R = 0$ indicating a disadvantaged group and $R = 1$ indicating an advantaged group. Let $Y$ denote total annual healthcare expenditures. A simple and intuitive summary of racial disparities is the difference in group-level averages, $\E[Y \mid R = 1] - \E[Y \mid R = 0]$. However, this marginal contrast can be misleading. Baseline characteristics such as age, sex, geographic region, and marriage, denoted by $C$, often differ across racial groups and are also associated with healthcare expenditures. 
In addition, early-life socioeconomic conditions, such as family socioeconomic status and childhood environment, denoted by $C_H$, may differ across racial groups and influence later-life characteristics and outcomes. These associations do not arise from race itself as an inherent or manipulable causal attribute. Rather, they reflect the influence of broader structural and historical processes, denoted by $H$, including discrimination, segregation, and related policy environments, that shape both racial classification and socioeconomic conditions \citep{vanderweele_causal_2014, jackson2018decomposition, bailey_how_2021, howe_recommendations_2022}. In our empirical analysis, we approximate $C_H$ using family poverty status as a proxy for early-life socioeconomic conditions. We define $X = \{C, C_H\}$ as the set of observed covariates.

As a result, the unadjusted group difference may conflate disparities in outcomes with differences in observed covariate distributions arising from these shared structural determinants. To address this, we define the total racial disparity as a covariate-standardized difference in expected outcomes. 
Conceptually, this measure asks: what would be the difference in average expenditures between racial groups if they had the same distribution of observed covariate characteristics $X$? By standardizing over $X$, this approach removes differences attributable to variation in baseline characteristics, allowing the disparity to be interpreted as differences that persist after accounting for these factors. In contrast, the naive comparison $\E[Y \mid R=1] - \E[Y \mid R=0]$ reflects both differences in outcomes and differences in covariate distributions, and therefore does not isolate disparities from differences in baseline composition. 

\begin{definition}\label{def:total_disparity}
    The \textit{total racial disparity}, denoted by $\rho_\text{total}$, is defined as the difference in expected healthcare expenditures between racial groups, standardized over the distribution of observed covariates $X$. It is given by
    \begin{align}
        \rho_\text{total} = \int y \, \{dP(y \mid R=1, x) - dP(y \mid R=0, x)\} \, dP(x) \ . 
    \end{align}
\end{definition}

By aggregating conditional differences in outcomes across levels of $X$, weighted by the covariate distribution $P(X)$, $\rho_\text{total}$ compares outcome distributions under a common covariate distribution. This standardized estimand is particularly relevant for policy because it distinguishes disparities driven by differences in population composition from those that persist under a common distribution of baseline characteristics, thereby highlighting differences that are not explained by observed covariates.

If racial group membership were truly exogenous (as in Figure~\ref{fig:graphs}(a)), the unadjusted difference $\E[Y \mid R = 1] - \E[Y \mid R = 0]$ would coincide with this standardized measure. In practice, associations between $R$ and observed covariates $X = \{C, C_H\}$ arise from shared structural and historical processes, as illustrated in Figure~\ref{fig:graphs}(b) and are represented in the observed-data model by a bidirected arrow in Figure~\ref{fig:graphs}(c), making standardization essential.

We emphasize that in Figure~\ref{fig:graphs}, race is not interpreted as an inherent or manipulable causal attribute, but rather as an index of racialized social position. Accordingly, arrows emanating from $R$ should be understood as shorthand for differences associated with this social position, rather than as direct causal effects of race itself. Associations between $R$ and downstream variables reflect the influence of broader structural and historical processes collected in $H$, including discrimination, segregation, and related policy environments, which shape both racial classification and socioeconomic conditions. Because these upstream processes are not directly observed in MEPS, they are not explicitly modeled in the empirical analysis, but are instead represented conceptually through $H$ in the diagram.

\begin{figure}[t] 
	\begin{center}
    \scalebox{0.8}{
    \begin{tikzpicture}[>=stealth, node distance=1.65cm,
       mycircle/.style={draw, ellipse, align=center, minimum size=0.75cm}]
        \tikzstyle{format} = [thick, circle, minimum size=1.0mm, inner sep=2pt]
        \tikzstyle{square} = [draw, thick, minimum size=4.5mm, inner sep=2pt]
        
    \begin{scope}[xshift=0cm, yshift=0cm]
		\path[->, thick]
		
		node[] (a) {$R$}
		node[above right of=a, yshift=0.cm] (c) {$X$}
		node[right of=a, xshift=0.75cm] (y) {$Y$}
		
		(c) edge[black] (y) 
		(a) edge[black] (y)

        node[below of=a, xshift=1.25cm, yshift=1.0cm] (t1) {(a)} ;
		
	\end{scope}
    \begin{scope}[xshift=6.cm, yshift=0cm]
		\path[->, thick]
		
		node[] (a) {$R$}
        node[left of =a, xshift=0.cm, mycircle, dashed] (h) {$H$}
		node[above of=a, yshift=-0.75cm] (ch) {$C_H$}
        node[above of=a, yshift=0.cm] (c) {$C$}
		node[right of=a, xshift=0.5cm] (y) {$Y$}
		
		(h) edge[black] (a) 
        (h) edge[black] (ch) 
        (h) edge[black, <->] (c) 
		% (c) edge[black, <->] (a)
        (ch) edge[black] (y) 
		(c) edge[black] (y) 
		(a) edge[black] (y)

        node[below of=a, xshift=0.cm, yshift=1.cm] (t2) {(b)} ;
		
	\end{scope}
    \begin{scope}[xshift=10.cm, yshift=0cm]
		\path[->, thick]
		
		node[] (a) {$R$}
		node[above right of=a, yshift=0.cm] (x) {$X$}
		node[right of=a, xshift=0.75cm] (y) {$Y$}
		
		% (u) edge[gray, dashed] (a) 
		% (u) edge[gray, dashed] (x) 
		(x) edge[black, <->] (a) 
		(x) edge[black] (y) 
		(a) edge[black] (y)

        node[below of=a, xshift=1.25cm, yshift=1.cm] (t3) {(c)} ;
		
	\end{scope}
    \begin{scope}[xshift=14.cm, yshift=0cm]
		\path[->, thick]
		
		node[] (a) {$R$}
		node[right of=a, xshift=0.75cm] (m) {$M$}
		node[above left of=m, xshift=0cm, yshift=0.0cm] (x) {$X$}
		node[right of=m, xshift=0.75cm] (y) {$Y$}

		(x) edge[black, <->] (a) 
		(x) edge[black] (m) 
		(x) edge[black, bend right=0] (y) 
		(a) edge[black] (m) 
		(m) edge[black] (y) 
        (a) edge[black, out=15, in=135] (y)  

        node[below of=m, xshift=0cm, yshift=1.cm] (t4) {(d)} ;
		
	\end{scope}
	\end{tikzpicture}
	}
    % \vspace{-0.2cm}
	\caption{Conceptual diagrams illustrating assumed relationships among racial group membership $R$, outcome $Y$, baseline covariates $C$, early-life conditions $C_H$, observed covariates $X = \{C, C_H\}$, mediators $M$, and unobserved structural processes $H$.  
    (a) A simplified representation in which $R$ is associated with $Y$, and $X$ influences $Y$ but is not affected by $R$; 
    (b) A representation in which unobserved structural and historical processes $H$ influence both racial classification $R$ and early-life socioeconomic conditions $C_H$. The bidirected arrow between $H$ and $C$ represents marginal correlations; 
    (c) A reduced representation of (b) obtained by marginalizing over $H$ and grouping observed covariates as $X = \{C, C_H\}$. This motivates the need to standardize over $X$ when defining disparities, in order to separate differences in outcomes from differences in covariate distributions; and 
    (d) A mediation structure in which $M$ lies on pathways linking $R$ to $Y$, with $X$ influencing both $M$ and $Y$. This representation motivates the decomposition of total disparity into mediator-attributable components.}
	\label{fig:graphs}
	\end{center}
\end{figure}

We note that the total disparity $\rho_\text{total}$ in Definition~\ref{def:total_disparity} is descriptive. It captures structural inequities in healthcare spending that reflect racialized differences in social positioning, access to resources, and accumulated disadvantage. It is fully defined by the observed data distribution and does not rely on counterfactuals or interpret race as a manipulable exposure. In contrast to causal effect estimands, such as the average treatment effect, which are ill-defined when the exposure is non-manipulable, this formulation offers a meaningful and interpretable measure of disparity. However, as a summary measure, $\rho_\text{total}$ does not reveal how differences in mediating mechanisms contribute to unequal outcomes. To better understand the pathways through which disparity arises, we consider a decomposition based on a mediating variable $M$ (Figure~\ref{fig:graphs}(d)).

For simplicity, we begin with a single mediator and later generalize to settings with multiple mediators. As before, let $X$ denote observed covariates. Suppose that $M$ is a variable such as socioeconomic status that differs in distribution across racial groups and influences healthcare expenditures. We define the \textit{mediator-attributable disparity}, denoted by $\rho_{R \rightarrow M \rightarrow Y}$, as the component of $\rho_{\text{total}}$ that is explained by differences in the distribution of $M$ across racial groups, while the outcome-generating process remains as observed in the disadvantaged group: 
\begin{align*}
    \rho_{R \rightarrow M \rightarrow Y} &= \int y \, dP(y \mid R=0, m, x) \, \big\{ dP(m \mid R=1, x) - dP(m \mid R=0, x)  \big\} \, dP(x) \ . 
\end{align*} 
Conceptually, this quantity isolates the portion of the disparity attributable to differences in the distribution of $M$ by asking how outcomes would change if the disadvantaged group had the same distribution of $M$ as the advantaged group, while all other aspects of the data-generating process remain unchanged.

The mediator-attributable disparity quantifies how healthcare expenditures for the disadvantaged group ($R = 0$) would change if, within each level of covariates $X = x$, their distribution of $M$ were replaced by that of the advantaged group ($R = 1$), while keeping their outcome-generating process fixed. Here, ``holding the outcome-generating process fixed'' means that the conditional relationship between $Y$ and $(M, X)$ is evaluated using the distribution observed in the disadvantaged group, $P(y \, \mid \, R=0, m, x)$, so that any change in outcomes arises solely from shifting the distribution of $M$, rather than from changes in how $M$ or $X$ relate to $Y$. A portion of $\rho_{\text{total}}$ would remain even after this alignment, highlighting disparities not attributable to $M$. We refer to this as the \textit{residual disparity}, defined as $\rho_{\text{res}, R \rightarrow M \rightarrow Y} = \rho_{\text{total}} - \rho_{R \rightarrow M \rightarrow Y}$, which equals:
\begin{align*}
    \rho_{\text{res}, R \rightarrow M \rightarrow Y} &= \int y \, \big\{ dP(y \mid R=1, m, x) - dP(y \mid R=0, m, x) \big\} \, dP(m \mid R=1, x) \, dP(x) \ .
\end{align*} 

Both components have direct policy relevance. The mediator-attributable disparity highlights the extent to which racial disparities might be reduced through interventions that shift the distribution of $M$. The residual disparity captures structural inequities that would persist even after equalizing $M$, including the influence of unmeasured or downstream factors such as discrimination, bias, or other dimensions of social inequality not captured by the mediator. 

\begin{figure}[t]
\centering
\scalebox{0.8}{
\begin{tikzpicture}[
    node distance=1cm and 0.5cm,
    mynode/.style={draw, rectangle, align=center, minimum width=2cm, minimum height=1cm},
    mycircle/.style={draw, ellipse, align=center, minimum size=1cm},
    myoval/.style={draw, ellipse, align=center, minimum width=2.5cm, minimum height=1cm},
    myarrow/.style={->, >=Stealth, thick},
    mybiarrow/.style={<->, >=Stealth, thick},
    myunarrow/.style={-, >=Stealth, thick},
    mylabel/.style={text width=2cm, align=center},
    ]

    % Nodes
    \node[myoval] (race) {Race};
    
    % unobserved historical factor
     \node[mycircle, dashed, left=of race, xshift=-0.5cm] (H) {$H$};
    % big container box
      \node[draw, rectangle, minimum width=4cm, minimum height=5cm, 
      below=of race,  yshift=0.5cm, label=below:{\textbf{Covariates} ($X$)}] (covariates) {};
        % baseline covariates
      \node[mynode, below=of race ,  yshift=0.3cm,
      label=below:{\textbf{Baselines} ($C$)}] (baselines) {Age\\Sex\\Region\\Marriage};
      % family SES inside block
      \node[mynode, below=of baselines,  yshift=0.2cm , 
      label=below:{\textbf{Family SES} ($C_H$)}] (family) {Family poverty};

    \node[mynode, right=of race, xshift=1.8cm, yshift=0.0cm, label=above:{\textbf{SES} ($M_1$)}] (SES) {Income\\Education\\Employment};
    
    \node[mynode, below=of SES, xshift=1.8cm, yshift=-0.5cm,  label=below:{\textbf{Insurance} ($M_2$)}] (ins) {Insured};
    
    \node[mynode, right=of SES, xshift=2.8cm, yshift=0.cm, label=above:{\hspace{1.5cm}\textbf{ Health Behaviors} ($M_3$)}] (behavior) {Smoke\\Exercise};
    
    \node[mynode, below=of behavior, xshift=3.5cm, label=below:{\textbf{Health Status} ($M_4$)}] (status) {BMI\\Health perception\\Functional status\\Chronic conditions};
    
    \node[myoval, right=of behavior, xshift=3cm] (expenditure) {Expenditures};
    
    % Arrows
    \draw[myarrow] (race)--(SES);
    \draw[myarrow] (race)--(ins);
    \draw[myarrow] (race)--(status);
    \draw[myarrow] (race) to[bend right=-30] (behavior);
    \draw[myarrow] (race) to[bend left=30] (expenditure);

    \draw[myarrow] (SES) -- (ins);
    \draw[myarrow] (SES) -- (behavior);
    \draw[myarrow] (SES) -- (status);
    \draw[myarrow] (SES) to[bend right=-30] (expenditure);

    \draw[myarrow] (ins) -- (behavior);
    \draw[myarrow] (ins) -- (status);
    \draw[myarrow] (ins) to[out=20, in=185] (expenditure);
    
    \draw[myarrow] (behavior) -- (status);
    \draw[myarrow] (behavior) -- (expenditure);
    
    \draw[myarrow] (status) -- (expenditure);

    \draw[myarrow] (H) -- (race);
    \draw[myarrow] (H) to[bend right = 30] (family);
    \draw[mybiarrow] (H) to[bend right = 10] (baselines);
   % \draw[mybiarrow] (baselines) to[bend right = 60] (family);

    \draw[mybiarrow, black] (baselines) to[bend left=30] (race);
    \draw[myarrow] (covariates) -- (SES);
    \draw[myarrow] (covariates) -- (ins);
    \draw[myarrow] (covariates) -- (behavior);
    \draw[myarrow] (covariates) to[out=-31, in=205] (status);
    \draw[myarrow] (covariates) to[out=-28, in=185, looseness=1.5] (expenditure);

\end{tikzpicture}
}
\vspace{-1.5cm}
\caption{Graphical representation of the relationships among racial group membership, observed covariates, mediating factors, and healthcare expenditures, highlighting pathways via socioeconomic status ($M_1$), insurance access ($M_2$), health behaviors ($M_3$), and health status ($M_4$), as described in Section~\ref{sec:MEPS}. Observed covariates ($X$), including demographic characteristics ($C$) and early-life socioeconomic conditions ($C_H$), may influence both mediators and outcomes. Associations between race and downstream variables should be interpreted as reflecting differences associated with racialized social position shaped by broader structural and historical processes, denoted by $H$, rather than inherent causal effects of race itself. These upstream structural processes are not directly observed in MEPS and are therefore not separately identified in the empirical decomposition.}
\label{fig:race_dag}
\end{figure}

In real-world settings, racial disparities typically emerge through complex mechanisms involving multiple, interdependent mediators. A more granular decomposition is needed to understand how specific factors contribute to observed differences in outcomes. Building on the single-mediator framework, we now consider four sequentially ordered mediators: socioeconomic status ($M_1$), insurance access ($M_2$), health behaviors ($M_3$), and health status ($M_4$), as illustrated in Figure~\ref{fig:race_dag}. Each of these mediators may differ in distribution across racial groups and may affect healthcare expenditures either directly or indirectly through downstream pathways. 

Figure~\ref{fig:race_dag} provides a simplified representation of these relationships. As discussed above, racial group membership is not interpreted as an inherent or manipulable causal attribute, but rather as an index of racialized social position. Associations between race and mediators such as socioeconomic status, insurance, and health status reflect the influence of broader structural and historical processes, including discrimination, segregation, and related policy environments, which are not directly observed in the data.

The considered ordering of mediators reflects a conceptual framework in which upstream socioeconomic conditions shape access to insurance and health-related behaviors, which in turn influence health status and subsequent healthcare utilization. This ordering is motivated by prior empirical and theoretical work on the social determinants of health, as discussed in Appendix Section~\ref{app:order}. We emphasize that the assumed ordering is a modeling assumption that cannot be empirically verified from the observed data, and alternative orderings or feedback relationships may be plausible in practice. Because our proposed decomposition is defined with respect to this ordering, different specifications correspond to different pathway interpretations and may yield different estimates. To assess the sensitivity of our findings to this assumption, we consider a plausible alternative ordering and report the corresponding results in Appendix Section~\ref{app:order}. The substantive conclusions regarding the primary contributors to disparities remain broadly consistent across specifications.

We adopt a reference-based decomposition strategy in which the disadvantaged group ($R=0$) serves as the baseline for comparison. For each mediator, we isolate its contribution by asking how outcomes for the disadvantaged group would change if only the distribution of that mediator were shifted to match that of the advantaged group, while all other mediators and the outcome-generating process are held fixed at their levels under the disadvantaged group. Operationally, for a given mediator $M_k$, this involves replacing its conditional distribution $P(m_k \mid \cdot, R=0)$ with $P(m_k \mid \cdot, R=1)$, while leaving the distributions of all other mediators and the conditional outcome model $P(y \mid \cdot, R=0)$ unchanged. For example, if $M_k$ represents insurance access, this quantity captures how average expenditures among the disadvantaged group would change if they had the same distribution of insurance coverage as the advantaged group, while all other factors remain as observed for the disadvantaged group.

To formalize this decomposition and assess how much of the total disparity $\rho_\text{total}$ can be attributed to individual mediators, we define a family of disparity components indexed by $k = 1, \dots, 4$, corresponding to the mediators in the specified ordering.
% The \textit{$k$-th mediator-attributable disparity} captures the portion of $\rho_\text{total}$ explained by differences in the conditional distribution of $M_k$ across racial groups, given covariates and earlier mediators. To isolate the contribution of $M_k$, we hold the distributions of all other mediators and the outcome-generating process fixed at their levels under the disadvantaged group ($R = 0$). 
Let $\overline{M}_k = (M_1, \dots, M_k)$ denote the first $k$ mediators in the sequence and let $\overline{m}_k$ be a realization of these variables. For notational convenience, let $\overline{M}_0$ and $\overline{m}_0$ denote the empty set. Using this notation, we now formalize the contribution of each mediator to the total disparity under the reference-based decomposition described above. 

\begin{definition}
The \textit{$k$-th mediator-attributable disparity}, denoted by $\rho_{R \rightarrow M_k \leadsto Y}$, is the portion of the total disparity attributable to differences in the conditional distribution of the $k$-th mediator across racial groups. It is given by 
\begin{align}
    & \rho_{R \rightarrow M_k \leadsto Y}  
    = \label{eq:rho_mediator-attributable} \int \!\! y \, dP(y \! \mid \! \overline{m}_4, R=0,x) \, \Big\{dP(m_k \! \mid \! \overline{m}_{k-1}, R=1,x) - dP(m_k \! \mid \! \overline{m}_{k-1}, R=0,x) \Big\} \!  \notag \\
    & \hspace{5cm} \prod_{\substack{j=1 \\  j \neq k}}^4 \! dP(m_j \! \mid \! \overline{m}_{j-1}, R=0,x) \, dP(x) \, .  
\end{align}
\end{definition}%
\noindent 
The product over $j \neq k$ reflects that all other mediators are integrated over their distributions under the disadvantaged group.

This estimand captures the reduction in racial disparity that would result from shifting the distribution of $M_k$ for the disadvantaged group to match that of the advantaged group, conditional on covariates and earlier mediators. It isolates the contribution of $M_k$ to the total disparity without treating race as a manipulable cause. As such, $\rho_{R \rightarrow M_k \leadsto Y}$ provides an interpretable summary of the extent to which disparities are associated with differences in the distribution of $M_k$.

The portion of the disparity that remains after equalizing the distribution of $M_k$ reflects disparities not explained by that mediator and is captured by the following residual term: 

\begin{definition}
The \textit{residual disparity relative to the $k$-th mediator} is defined as $\rho_{\text{res}, R \rightarrow M_k \leadsto Y} =   \rho_\text{total} - \rho_{R \rightarrow M_k \leadsto Y}$. 
\end{definition}

Although we can compute disparity components attributable to each mediator individually, the total disparity $\rho_\text{total}$ is not equal to the sum of the four $\rho_{R \rightarrow M_k \leadsto Y}$ terms. Each component isolates the contribution of shifting the distribution of one mediator at a time, holding the other distributions fixed, and does not capture the combined impact of shifting all mediators' distributions simultaneously. To complement these component-wise contributions, we define an additional quantity that captures disparity in outcome expectations between racial groups when the full vector of mediators $(M_1, M_2, M_3, M_4)$ is held fixed at its distribution under the disadvantaged group ($R = 0$). 

\begin{definition}
The \textit{outcome-attributed disparity}, denoted by $\rho_{R \rightarrow Y}$, is defined as: 
\begin{align}
     & \rho_{R \rightarrow Y} {=} \int y  \left\{ dP(y \mid \overline{m}_4, R=1,x) {-} dP(y \mid \overline{m}_4, R=0,x) \right\} \notag \\
     & \hspace{5cm} \prod_{k=1}^4 dP(m_k \mid \overline{m}_{k-1},R=0,x)  dP(x) \ . 
     \label{eq:rho_unexplained}
\end{align}
\end{definition}

This estimand captures the portion of racial disparity that would persist even after equalizing the distributions of all observed mediators. It reflects differences in how identical mediator profiles are translated into outcomes across racial groups. Such disparities may arise from unmeasured mediators, differences in care quality, provider bias, or other structural forces that influence the outcome-generating process beyond what is captured by the included variables. While not directly intervenable through mediator-targeted policies, this quantity highlights the potential impact of systemic inequities in healthcare delivery and calls attention to the need for institutional reforms aimed at promoting fairness in clinical decision-making and care provision.

To quantify the disparity attributable to joint differences in mediator distributions, we define the following residual term:

\begin{definition}
The \textit{residual disparity relative to the outcome} is defined as $\rho_{\text{res}, R \rightarrow Y} = \rho_\text{total} - \rho_{R \rightarrow Y}$. 
\end{definition}

This quantity corresponds to the disparity reduction that would result from simultaneously shifting the distributions of all four mediators to those of the advantaged group. It coincides with the cumulative mediator-attributable disparity. 

While the decomposition of the total disparity into mediator/outcome-attributable and residual components can be interpreted through the lens of causal mediation (under certain identification assumptions; see \citealp{avin2005identifiability}), we emphasize that it does not rely on positing counterfactual interventions on race. Race is not a manipulable treatment in the conventional sense, but a socially constructed attribute shaped by historical, structural, and cultural forces that influence lived experience and access to resources \citep{vanderweele_causal_2012}. Rather than attempting to define or estimate the effect of race itself, we adopted a perspective that focuses on modifiable mediators. Building on work by VanderWeele and Robinson \citep{vanderweele_causal_2014} and Jackson and VanderWeele \citep{jackson2018decomposition}, our approach frames racial disparities as differences in outcome distributions that may be partially reduced through interventions on downstream mechanisms such as socioeconomic status, insurance access, health behaviors, or health status. By shifting attention from the causal status of race to the policy relevance of mediators, this framework enables empirical insights into the mechanisms that sustain health inequities and the levers through which they might be addressed. To clarify the interpretation of each component, we provide explicit descriptions of each disparity term. 

% SES-mediated disparity
$\rho_{R \rightarrow M_1 \leadsto Y}$ (SES-mediated disparity): 
This component reflects the extent to which differences in SES contribute to the total racial disparity. It quantifies the reduction in disparity that would occur if, within levels of covariates, SES for the disadvantaged group were equalized to that of the advantaged group, while downstream mediators (insurance access, health behaviors, and health status) evolve as observed. A large SES-mediated disparity suggests that addressing educational and economic barriers could meaningfully reduce inequities. 

% Insurance-mediated disparity 
$\rho_{R \rightarrow M_2 \leadsto Y}$ (Insurance-mediated disparity):
This component captures the portion of the disparity attributable to differences in insurance access. It measures the reduction in disparity that would result if, conditional on covariates and SES, insurance access for the disadvantaged group were shifted to match that of the advantaged group, while allowing health behaviors and health status respond as observed. A large disparity component through insurance access suggests that expanding coverage (e.g., via Medicaid) may help reduce inequities. 

% Health behavior-mediated disparity 
$\rho_{R \rightarrow M_3 \leadsto Y}$ (Health behavior-mediated disparity):
This component represents the portion of the disparity explained by differences in health behaviors. It quantifies the reduction in disparity that would follow from equalizing health behaviors across racial groups, conditional on covariates, SES, and insurance access, while allowing health status to evolve naturally. A large contribution through health behaviors suggests that promoting healthier behaviors may help reduce disparities.  

% Health status-mediated disparity
$\rho_{R \rightarrow M_4 \leadsto Y}$ (Health status-mediated disparity):
This component isolates the contribution of health status to the total disparity. It reflects the reduction in disparity that would occur if, for individuals with the same covariates and values of the first three mediators, health status were equalized between racial groups. A large contribution through health status suggests that improving chronic disease management and physical health may help reduce inequities. Note that, under this definition, the residual disparity relative to health status, $\rho_{\text{res}, R \rightarrow M_4 \leadsto Y}$, is conceptually related to IOM-concordant definitions of healthcare disparities, which characterize disparities as differences in treatment across racial or ethnic groups that are not explained by underlying health conditions or patient preferences \citep{nelson2002unequal, nelson2003unequal}. In particular, this residual captures the portion of the disparity that would remain if racial groups had the same distribution of health status, holding other mediators fixed \citep{mcguire2006implementing}.

% Outcome-attributed disparity
$\rho_{R \rightarrow Y}$ (Outcome-attributed disparity):
This component captures the portion of the disparity in healthcare expenditures that would persist if, for each level of covariates, the distributions of all four mediators (SES, insurance access, health behaviors, and health status) were set to those of the disadvantaged group ($R = 0$), but outcomes were generated under the advantaged group's outcome model ($R = 1$). It reflects differences in how identical mediator profiles are associated with outcomes across racial groups, potentially arising from unmeasured factors, provider bias, or structural inequities in care delivery. While not directly intervenable through mediator shifts, this quantity highlights disparities embedded in the outcome-generating process that are not accounted for by the observed mediators. 

Our definitions of disparity components follow a \textit{reference-zero} decomposition strategy, in which each mediator-attributable disparity is computed by shifting the distribution of one mediator at a time, setting it to the advantaged group's distribution ($R = 1$), while holding all other mediator distributions and the outcome mechanism fixed at their observed levels in the disadvantaged group ($R = 0$). This approach allows us to quantify how much disparity would be reduced under targeted interventions on specific mediators. As noted earlier, these components are not mutually exclusive and do not sum to the total disparity. Rather than decomposing the total disparity additively, we isolate the marginal contribution of each mediator relative to a shared reference distribution. For comparison, we also explore a sequential decomposition strategy, detailed in Appendix Section~\ref{app:pse_decomp}, in which disparities are allocated cumulatively as mediators are progressively equalized across groups \citep{daniel_causal_2015, steen_flexible_2017, zhou_semiparametric_2022}. 

While our framework does not define disparities through counterfactual interventions on race, the components introduced above correspond, under standard causal identification assumptions, to identifiable path-specific effects (PSEs) in a general causal setting \citep{avin2005identifiability}. In mediation analysis, PSEs isolate how a treatment influences an outcome through specific subsets of pathways in a causal graph. These may include direct effects as well as indirect pathways through mediators and their descendants. In the causal model corresponding to Figure~\ref{fig:race_dag}, the identification functional for the direct path $\{R \rightarrow Y\}$ coincides with our outcome-attributed disparity $\rho_{R \rightarrow Y}$. Similarly, each component $\rho_{R \rightarrow M_k \leadsto Y}$ corresponds to a PSE along the set of paths from $R$ through $M_k$ to $Y$: $\{R \rightarrow M_k \rightarrow Y\}$ and $\{R \rightarrow M_k \rightarrow \ldots \rightarrow Y\}$, denoted compactly as $\{R \rightarrow M_k \leadsto Y\}$. These path-specific effects follow the framework of Shpitser and Tchetgen Tchetgen \citep{shpitser2016causal}, which ensures identifiability under edge consistency and avoids issues such as the recanting witness problem. Formal definitions and identification assumptions are provided in Appendix Section~\ref{app:causal_pse}. 

%  ---  ---  ---  ---  ---  ---  ---  ---  ---  ---  ---  ---  --- --
\subsection{Estimation techniques and multiply robust estimators}
\label{subsec:method_est}
%  ---  ---  ---  ---  ---  ---  ---  ---  ---  ---  ---  ---  --- --

Estimating the disparity components defined in Section~\ref{subsec:method_def} is challenging for several reasons. First, the required nuisance functions, including outcome regressions and mediator models, may be complex and nonlinear, so simple parametric plug-in estimators can be sensitive to model misspecification. Second, when these nuisance functions are estimated using flexible machine learning methods, naive plug-in estimators may exhibit non-negligible first-order bias, which can invalidate standard root-$n$ inference. Third, the identification formulas involve conditional distributions of multiple ordered mediators, making direct estimation computationally burdensome, especially with mixed discrete and continuous mediators. In such settings, evaluation of the identifying functionals may require high-dimensional numerical integration or Monte Carlo approximation under a working model, which can be unstable or computationally intensive.  Finally, in our application, healthcare expenditures are zero-inflated and highly right-skewed, which further complicates estimation of the outcome regression. 

Our estimation strategy is designed to address these challenges. We use flexible regression methods to accommodate nonlinear relationships, and we construct one-step corrected estimators based on influence functions to reduce first-order bias and retain valid asymptotic inference under suitable convergence conditions. In addition, the influence-function representation allows us to avoid direct estimation of high-dimensional mediator densities and associated numerical integration. The trade-off is increased modeling and notational complexity, as the procedure requires estimation of several nuisance functions and relies on regularity and rate conditions for asymptotic guarantees, while avoiding the substantial computational burden of direct density estimation and numerical integration. 

To simplify the estimation discussion, we express the total, mediator-attributable, and unexplained disparities as: 
$\rho_\text{total} = \gamma_\text{adv} - \gamma_\text{dis}$,  $\rho_{R \rightarrow M_k \leadsto Y} = \gamma_{R \rightarrow M_k \leadsto Y}  - \gamma_\text{dis}$, and $\rho_{R \rightarrow  Y} = \gamma_{R \rightarrow Y}  - \gamma_\text{dis}$, where 
{\small
\begin{align}
\label{eq:pse_effs_ID}
    &\gamma_\text{adv} 
    = \int y \, dP(y \mid R=1,x) \, dP(x) \ , \quad \gamma_\text{dis} = \int y \, dP(y \mid R=0,x) \, dP(x)   \notag
    \\
    &\gamma_{R \rightarrow Y} 
    = \int y \,  dP(y \mid \overline{m}_4, R=1,x) \prod_{k=1}^4 dP(m_k \mid \overline{m}_{k-1},R=0,x) \, dP(x) 
     \\
    &\gamma_{R \rightarrow M_k \leadsto Y} 
    = \int y \, dP(y \mid \overline{m}_4, R=0,x) \, dP(m_k \mid \overline{m}_{k-1}, R=1,x) \!\! \prod_{\substack{j=1, \\ j \neq k}}^4 \!\! dP(m_j \mid \overline{m}_{j-1}, R=0,x) \, dP(x) \ . \notag
\end{align}
}

There is a substantial literature on robust and flexible estimation of covariate-adjusted functionals, such as $\gamma_\text{adv}, \gamma_\text{dis}$, within non/semiparametric models \citep{van2000asymptotic, bang05doubly, tsiatis2007semiparametric, van2011targeted, double17chernozhukov}. More recent work has extended these tools to estimands involving one or more mediators, such as $\gamma_{R \rightarrow Y}$ and $\gamma_{R \rightarrow M_k \leadsto Y}$ \citep{tchetgen_semiparametric_2012, miles_semiparametric_2020, benkeser_nonparametric_2021, zhou_semiparametric_2022}. Here, we develop one-step corrected plug-in estimators using nonparametric influence functions for the functionals in \eqref{eq:pse_effs_ID}. Our approach closely follows the estimation framework for the natural path-specific effects developed by \cite{zhou_semiparametric_2022}. 

Given $n$ i.i.d. observations $\{O_i = (Y_i, \overline{M}_{4, i}, R_i, X_i): i = 1, \ldots n\}$ drawn from distribution $P$, the parameters in \eqref{eq:pse_effs_ID} can in principle be estimated by plug-in substitution using estimates of the nuisance functions, including the outcome mean regression and conditional densities of the mediators, along with the empirical distribution of covariates $X$. 
As noted above, however, this approach is vulnerable to first-order bias and can be computationally demanding because it requires estimation of conditional distributions for mixed-type multivariate mediators. In what follows, we derive one-step corrected plug-in estimators designed to address these limitations. We particularly focus on estimation of $\gamma_{R \rightarrow Y}$ and $\gamma_{R \rightarrow M_k \leadsto Y}$, since $\gamma_\text{adv}$ and $\gamma_\text{dis}$ are standard covariate-adjusted functionals \citep{robins1986new, pearl_causality_2009}, and their estimation has been widely studies in prior work \citep{bang05doubly, tsiatis2007semiparametric, van_der_laan_super_2007, van2000asymptotic, double17chernozhukov}.   

To address the \textit{first issue} regarding first-order bias, we can analyze the stochastic properties of the plug-in estimator by utilizing a linear expansion. For an integrable function $f$ defined on the observed data $O$, let $P f \coloneqq \int f(o) dP(o)$ denote the expectation under the true distribution $P$, and let $P_n f \coloneqq \frac{1}{n} \sum_{i = 1}^n f(O_i)$ represent the empirical average based on the sample. The linear expansion of the plug-in estimator for parameter $\gamma$, denoted by $\gamma^\text{plug-in}(\hat{Q})$ (where $\hat{Q}$ is the collection of nuisance estimates)  is given by: $\gamma^{\text{plug-in}}(\hat{Q}) = \gamma(Q) - P\Phi(\hat{Q}) + R_2(\hat{Q}, Q),$ where $\Phi$ denotes the gradient (or influence function) of the parameter, and $ R_2(\hat{Q}, Q)$ denotes the remainder terms of second and higher orders from the linear approximation. The term $-P\Phi(\hat{Q})$ is the plug-in's first-order bias, due to substituting $\hat{Q}$ for the true nuisance parameters in $\Phi(Q)$. Although $\Phi$ has zero expectation under $P$ (i.e., $P\Phi = 0$), this bias  may still be significant. By deriving the nonparametric influence functions for the counterfactual means, we apply a one-step correction that debiases the plug-in estimator by adjusting for an estimate of its first-order bias (i.e., $-P_n\Phi(\hat{Q})$), yielding the estimator $\gamma^{+}(\hat{Q}) = \gamma^{\text{plug-in}}(\hat{Q}) + P_n\Phi(\hat{Q})$ \citep{bickel1993efficient, van2000asymptotic, double17chernozhukov}.   

We note that an influence function provides a first-order approximation to how the target parameter changes under small perturbations of the data-generating distribution. It plays a central role in semiparametric estimation because it characterizes both the bias of plug-in estimators and the form of efficient estimators \citep{bickel1993efficient, van2000asymptotic, double17chernozhukov}.

To address the \textit{second issue} regarding density estimation and numerical integration, we parameterize the nonparametric influence functions to bypass these tasks. To simplify notation, we set $(r_0, r_1, r_2, r_3, r_4) = (1, 0, 0, 0, 0)$ when estimating $\gamma_{R \rightarrow Y}$, and $(r_0, r_1, r_2, r_3, r_4) = (0, {\bf 1_k})$ when estimating $\gamma_{R \rightarrow M_k \leadsto Y}$, where ${\bf 1_k}$ denotes an indicator  vector of length four with 1 in the $k$-th position and 0s elsewhere. We rely on the following key nuisance functional components: 
(i) the propensity score $P(R=1 \mid X)$, denoted as $\pi(X)$; 
(ii) the binary regressions $P(R=1 \mid \overline{M}_k, X)$ denoted as $g_{k}(\overline{M}_k, X)$; 
(iii) the outcome regressions $\E[Y \mid \overline{M}_k, r_0, X]$ denoted as $\mu_k(\overline{M}_k, r_0, X)$;  
(iv) the sequential regressions $\mathscr{B}_{k}(\overline{M}_{k-1}, r_{k}, X) = \E[\mu_k(\overline{M}_{k}, r_0, X) \mid \overline{M}_{k-1}, r_{k}, X]$, $\mathscr{C}_{\mathscr{B}_k}(r_1,X) = \E[\mathscr{B}_k(\overline{M}_{k-1}, r_{k}, X)\mid r_1,X]$, and $\mathscr{C}_{\mu_4}(r_1,X) = \E[\mu_4(\overline{M}_{4}, r_{0}, X)\mid r_1,X]$;  
and (v) the marginal distribution of covariates, $P_{X}$.  
Let $Q = \{\pi, \{g_k, \mu_k, \mathscr{B}_k, \mathscr{C}_{\mathscr{B}_k} : \forall k \}, \mathscr{C}_{\mu_4} \}$ collect all the nuisances. The influence functions for $\gamma_{R \rightarrow Y}$ and $\gamma_{R \rightarrow M_k \leadsto Y}$, denoted by $\Phi_{R \rightarrow Y}(Q)$ and $\Phi_{R \rightarrow M_k \leadsto Y}(Q)$, respectively, are given as follows. 
Each influence function below can be interpreted as a sum of components that adjust for different sources of bias arising from nuisance function estimation. Broadly, these terms combine outcome regression components, inverse probability weighting terms based on propensity scores and mediator models, and augmentation terms that ensure robustness to misspecification of certain nuisance functions. Together, these components yield estimators with desirable robustness and efficiency properties.
{\small
\begin{align}
   &\Phi_{R \rightarrow Y}(Q)(O_i)  \label{eq:eif_direct} \\
  &\hspace{0.5cm} =  
   \frac{R_i}{1-\pi(X_i)} \ \frac{1-g_{4}(\overline{M}_{4, i}, X_i)}{g_{4}(\overline{M}_{4, i},X_i)} \ \left\{Y_i - \mu_4(\overline{M}_{4, i},R=1,X_i)\right\}  \notag \\
   &\hspace{1.05cm} + \frac{1-R_i}{1-\pi(X_i)}\left\{ \mu_4(\overline{M}_{4, i},R=1,X_i) - \mathscr{C}_{\mu_4}(R=0,X_i) \right\} 
   + \mathscr{C}_{\mu_4}(R=0,X_i) - \gamma_{R \rightarrow Y} \ , 
    \notag   
    \\
    &\Phi_{R \rightarrow M_k \leadsto Y}(Q)(O_i)  \label{eq:eif_m4}  \\
    &\hspace{0.5cm}= \frac{1-R_i}{1-\pi(X_i)} \ \frac{g_{k}(\overline{M}_{k, i}, X_i)}{1-g_{k}(\overline{M}_{k, i},X_i)} \ \frac{1-g_{k-1}(\overline{M}_{k-1,i},X_i)}{g_{k-1}(\overline{M}_{k-1,i}, X_i)} \left\{Y_i - \mu_k(\overline{M}_{k, i},R=0,X_i)\right\} 
     \notag \\
    &\hspace{1.05cm} + \frac{R_i}{1-\pi(X_i)}\ \frac{1-g_{k-1}(\overline{M}_{k-1, i},X_i)}{g_{k-1}(\overline{M}_{k-1, i},X_i)} \left\{\mu_k(\overline{M}_{k, i},R=0,X_i) - \mathscr{B}_k(\overline{M}_{k-1, i},R=1,X_i)\right\}
    \notag \\
    &\hspace{1.05cm} + \frac{1-R_i}{1-\pi(X_i)}\left\{ \mathscr{B}_k(\overline{M}_{k-1, i},R=1,X_i) - \mathscr{C}_{\mathscr{B}_k}(r_1,X_i) \right\} 
    + \mathscr{C}_{\mathscr{B}_k}(r_1,X_i)- \gamma_{R \rightarrow M_k \leadsto Y} \ . \notag
    % \ , \ k=1, 2, 3, 4 \ .  
\end{align}
}
See detailed derivations in Appendix Section~\ref{app:proofs_est}. 

Although the expressions in \eqref{eq:eif_direct} and \eqref{eq:eif_m4} are algebraically complex, their structure follows a common pattern. Each influence function contains:
(i) a weighted residual term, which captures discrepancies between observed outcomes and predicted values from the outcome regression;
(ii) augmentation terms that adjust for differences between intermediate regression functions; and
(iii) a centering term that ensures the overall expression has mean zero at the true distribution.
The weighted residual components involve inverse probability weights constructed from the propensity score and mediator models. The augmentation terms serve to reduce bias when certain nuisance functions are misspecified. As a result, the estimator exhibits a multiply robust structure as formalized in Corollary~\ref{cor:robustness}.

Given the observed sample, we can use flexible statistical and machine learning models to estimate regressions ${\pi}, {g}_k, {\mu_k}$, while $\mathscr{B}_k, \mathscr{C}_{\mathscr{B}_k}, \mathscr{C}_{\mu_4}$ can be estimated via a sequential regression scheme. Estimation of $\mathscr{B}_k$ involves constructing a pseudo-outcome variable $\hat{\mu}_k(\overline{M}_{k, i}, r_0,X_i)$, setting $R_i = r_0$ for all observations. This pseudo-outcome is then regressed on $\overline{M}_{k-1}, X$ using only data points where $R_{i} = r_k$, yielding estimate $\hat{\mathscr{B}}_k$. Estimation of $\mathscr{C}_{\mathscr{B}_k}$ involves constructing a pseudo-outcome variable $\hat{\mathscr{B}}_k(\overline{M}_{k-1, i}, r_k,X_i)$, setting $R_i = r_k$ for all observations. This pseudo-outcome is then regressed on $X$ using only data points where $R_{i} = r_1$, yielding estimate $\hat{\mathscr{C}}_{\mathscr{B}_k}$. Finally, $\mathscr{C}_{\mu_4}$ can be estimated via first constructing the  pseudo-outcome variable $\hat{\mu}_4(\overline{M}_{4, i}, r_0,X_i)$, setting $R_i = r_0$ for all observations, and then regressing this pseudo-outcome  on $X$ using only data points where $R_{i} = r_1$, yielding estimate $\hat{\mathscr{C}}_{\mu_4}$. Let $\hat{Q}$ collect the nuisance estimates. Our one-step estimators of $\gamma_{R \rightarrow Y}$ and $\gamma_{R \rightarrow M_k \leadsto Y}$, defined in \eqref{eq:pse_effs_ID}, are given as follows:
{\small
\begin{align}
   &\gamma^{+}_{R \rightarrow Y}(\hat{Q}) 
   = \frac{1}{n}\sum_{i = 1}^n \Big\{ 
   \frac{R_i}{1-\hat{\pi}(X_i)} \ \frac{1-\hat{g}_{4}(\overline{M}_{4, i}, X_i)}{\hat{g}_{4}(\overline{M}_{4, i},X_i)} \ \left\{Y_i - \hat{\mu}_4(\overline{M}_{4, i},R=1,X_i)\right\}  
   \notag \\
   &\hspace{2.15cm} + \frac{1-R_i}{1-\hat{\pi}(X_i)}\big\{ \hat{\mu}_4(\overline{M}_{4, i},R=1,X_i) - \hat{\mathscr{C}}_{\mu_4}(R=0,X_i) \big\} 
    + \hat{\mathscr{C}}_{\mu_4}(R=0,X_i)\Big\} \ , \label{eq:pse_effs_est_direct}  
    \\
    &\gamma^{+}_{R \rightarrow M_k \leadsto Y}(\hat{Q}) \notag \\
    &\hspace{0.5cm}= \frac{1}{n}\sum_{i = 1}^n \Big\{ \frac{1-R_i}{1-\hat{\pi}(X_i)} \ \frac{\hat{g}_{k}(\overline{M}_{k, i}, X_i)}{1-\hat{g}_{k}(\overline{M}_{k, i},X_i)} \ \frac{1-\hat{g}_{k-1}(\overline{M}_{k-1,i},X_i)}{\hat{g}_{k-1}(\overline{M}_{k-1,i}, X_i)} \left\{Y_i - \hat{\mu}_k(\overline{M}_{k, i},R=0,X_i)\right\} 
     \notag \\
    &\hspace{1.15cm} + \frac{R_i}{1-\hat{\pi}(X_i)}\ \frac{1-\hat{g}_{k-1}(\overline{M}_{k-1, i},X_i)}{\hat{g}_{k-1}(\overline{M}_{k-1, i},X_i)} \big\{\hat{\mu}_k(\overline{M}_{k, i},R=0,X_i) - \hat{\mathscr{B}}_k(\overline{M}_{k-1, i},R=1,X_i)\big\}
    \notag \\
    &\hspace{1.15cm} + \frac{1-R_i}{1-\hat{\pi}(X_i)}\big\{ \hat{\mathscr{B}}_k(\overline{M}_{k-1, i},R=1,X_i) - \hat{\mathscr{C}}_{\mathscr{B}_k}(r_1,X_i) \big\} 
    + \hat{\mathscr{C}}_{\mathscr{B}_k}(r_1,X_i)\Big\} \ .  
    % \ , \ k=1, 2, 3, 4 \ .  
    \label{eq:pse_effs_est_m4} 
\end{align}
}

Let $\gamma^{+}(\hat{Q})$ denote either $\gamma^{+}_{R \rightarrow Y}(\hat{Q})$ in \eqref{eq:pse_effs_est_direct} or $\gamma^{+}_{R \rightarrow M_k \leadsto Y}(\hat{Q})$ in \eqref{eq:pse_effs_est_m4}. Asymptotic properties of $\gamma^{+}(\hat{Q})$ can be established through analyzing a linear expansion: $\gamma^{+}(\hat{Q}) - \gamma(Q) =  P_n(\Phi(Q)) + (P_n - P)(\Phi(\hat{Q}) - \Phi(Q)) + R_2(\hat{Q},Q)$. 
The term $P_n(\Phi(Q))$ is $O_P(n^{-1/2})$ (under central limit theorem), and the term $ (P_n - P)(\Phi(\hat{Q}) - \Phi(Q))$ is $o_P(n^{-1/2})$ (under regularity conditions detailed in Appendix Section~\ref{app:proofs_inf}). Thus, $\gamma^{+}(\hat{Q})$ is asymptotically linear if $R_2(\hat{Q}, Q) = o_P(n^{-1/2})$. The following theorem formally states sufficient requirements for the one-step corrected plug-in estimators to be asymptotically linear. Detailed derivations of the remainder terms are provided in Appendix Section~\ref{app:proofs_inf}.
\begin{theorem}
\label{thm:asymp_lin}
    Assume the  following $L^2(P)$ convergence rates for the nuisance estimates:
    $\lVert \hat{\pi} - \pi \rVert = o_P(n^{-\frac{1}{a}})$, 
    $\lVert \hat{g}_{k} - g_{k} \rVert = o_P(n^{-\frac{1}{b_k}})$, 
    $\lVert \hat{\mathscr{C}}_{\mu_4} - \mathscr{C}_{\mu_4} \rVert = o_P(n^{-\frac{1}{c}})$, 
    $\lVert \hat{\mathscr{C}}_{\mathscr{B}_k} - \mathscr{C}_{\mathscr{B}_k} \rVert = o_P(n^{-\frac{1}{d_k}})$, 
    $\lVert \hat{\mathscr{B}}_k - \mathscr{B}_k \rVert = o_P(n^{-\frac{1}{l_k}})$, 
    $\lVert \hat{\mu}_k - \mu_k \rVert = o_P(n^{-\frac{1}{m_k}})$ for $k=1,2,3,4$. 
    Under regularity conditions detailed in Appendix Section~\ref{app:proofs_inf}, 
    \begin{enumerate}
        \item \, if $\frac{1}{a}+\frac{1}{c} \geq \frac{1}{2}$ and $\frac{1}{b_4}+\frac{1}{m_4} \geq \frac{1}{2}$, then 
        $\sqrt{n}\big(\gamma^{+}_{R \rightarrow Y}(\hat{Q}) - \gamma_{R \rightarrow Y}(Q)\big)$ is asymptotically normal with variance equal to $\E[\Phi_{R \rightarrow Y}^2(Q)]$; 

        %\vspace{0.15cm} 
        \item \, if $\frac{1}{a}+\frac{1}{d_k} \geq \frac{1}{2}$, $\frac{1}{b_{k-1}}+\frac{1}{l_k} \geq \frac{1}{2}$ and $\frac{1}{b_k}+\frac{1}{m_k} \geq \frac{1}{2}$, $k=1,2,3,4$, then 
        $\sqrt{n}\big(\gamma^{+}_{R \rightarrow M_k \leadsto Y}(\hat{Q}) - \gamma_{R \rightarrow M_k \leadsto Y}(Q)\big)$ is asymptotically normal with variance equal to $\E[\Phi_{R \rightarrow M_k \leadsto Y}^2(Q)]$. 
    \end{enumerate}
\end{theorem}%
See a proof in Appendix Section~\ref{app:proofs_inf}. Given that $\pi \equiv g_{0}$, $\mathscr{B}_1 \equiv \mathscr{C}_{\mathscr{B}_1}$, we have $a = b_0$ and $d_1=l_1$. 

The convergence rate conditions in Theorem~\ref{thm:asymp_lin} can be interpreted as requiring that certain pairs of nuisance functions are estimated with sufficient accuracy so that their combined estimation error is negligible at the $n^{-1/2}$ scale. These conditions are weaker than requiring all nuisance functions to be estimated at root-$n$ rates, which would be unrealistic in high-dimensional settings. For example, the condition $\frac{1}{a} + \frac{1}{c} \geq \frac{1}{2}$ implies that the product of the estimation errors for $\pi$ and $\mathscr{C}_{\mu_4}$ converges faster than $n^{-1/2}$. This allows one nuisance function to be estimated relatively slowly, provided the other is estimated more accurately. 

Concretely, these conditions are satisfied by many combinations of convergence rates. For instance, if both nuisance functions converge at rate $n^{-1/4}$, then $\frac{1}{4} + \frac{1}{4} = \frac{1}{2}$, satisfying the requirement. Similarly, if one nuisance is estimated at rate $n^{-1/3}$ and another at $n^{-1/6}$, the condition is also satisfied. In contrast, if both nuisance functions converge too slowly, for example at rates slower than $n^{-1/4}$, the condition may fail, and the estimator may no longer achieve root-$n$ consistency. 

These types of rate requirements are standard in semiparametric estimation with machine learning, and are often achievable in practice using flexible methods such as random forests, gradient boosting, or ensemble learners, particularly when combined with cross-fitting \citep{double17chernozhukov}. In our application, the relatively large sample sizes and use of super learner facilitate estimation of nuisance functions at rates that are plausibly sufficient for these conditions to hold, although the rates themselves are not directly verifiable from the data. 

If the rate conditions are violated, the one-step estimators may still be consistent, but their asymptotic normality and associated confidence intervals may no longer be valid. In such cases, inference based on the influence function may be unreliable, highlighting the importance of flexible and well-performing nuisance estimation in practice. 

The $L^2(P)$ convergence assumptions in Theorem~\ref{thm:asymp_lin} establish that $R_2(\hat{Q}) = o_P(n^{-1/2})$, even when flexible models with slower convergence rates than $n^{-1/2}$ are used for nuisance functional estimations. Moreover, Theorem~\ref{thm:asymp_lin} implies certain robustness behaviors for consistency of $\gamma^{+}(\hat{Q})$, formalized in the following corollary.

\begin{corollary}
\label{cor:robustness}
    Under regularity conditions detailed in Appendix Section~\ref{app:proofs_inf}, the one-step estimators in \eqref{eq:pse_effs_est_direct} and \eqref{eq:pse_effs_est_m4} are consistent if at least one of the following sets of nuisance estimates is consistent:
    \begin{enumerate}
        \item \, For $\gamma^{+}_{R \rightarrow Y}(\hat{Q})$: if either (i) $\hat{\pi}$ and $\hat{g}_4$; (ii) $\hat{\pi}$ and $\hat{\mu}_4$; or (iii) $\hat{\mathscr{C}}_{\mu_4}$ and $\hat{\mu}_4$, are consistently estimated. 

        \item \, For $\gamma^{+}_{R \rightarrow M_k \leadsto Y}(\hat{Q})$, $k=1,2,3,4$: if either (i) $\hat{\pi}$, $\hat{g}_{k-1}$, and $\hat{g}_k$; (ii) $\hat{\pi}$, $\hat{g}_{k-1}$, and $\hat{\mu}_k$; (iii) $\hat{\pi}$, $\hat{\mathscr{B}}_k$, and $\hat{\mu}_k$; or (iv) $\hat{\mathscr{C}}_{\mathscr{B}_k}$, $\hat{\mathscr{B}}_k$, and $\hat{\mu}_k$, are consistently estimated. 
    \end{enumerate}
\end{corollary}

See a proof in Appendix Section~\ref{app:proofs_inf}.

Given that $\pi \equiv g_{0}$ and $\mathscr{B}_1 \equiv \mathscr{C}_{\mathscr{B}_1}$, when $k=1$, the third set of nuisance estimates for consistency of $\gamma^{+}_{R \rightarrow M_k \leadsto Y}(\hat{Q})$ is a superset of the fourth condition, making it redundant. Corollary \ref{cor:robustness} suggests that $\gamma^{+}(\hat{Q})$ can achieve consistency even if certain parts of the underlying observed joint distribution are misspecified.  

Corollary~\ref{cor:robustness} establishes a multiply robust structure for the proposed estimators. Unlike standard doubly robust estimators, which remain consistent if one of two nuisance components is correctly specified, our estimators remain consistent under multiple distinct combinations of correctly specified nuisance functions. Intuitively, this robustness arises because the influence function representation combines several augmentation terms, each of which can compensate for misspecification in other components. As a result, errors in certain nuisance models can be offset by correct specification of others, yielding multiple combinations of nuisance functions under which consistency is achieved. Table~\ref{tab:misspec_combined} summarizes the different robustness pathways for each estimand by enumerating combinations of correctly specified and misspecified nuisance functions. This representation makes explicit the multiple routes through which consistency can be achieved and facilitates comparison across estimands.

From a practical perspective, this property is particularly valuable in complex settings with multiple mediators, where it is unrealistic to assume that all nuisance models are correctly specified. The multiply robust structure provides protection against model misspecification, ensuring that valid estimation can still be achieved even when some components are incorrectly modeled. This extends standard doubly robust approaches, which typically rely on correct specification of either a propensity model or an outcome regression, by allowing for a richer set of valid specification combinations involving mediator and sequential regression components.

% \ding{51}  % ✓ check mark
% \ding{55}   % ✗ cross mark
\begin{table}[!th]
    \centering
    \caption{Misspecification scenarios for nuisance functions and resulting consistency conditions. A check mark (\ding{51}) indicates correct specification, while a cross (\ding{55}) denotes misspecification. Each column represents a distinct combination of nuisance functions under which the estimator remains consistent, illustrating the multiply robust structure of the proposed estimators. Unlike standard doubly robust methods, which rely on a single pair of correctly specified models, these estimators admit multiple combinations of nuisance functions that ensure consistency.}
    \label{tab:misspec_combined}

    \begin{subtable}[t]{\linewidth}
        \centering
        \caption{$\gamma^+_{R \rightarrow Y}$}
        \label{tab:misspec_sim2_a}
        \begin{tabular*}{\linewidth}{@{\extracolsep{\fill}}c c c c c}
            \toprule
            Function & Condition 1 & Condition 2 & Condition 3 & Correct \\
            \midrule
            $\pi$                 & \ding{51} & \ding{51} & \ding{55} & \ding{51}  \\
            $g_4$                 & \ding{51} & \ding{55} & \ding{55} & \ding{51}  \\
            $\mu_4$               & \ding{55} & \ding{51} & \ding{51} & \ding{51}  \\
            $\mathcal{C}_{\mu_4}$ & \ding{55} & \ding{55} & \ding{51} & \ding{51}  \\
            \bottomrule
        \end{tabular*}
    \end{subtable}
    
    \vspace{1.5em}
    \begin{subtable}[t]{\linewidth}
        \centering
        \caption{$\gamma^+_{R \rightarrow M_1 \leadsto Y}$}
        \label{tab:misspec_sim2_b}
        \begin{tabular*}{\linewidth}{@{\extracolsep{\fill}}c c c c c}
            \toprule
            Function & Condition 1 & Condition 2 & Condition 3 & Correct \\
            \midrule
            $\pi$           & \ding{51} & \ding{51} & \ding{55} & \ding{51} \\
            $g_1$           & \ding{51} & \ding{55} & \ding{55} & \ding{51} \\
            $\mu_1$         & \ding{55} & \ding{51} & \ding{51} & \ding{51} \\
            $\mathcal{B}_1$ & \ding{55} & \ding{55} & \ding{51} & \ding{51} \\
            \bottomrule
        \end{tabular*}
    \end{subtable}
    
    \vspace{1.5em}
    \begin{subtable}[t]{\linewidth}
        \centering
        \caption{$\gamma^+_{R \rightarrow M_k \leadsto Y}\ (k=2,3,4)$}
        \label{tab:misspec_sim2_c}
        \begin{tabular*}{\linewidth}{@{\extracolsep{\fill}}c c c c c c}
            \toprule
            Function & Condition 1 & Condition 2 & Condition 3 & Condition 4 & Correct \\
            \midrule
            $\pi$                         & \ding{51} & \ding{51} & \ding{51} & \ding{55} & \ding{51} \\
            $g_{k-1}$                     & \ding{51} & \ding{51} & \ding{55} & \ding{55} & \ding{51} \\
            $g_k$                         & \ding{51} & \ding{55} & \ding{55} & \ding{55} & \ding{51} \\
            $\mu_k$                       & \ding{55} & \ding{51} & \ding{51} & \ding{51} & \ding{51} \\
            $\mathcal{B}_k$               & \ding{55} & \ding{55} & \ding{51} & \ding{51} & \ding{51} \\
            $\mathcal{C}_{\mathcal{B}_k}$ & \ding{55} & \ding{55} & \ding{55} & \ding{51} & \ding{51} \\
            \bottomrule
        \end{tabular*}
    \end{subtable}
\end{table}

One-step corrected plug-in estimates of $\rho_{R \rightarrow M_k \leadsto Y}$ and $\rho_{R \rightarrow Y}$, defined in \eqref{eq:rho_mediator-attributable} and \eqref{eq:rho_unexplained}, can be obtained via one-step corrected plug-in estimates of $\gamma_{R \rightarrow Y}$, $\gamma_{R \rightarrow M_k \leadsto Y}$, and $\gamma_\text{dis}$. Such an estimator for $\gamma_\text{dis}$ is known as the \textit{augmented inverse probability weighted} estimator, which we denote by $\gamma^{+}_\text{dis}(\hat{Q})$, where $\hat{Q}$ is a slight abuse of notation that refers to estimates of the propensity score and the outcome regression \citep{robins1994estimation}. Thus, we can write: 
\begin{align}
    \rho^{+}_{R \rightarrow Y}(\hat{Q}) &=  \gamma^{+}_{R \rightarrow Y}(\hat{Q}) - \gamma^{+}_\text{dis}(\hat{Q}) \ , \quad 
    \rho^{+}_{R \rightarrow M_k \leadsto Y}(\hat{Q}) = \gamma^{+}_{R \rightarrow M_k \leadsto Y}(\hat{Q}) - \gamma^{+}_\text{dis}(\hat{Q}) \ . \label{equ:rho_define}
\end{align}

%####################################################
\section{Simulation studies}
\label{sec:simulation}
%####################################################

We evaluate the finite-sample behavior and robustness of our proposed estimators, described in Section~\ref{subsec:method_est}, through two sets of simulation studies. The first study mimics the structure of our real-data application to assess whether the estimators attain their theoretical properties in moderate samples. We compare finite-sample performance, robustness, and empirical confidence interval coverage under both super learner-based and generalized linear models (GLMs)-based nuisance estimation, highlighting the practical motivation for adopting super learner in the real-data analysis in Section~\ref{sec:data_analysis}.  The second study examines the robustness of the estimators under model misspecification. For both studies, we generate data sets of sizes 250, 500, 1000, 2000, 4000, and 8000, with 1000 replications per sample size. 

\vspace{0.2cm}
%  ---  ---  ---  ---  ---  ---  ---  ---  ---  ---  ---  ---  ---  ---  --- 
{\bf Simulation 1: Finite sample performance and theoretical guarantees.}
% \label{subsec:simu_performance}
%  ---  ---  ---  ---  ---  ---  ---  ---  ---  ---  ---  ---  ---  ---  --- 

Here, we evaluate the finite-sample performance and root-$n$ consistency of our estimators, as established in Theorem~\ref{thm:asymp_lin}, using both super learners and GLMs for nuisance function estimation. We generated data with three covariates, one binary treatment, four ordered mediators, one univariate ($M_2$) and three multivariate ($M_1$, $M_3$, and $M_4$), and a zero-inflated, right-skewed outcome, incorporating nonlinearities. See Appendix Section~\ref{app:simu_meps} for the detailed data generation process.  

We first compute the true parameter values and corresponding variances by generating a large data set and deriving the true forms of the density ratios and sequential regressions, leveraging knowledge of the ground truth. To evaluate our estimators, we fit all the nuisance functions using two approaches: a flexible super learner ensemble, including the same candidate learners used in the empirical analysis (\texttt{mean}, \texttt{glm}, \texttt{glm.interaction},  \texttt{gam}, \texttt{glmnet}, \texttt{earth}, \texttt{ksvm}, \texttt{xgboost},  \texttt{randomForest},  \texttt{dbarts}), and a GLM without interactions or higher-order terms. 

We assess the finite-sample performance of the estimators based on bias, standard deviation (SD), mean squared error (MSE), 95\% confidence interval (CI) coverage, and average CI width. Table~\ref{tab:simu_meps} in Appendix Section~\ref{app:simu_meps} summarizes these results, showing that the super learner approach achieves low bias, reduced SD and MSE, and reliable coverage, whereas the GLM-based estimators exhibit substantial bias. 

\begin{figure}[!t]
    \centering
    \includegraphics[width=15cm]{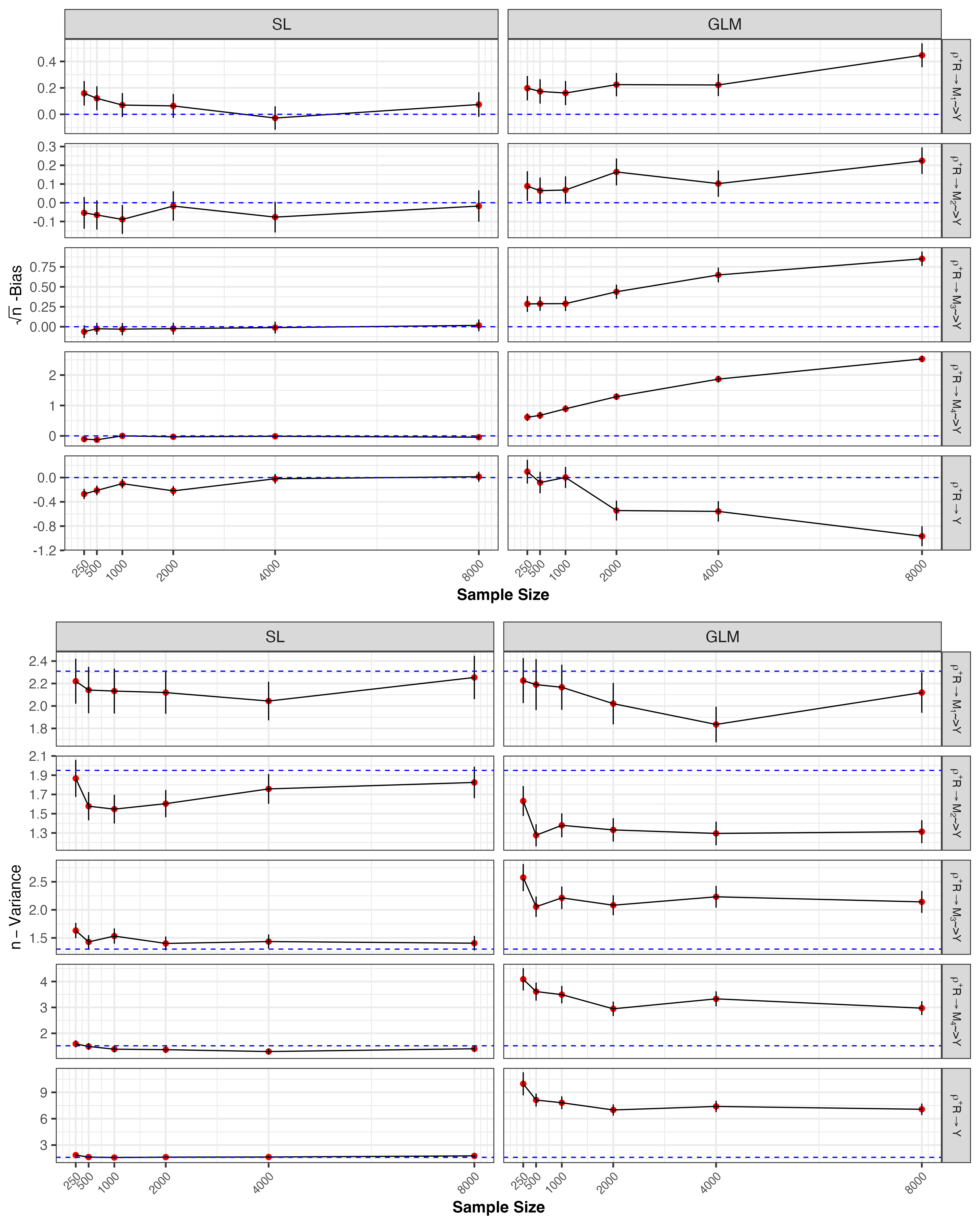}
    \caption{Comparative simulation results assessing the $\sqrt{n}$-consistency of the one-step corrected plug-in estimator using super learner versus GLM for nuisance estimation.}
    \label{fig:simu_meps}
\end{figure}

We further examine the asymptotic properties of the estimators by evaluating the root-$n$-scaled bias and the $n$-scaled variance. Figure~\ref{fig:simu_meps} shows that, when using super learners for all nuisance estimations, the root-$n$-scaled bias for all effects converges to zero and the $n$-scaled variance converges to the true variance, whereas the GLM-based approach fails to converge.

These findings confirm the reliability of our empirical results, support the use of a two-part modeling strategy for zero-inflated and right-skewed outcomes and highlight the advantage of super learners in capturing complex relationships, particularly as sample size increases. Therefore, they  provide strong empirical justification for applying our proposed framework with super learner-based nuisance estimation in the MEPS analysis.

\vspace{0.2cm}
%  ---  ---  ---  ---  ---  ---  ---  ---  ---  ---  ---  ---  --- -
{\bf Simulation 2: Robustness to model misspecification.} 
% \label{subsec:misspecification}
%  ---  ---  ---  ---  ---  ---  ---  ---  ---  ---  ---  ---  ---  --- --
% 
Here, we evaluate the robustness of the estimators to model misspecification, as established in Corollary~\ref{cor:robustness}. Data are generated with four uniform covariates, a binary treatment, four ordered univariate continuous mediators (each normally distributed), and a normally distributed outcome. See Appendix Section~\ref{app:simu_misspecific} for the detailed data generation process. 

One-step estimators for counterfactual means (i.e., $\gamma^+_{R \rightarrow Y}$ and $\gamma^+_{R \rightarrow M_k \leadsto Y}$) are constructed using estimates of the nuisance functions $Q = \{\pi, \{g_k, \mu_k, \mathscr{B}_k, \mathscr{C}_{\mathscr{B}_k} : \forall k \}, \mathscr{C}_{\mu_4} \}$. 
We evaluate the consistency of $\hat{\gamma}^+_{R \rightarrow Y}$ under three conditions: (i) only $\hat{\pi}$ and $\hat{g}_4$ are consistent; (ii) only $\hat{\pi}$ and $\hat{\mu}_4$ are consistent; (iii) only $\hat{\mathscr{C}}_{\mu_4}$ and $\hat{\mu}_4$ are consistent. 
Similarly, the consistency of $\hat{\gamma}^+_{R \rightarrow M_1 \leadsto Y}$ is evaluated under three conditions:(i) only $\hat{\pi}$ and $\hat{g}_1$ are consistent; (ii) only $\hat{\pi}$, and $\hat{\mu}_1$ are consistent; (iii) only $\hat{\mathscr{B}}_1$ and $\hat{\mu}_1$ are consistent. 
For $k=2,3,4$, the consistency of $\hat{\gamma}^{+}_{R \rightarrow M_k \leadsto Y}(\hat{Q})$, $k=2,3,4$ is evaluated under four conditions: (i) only $\hat{\pi}$, $\hat{g}_{k-1}$, and $\hat{g}_k$ are consistent; (ii) only $\hat{\pi}$, $\hat{g}_{k-1}$ and $\hat{\mu}_k$ are consistent; (iii) only $\hat{\pi}$, $\hat{\mathscr{B}}_k$ and $\hat{\mu}_k$ are consistent; and (iv) only $\hat{\mathscr{C}}_{\mathscr{B}_k}$, $\hat{\mathscr{B}}_k$, and $\hat{\mu}_k$ are consistent.  

The nuisance functions can be consistently estimated using GLMs. To introduce model misspecification, we apply nonlinear transformations to the covariates, as described in Appendix Section~\ref{app:simu_misspecific}. We also consider two additional scenarios in which all nuisance functions are misspecified and estimated using either GLMs or super learners.

Figure~\ref{fig:misspecific} in Appendix Section~\ref{app:simu_misspecific} illustrates that the one-step estimators achieve root-$n$ consistency under the specific model misspecification conditions outlined above, underscoring their robustness. In contrast, estimators based solely on misspecified GLM nuisance estimates fail to maintain the root-$n$-scaled bias property. Notably, the super learner approach offers a significant advantage, achieving root-$n$ consistency even when all nuisance functions are misspecified, particularly as sample size increases.

%######################################
\section{Empirical analysis of the MEPS data}
\label{sec:data_analysis}
%######################################

We now apply our methodological framework to the MEPS data described in Section~\ref{sec:MEPS}. We note that the analyses are conducted separately within each cross-sectional dataset (2009 and 2016). Comparisons across years are used to assess how disparities and their associated pathways differ over time, rather than to identify mediation through temporal variation.

%  ---  ---  ---  ---  ---  ---  ---  ---
\subsection{Implementation details}
\label{subsec:implement}
%  ---  ---  ---  ---  ---  ---  ---  ---

To estimate the disparity components of interest using the estimators outlined in \eqref{eq:pse_effs_est_direct}, \eqref{eq:pse_effs_est_m4}, and \eqref{equ:rho_define}, we fit each nuisance function-valued parameter in $Q = \{\pi, \{g_k, \mu_k, \mathscr{B}_k, \mathscr{C}_{\mathscr{B}_k} : \forall k \}, \mathscr{C}_{\mu_4} \}$, as described in Section~\ref{subsec:method_est}, using super learners. This ensemble learning method combines flexible statistical and machine learning models via cross-validation to mitigate model misspecification and improve predictive accuracy \citep{van_der_laan_super_2007,polley_super_2010}. We include \texttt{mean}, \texttt{glm}, \texttt{glm.interaction},  \texttt{gam}, \texttt{glmnet}, \texttt{earth}, \texttt{ksvm}, \texttt{xgboost},  \texttt{randomForest},  \texttt{dbarts} as candidate learners. 

\begin{figure}[t]
    \centering
    \includegraphics[width=0.43\textwidth]{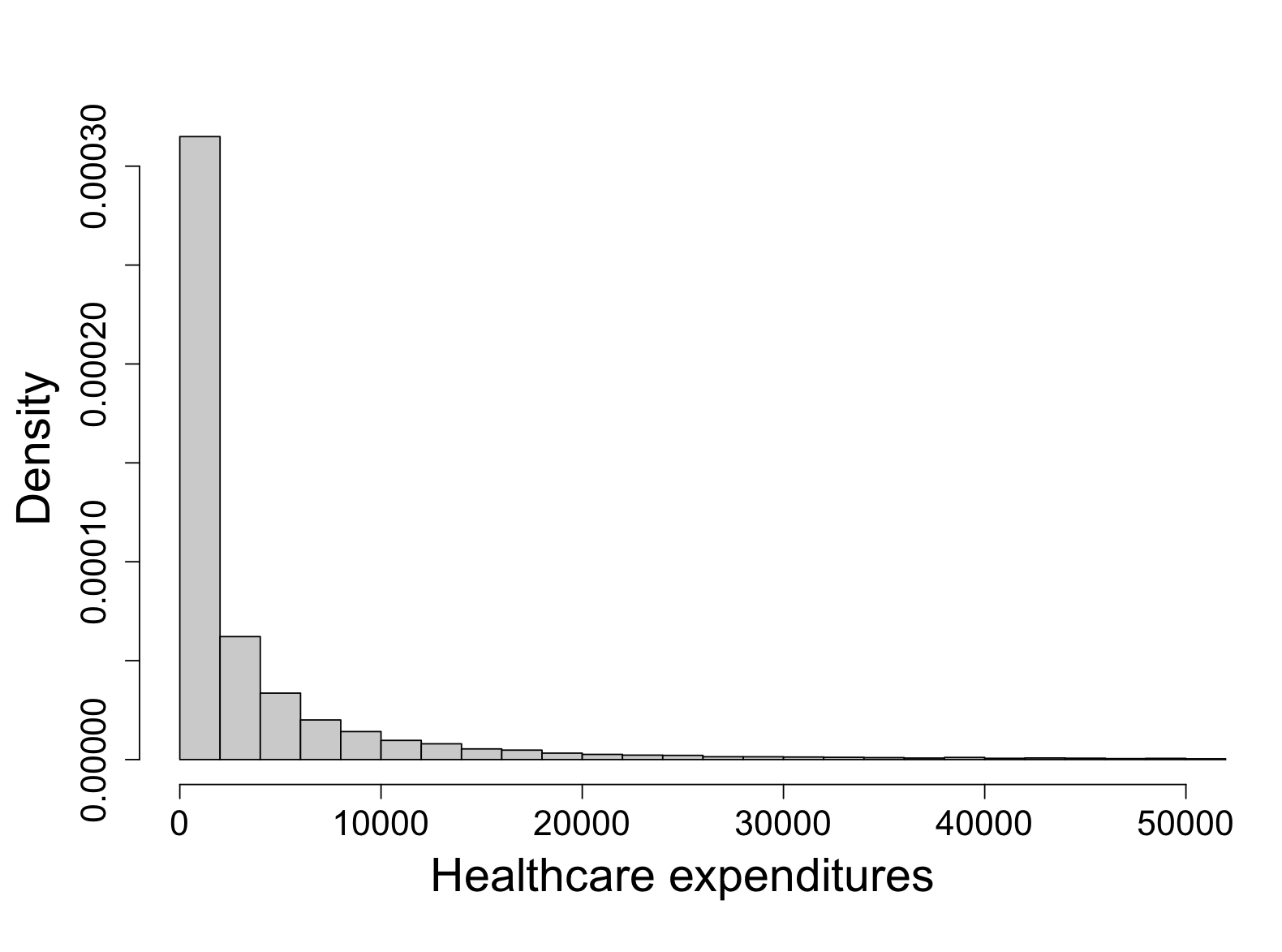}
    % \hfill 
    \hspace{0.2cm}
    \includegraphics[width=0.35\textwidth]{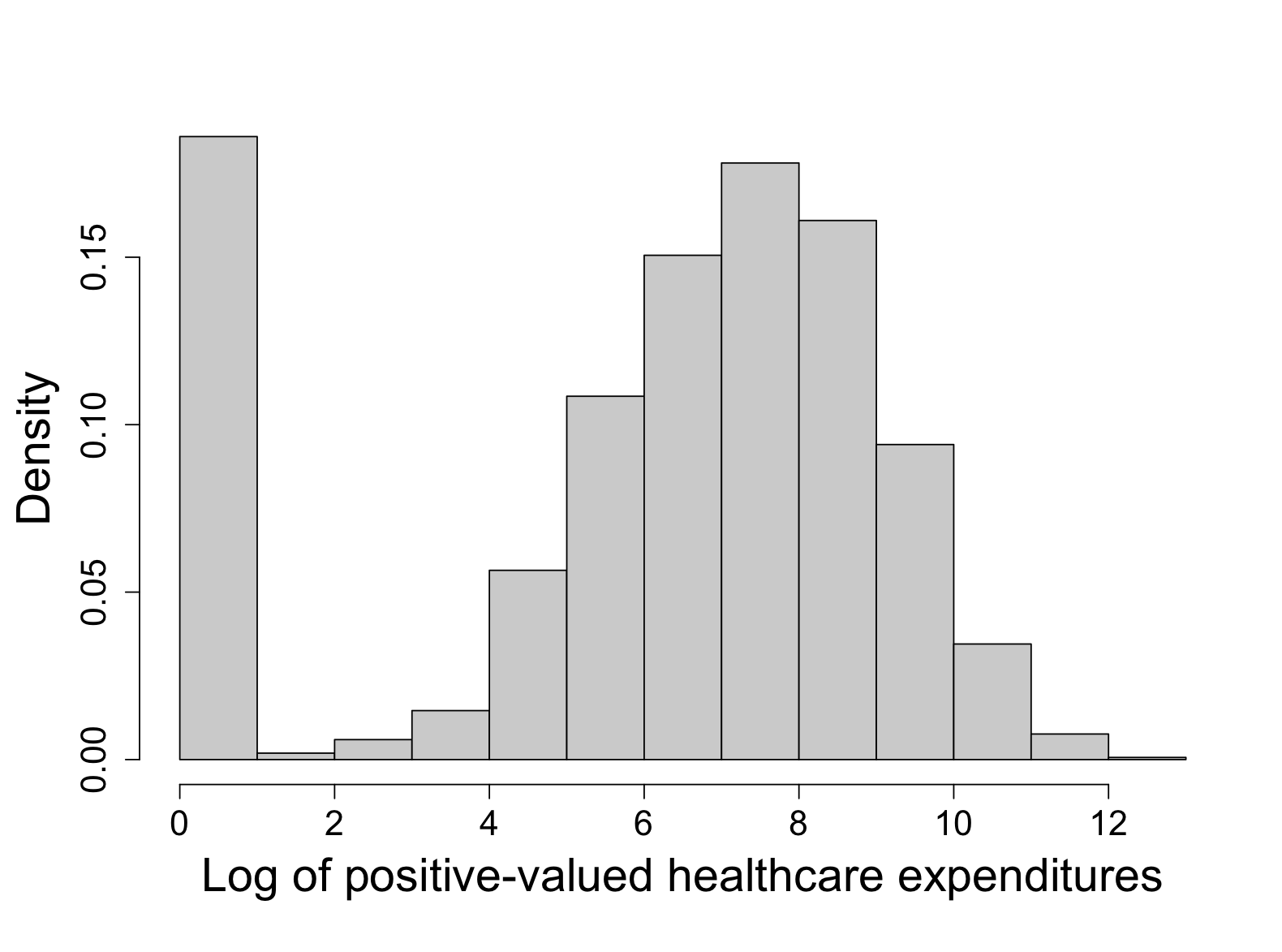}
    \caption{Empirical distribution of healthcare expenditures in the 2009 MEPS data. The left panel shows the raw distribution of total expenditures, highlighting substantial right-skewness and a mass at zero. The right panel displays the distribution of log-transformed positive expenditures, illustrating the reduced skewness after transformation.} 
    \label{fig:expenditure}
\end{figure}

When estimating outcome mean regressions $\mu_k(\overline{M}_{k}, r_0, X)$ using MEPS data, challenges arise from zero-inflated and right-skewed distribution of healthcare expenditures (see Figure~\ref{fig:expenditure}, the proportion of zero expenditures was 19.0\%, and the skewness of the positive expenditures was 7.024). Linear regression is highly sensitive to extreme positive values, whereas a log-transformed GLM does not adequately accommodate the point mass at zero. In health economics, a widely used solution is the two-part model \citep{belotti_twopm_2015, cook_comparing_2010, an_health_2015, simmons_racial_2019}, which treats the outcome as a mixture of two components: 

\vspace{-0.25cm}
\begin{itemize}
    \item[(i)] Probability model: the probability of any healthcare expenditure, $P(Y >0 \, \mid \, \overline{M}_{k}, R, X)$, and   
    \item[(ii)] Positive outcome model: the conditional distribution of expenditures among individuals with positive spending,  $P(Y \, \mid \, Y>0, \overline{M}_{k}, R, X)$. 
\end{itemize}

\vspace{-0.25cm}
The conditional mean of the outcome is then obtained by combining these two components: 
$\mu_k(\overline{M}_{k}, r_0, X) = P(Y>0 \mid \overline{M}_{k}, r_0, X) \times \E[Y \mid Y>0, \overline{M}_{k}, r_0, X]$. A two-part modeling strategy for $\mu_k(\overline{M}_{k}, r_0, X)$ can be implemented using flexible learners for the binary part and generalized linear models (GLMs) with Gamma or Lognormal distributions for the positive part \citep{liu_statistical_2019}. Wu et al. \citep{wu_two-stage_2022-1} propose a two-stage super learner which combines GLMs with varying link functions. 

To further address the right-skewed nature of expenditures, we apply a log-transformation to the positive outcomes before adapting a two-part model for the outcome mean regression. Specifically, we redefine the observed outcome as $\mathbb{I}(Y > 0) \times \log Y$ and estimate $\E[\log Y \mid Y > 0, \overline{M}_k, r_0, X]$ in the second part of the two-part model, under the assumption of a normal error distribution. The predicted value for the $i$-th observation is then constructed as $\hat{\mu}_k(\overline{M}_{k,i}, r_0, X_i)= \hat{P}(Y > 0 \mid \overline{M}_{k, i}, r_0, X_i) \times \hat{\E}[\log Y \mid Y>0, \overline{M}_{k, i}, r_0, X_i]$.

We note that reporting values on the arithmetic mean scale (i.e., without log-transformation) can be overly sensitive to extreme values. By applying a log-transformation, we instead report the disparity measures on the geometric mean scale, which is less influenced by extremes and thus more appropriate for skewed data \citep{ barasa_measuring_2018}.  

As a result, all disparity estimates are reported on the geometric mean scale by exponentiating the estimands, i.e., $\exp(\rho_{R \rightarrow Y})$ and $\exp(\rho_{R \rightarrow M_k \leadsto Y})$. These can be interpreted as the ratio of geometric means of positive expenditures, adjusted for the probability of observing any expenditure.  For example, an estimated disparity of $1.20$ can be interpreted as the advantaged group having, on average, 20\% higher typical (geometric mean) healthcare expenditures than the disadvantaged group, after accounting for both the likelihood of any spending and the level of spending among users.

This approach simultaneously addresses zero-inflation and skewness, providing a robust and interpretable measure of disparity. Further details are provided in Appendix Section~\ref{app:meps_response}. We note that while the two-part model provides flexibility for handling zero-inflation and skewness, it relies on correct specification of both components and assumes that the log-transformation adequately stabilizes variability in positive expenditures. Misspecification of either part may affect estimation accuracy, although the use of flexible learners mitigates this concern in practice.

%  ---  ---  ---  ---  ---  ---  ---  ---
\subsection{Empirical results}
%  ---  ---  ---  ---  ---  ---  ---  ---

Table~\ref{tab:PSEs} reports estimates of the total disparity, mediator-attributable components, and the outcome-attributed disparity component, expressed as ratios of scaled geometric means of healthcare expenditures. 

% Total Effect
The total disparity ($\rho_\text{total}$) was statistically significant across all six racial group comparisons in 2009 (White vs. Black, White vs. Asian, White vs. Hispanic, Black vs. Asian, Black vs. Hispanic, and Asian vs. Hispanic). All point estimates exceeded 1, indicating that non-reference racial groups in each comparison had higher expected healthcare expenditures on the geometric mean scale. Whites consistently had the highest expenditures, likely reflecting systemic advantages in healthcare access and utilization \citep{cook_comparing_2010,bailey_structural_2017,dieleman_us_2021}.  
Among marginalized groups, Hispanics had the lowest expected  expenditures, underscoring structural inequities. In 2016, these disparities largely persisted, though the Black-Asian comparison was no longer significant. The White-Black gap widened, echoing national  trends reported by \cite{dickman_trends_2022}, while other comparisons showed modest declines. These evolving patterns may reflect shifts in socioeconomic conditions, policy environments, healthcare access, as well as differences in illness recognition, patient satisfaction, and healthcare experiences across racial groups, though further research is needed to identify the drivers of these changes.  

% SES 
The SES-mediated disparity ($\rho_{R \rightarrow M_1 \leadsto Y}$), where SES is defined by income, education and employment, was statistically significant across four racial group comparisons, except for White vs. Asian and Black vs. Asian, in both 2009 and 2016. This component reflects the disparity that would be reduced if the SES distribution (within levels of covariates) for one group were shifted to match that of the other. In 2009, if a Black or Hispanic population had SES distributions aligned with that of Whites, their scaled geometric mean expenditures would rise to 1.045 (95\% CI: 1.006--1.085) or 1.207 (95\% CI: 1.145--1.269), respectively. Similarly, aligning the SES distribution of  Hispanics with that of Blacks or Asians would result in a 21.0\% or 23.0\% increase, respectively. These findings suggest that SES plays a major role in racial disparities in healthcare spending. Whites tend to have relatively high SES levels, while Hispanics experience higher unemployment rates and lower levels of higher education and total income compared to Whites and Asians \citep{williams_understanding_2016}. These socioeconomic differences help explain the disparities captured by the SES-mediated component. In 2016, SES-mediated measures slightly increased relative to 2009, indicating a potentially growing role of income, education, and employment gaps in shaping healthcare expenditures. These patterns underscore the importance of SES as a key driver of racial disparities, both through direct economic effects on healthcare access and through its downstream influence on other mediators, including insurance access, health behaviors, and health status.

% insurance
The insurance-mediated disparity ($\rho_{R \rightarrow M_2 \leadsto Y}$) was statistically significant in all racial group comparisons except White vs. Black and Black vs. Asian, in 2009. This component reflects the disparity that would  be reduced if the distribution of insurance access, conditional on covariates and SES, were aligned across groups. If the insurance distribution of Asians were aligned with that of Whites, their scaled geometric mean expenditures would increase by 8.5\%. Similarly, aligning the insurance coverage of Hispanics with that of Whites, Blacks, or Asians, would raise their scaled geometric mean expenditures to 1.265 (95\% CI: 1.220--1.310), 1.432 (95\% CI: 1.371--1.492), or 1.255 (95\% CI: 1.205--1.305) times higher, respectively. These findings reflect the fact that, in 2009, Hispanics had the highest rate of being uninsured, more than three times that of Whites. 
By 2016, the insurance-mediated disparities disappeared in the White vs. Asian comparisons. Although the insurance-mediated disparities remained statistically significant in all comparisons involving Hispanics, the estimated ratios decreased to 1.245 (95\% CI: 1.204--1.285) for White vs. Hispanic, 1.333 (95\% CI: 1.279--1.387) for Black vs. Hispanic, and 1.244 (95\% CI: 1.199--1.289) for Asian vs. Hispanic respectively. This pattern coincides with a decline in observed uninsured rates across all racial groups, and especially a substantial narrowing of the gap between Asians and Whites. One contributing factor may be the Affordable Care Act, enacted in 2010 and fully implemented in 2014, which expanded coverage for economically disadvantaged and marginalized populations \citep{gaffney2017affordable, buchmueller_acas_2020-1}. Despite overall improvements in insurance coverage, Hispanics continued to experience the highest rate of uninsurance. At the same time, the expenditure gap between insured and uninsured groups widened, underscoring the increasing importance of insurance in healthcare disparities. Barriers to coverage among Hispanics may include unclear eligibility rules, enrollment difficulties, and language or literacy challenges \citep{hill_health_2024,wallace_disparities_2022}. Without insurance, individuals are more likely to delay or forgo care, while having coverage facilitates access and may raise overall expenditures through more timely and appropriate healthcare use \citep{frankovic_health_2023}. 

% health behavior 
The health behavior-mediated disparity ($\rho_{R \rightarrow M_3 \leadsto Y}$), where health behavior is defined by smoking status and physical activity, was relatively small overall. It was statistically significant only in the Asian vs. Hispanic comparison in 2009 (1.018, 95\% CI: 1.003--1.033) and non significant in all comparisons in 2016. This suggests that differences in these measured health behaviors contribute minimally to disparities in healthcare expenditures across racial groups. Several reasons  may explain this limited effect. First, health behaviors reflect physical endowments together with a cumulative set of experiences and circumstances that evolve over time within distinct social and physical contexts \citep{short2015social}. While race may be one contributing factor, other social determinants, including education, income, health beliefs, workplace conditions, neighborhood environments, and broader political-economic structures, also play important roles in shaping disparities \citep{mckinlay1979case, puka2023can}. In our framework, equalizing health behaviors while allowing upstream factors such as SES and insurance access to remain unchanged is therefore likely to capture only a limited portion of the overall disparities. Second, prior research indicates that behavioral patterns vary in complex and multidimensional ways across racial groups, with no clear binary distinction between ``healthy'' and ``unhealthy'' lifestyles \citep{cockerham2017comparison}. As observed in our data, Whites have a higher prevalence of smoking but also a higher level of physical activity. Such patterns suggest that different behaviors may offset one another across groups, thereby attenuating the overall contribution of health behaviors to racial disparities in healthcare expenditures. Third, the measurement of health behaviors in MEPS is limited. Important dimensions such as diet, sleep, and substance use are not fully captured, which may lead to an underestimation of the overall contribution of behavioral factors \citep{short2015social}. Therefore, these findings are consistent with broader evidence suggesting that individual behaviors alone account for only a small fraction of racial disparities, whereas structural and contextual factors play a more substantial role. At the same time, a more granular and in-depth examination of specific behavioral domains is needed to better understand the contribution of health behaviors to these disparities.

% health status
The health status-mediated disparity ($\rho_{R \rightarrow M_4 \leadsto Y}$) emerged as a substantial contributor to racial disparities in healthcare expenditures. Prior studies have shown that, compared to Whites, marginalized groups tend to report poorer self-rated health and experience higher rates of chronic conditions, often linked to lower SES, limited insurance access, and less favorable living conditions \citep{beydoun_racial_2016,ko_association_2020,wallace_disparities_2022}. These patterns would typically suggest that marginalized groups bear higher medical spending burdens relative to Whites \citep{charron-chenier_racial_2018}. However, when focusing solely on the differences in health status distributions across racial groups (holding SES, insurance access, and health behaviors fixed) our study reveals a different pattern. In both 2009 and 2016, the health status-mediated disparity was significant for all racial group comparisons except White vs. Black. For example, in 2016, aligning the health status of Black, Asian, or Hispanic populations with that of Whites (conditional on covariates and upstream mediators) would increase their scaled geometric mean expenditures by 6.1\% (not statistically significant, but close to significance; p-value = 0.063), 39.4\%, and 37.8\%, respectively. Likewise, aligning the health status of Asians or Hispanics  with that of Blacks would increase expenditures by factors of 1.289 and 1.169, respectively, whereas aligning the health status of Hispanics with that of Asians would reduce expenditures to 84.8\%. This apparent divergence from prior findings may reflect a higher prevalence of diagnosed disease among Whites, potentially due to greater access to screening and diagnostic services \citep{doubeni_association_2022}. It may also reflect biological, dietary, or other inherent group differences that influence disease risk but are not captured by socioeconomic or behavioral measures. 

% direct effect 
The outcome-attributed disparity ($\rho_{R \rightarrow Y}$), which captures differences in outcomes not mediated by observed variables, was statistically significant only for comparisons between Whites and marginalized racial groups in 2009. It was not significant in comparisons between any two marginalized groups. One likely explanation is that there could well be other mediating factors not considered here, leading the unexplained component to reflect the influence of unmeasured pathways. For instance, early life adversity (such as poverty, abuse, and traumatic stress, which vary by race) has been linked to poorer physical and mental health later in life, thereby influencing healthcare use and costs \citep{shonkoff_neuroscience_2009}. Another plausible explanation is structural racism. A systematic review has demonstrated that healthcare providers' implicit biases are associated with differences in treatment decisions, care quality, and patient outcomes \citep{hall_implicit_2015}. Such biases may also erode patient-provider communication and reduce trust, making marginalized patients less likely to follow medical recommendations \citep{cooper_associations_2012}. By 2016, the outcome-attributed disparity declined in the White vs. Black, White vs. Asian and White vs. Hispanic comparisons. However, it increased in comparisons involving Blacks vs. Hispanics, and Asians vs. Hispanics, with estimated ratios deviating significantly from 1. These shifts suggest that disparities not accounted for by SES, insurance, health behaviors, or health status became more pronounced in certain groups, underscoring the persistence of structural inequities and the evolving role of systemic bias in healthcare access and treatment.

% table for PSEs 
\begin{center}
\begin{table}[!th]
\caption{Disparity components across racial group comparisons, reported on the scaled geometric mean ratios; $p$-values smaller than 0.05 are boldfaced.}
 % or \footnotesize for slightly larger font
%\renewcommand{\arraystretch}{0.55} % Row spacing (default is 1)
\footnotesize
\renewcommand{\arraystretch}{0.9}
\setlength{\tabcolsep}{12.5pt} 
\centering
\begin{tabular}{lcccccc}
\toprule
& \multicolumn{3}{c}{\textbf{MEPS data in year 2009}} & \multicolumn{3}{c}{\textbf{MEPS data in year 2016}} \\
\cmidrule(lr){2-4} \cmidrule(lr){5-7}
Disparity & Value & 95\% CI & p-value & Value & 95\% CI & p-value \\ 
\midrule\addlinespace[2.5pt]
\multicolumn{7}{l}{Whites vs Blacks*} \\[2.5pt]
\midrule\addlinespace[2.5pt]
$\rho_{R \rightarrow M_1 \leadsto Y}$ & 1.045 & 1.006 --- 1.085 & {\bf 0.024} & 1.066 & 1.015 --- 1.117 & {\bf 0.011} \\
$\rho_{R \rightarrow M_2 \leadsto Y}$ & 1.011 & 0.981 --- 1.041 & 0.490 & 0.995 & 0.971 --- 1.020 & 0.710 \\
$\rho_{R \rightarrow M_3 \leadsto Y}$ & 0.984 & 0.962 --- 1.005 & 0.137 & 0.997 & 0.977 --- 1.017 & 0.769 \\
$\rho_{R \rightarrow M_4 \leadsto Y}$ & 1.014 & 0.953 --- 1.075 & 0.645 & 1.061 & 0.997 --- 1.124 & 0.063 \\
$\rho_{R \rightarrow Y}$ & 1.772 & 1.614 --- 1.930 & {\bf $<$0.001} & 1.768 & 1.607 --- 1.929 & {\bf $<$0.001} \\
$\rho_\text{total}$ & 1.901 & 1.680 --- 2.122 & {\bf $<$0.001} & 2.084 & 1.840 --- 2.329 & {\bf $<$0.001} \\
\midrule\addlinespace[2.5pt]
\multicolumn{7}{l}{Whites vs Asians*} \\[2.5pt]
\midrule\addlinespace[2.5pt]
$\rho_{R \rightarrow M_1 \leadsto Y}$ & 1.010 & 0.898 --- 1.122 & 0.861 & 1.035 & 0.946 --- 1.124 & 0.441 \\
$\rho_{R \rightarrow M_2 \leadsto Y}$ & 1.085 & 1.017 --- 1.154 & {\bf 0.015} & 1.003 & 0.959 --- 1.047 & 0.891 \\
$\rho_{R \rightarrow M_3 \leadsto Y}$ & 0.996 & 0.960 --- 1.033 & 0.840 & 0.995 & 0.970 --- 1.021 & 0.725 \\
$\rho_{R \rightarrow M_4 \leadsto Y}$ & 1.355 & 1.210 --- 1.499 & {\bf $<$0.001} & 1.394 & 1.275 --- 1.512 & {\bf $<$0.001} \\
$\rho_{R \rightarrow Y}$ & 2.316 & 1.999 --- 2.632 & {\bf $<$0.001} & 2.028 & 1.770 --- 2.287 & {\bf $<$0.001} \\
$\rho_\text{total}$ & 2.893 & 2.408 --- 3.378 & {\bf $<$0.001} & 2.561 & 2.184 --- 2.937 & {\bf $<$0.001} \\
\midrule\addlinespace[2.5pt]
\multicolumn{7}{l}{Whites vs Hispanics*} \\[2.5pt]
\midrule\addlinespace[2.5pt]
$\rho_{R \rightarrow M_1 \leadsto Y}$ & 1.207 & 1.145 --- 1.269 & {\bf $<$0.001} & 1.280 & 1.206 --- 1.353 & {\bf $<$0.001} \\
$\rho_{R \rightarrow M_2 \leadsto Y}$ & 1.265 & 1.220 --- 1.310 & {\bf $<$0.001} & 1.245 & 1.204 --- 1.285 & {\bf $<$0.001} \\
$\rho_{R \rightarrow M_3 \leadsto Y}$ & 1.031 & 0.993 --- 1.070 & 0.112 & 1.024 & 0.989 --- 1.059 & 0.175 \\
$\rho_{R \rightarrow M_4 \leadsto Y}$ & 1.274 & 1.205 --- 1.343 & {\bf $<$0.001} & 1.378 & 1.301 --- 1.456 & {\bf $<$0.001} \\
$\rho_{R \rightarrow Y}$ & 2.071 & 1.901 --- 2.240 & {\bf $<$0.001} & 1.830 & 1.676 --- 1.984 & {\bf $<$0.001} \\
$\rho_\text{total}$ & 3.705 & 3.318 --- 4.093 & {\bf $<$0.001} & 3.371 & 3.005 --- 3.737 & {\bf $<$0.001} \\
\midrule\addlinespace[2.5pt]
\multicolumn{7}{l}{Blacks vs Asians*} \\[2.5pt]
\midrule\addlinespace[2.5pt]
$\rho_{R \rightarrow M_1 \leadsto Y}$ & 1.064 & 0.940 --- 1.189 & 0.311 & 1.015 & 0.867 --- 1.163 & 0.846 \\
$\rho_{R \rightarrow M_2 \leadsto Y}$ & 1.079 & 0.980 --- 1.179 & 0.118 & 0.963 & 0.878 --- 1.048 & 0.392 \\
$\rho_{R \rightarrow M_3 \leadsto Y}$ & 0.977 & 0.902 --- 1.052 & 0.556 & 0.983 & 0.934 --- 1.032 & 0.493 \\
$\rho_{R \rightarrow M_4 \leadsto Y}$ & 1.313 & 1.112 --- 1.515 & {\bf 0.002} & 1.289 & 1.101 --- 1.476 & {\bf 0.003} \\
$\rho_{R \rightarrow Y}$ & 0.991 & 0.811 --- 1.171 & 0.922 & 0.921 & 0.777 --- 1.065 & 0.281 \\
$\rho_\text{total}$ & 1.466 & 1.115 --- 1.817 & {\bf 0.009} & 1.210 & 0.913 --- 1.506 & 0.165 \\
\midrule\addlinespace[2.5pt]
\multicolumn{7}{l}{Blacks vs Hispanics*} \\[2.5pt]
\midrule\addlinespace[2.5pt]
$\rho_{R \rightarrow M_1 \leadsto Y}$ & 1.210 & 1.150 --- 1.270 & {\bf $<$0.001} & 1.202 & 1.144 --- 1.260 & {\bf $<$0.001} \\
$\rho_{R \rightarrow M_2 \leadsto Y}$ & 1.432 & 1.371 --- 1.492 & {\bf $<$0.001} & 1.333 & 1.279 --- 1.387 & {\bf $<$0.001} \\
$\rho_{R \rightarrow M_3 \leadsto Y}$ & 1.028 & 1.000 --- 1.056 & 0.051 & 1.007 & 0.976 --- 1.038 & 0.651 \\
$\rho_{R \rightarrow M_4 \leadsto Y}$ & 1.209 & 1.144 --- 1.274 & {\bf $<$0.001} & 1.169 & 1.108 --- 1.229 & {\bf $<$0.001} \\
$\rho_{R \rightarrow Y}$ & 1.031 & 0.952 --- 1.109 & 0.442 & 0.896 & 0.822 --- 0.969 & {\bf 0.005} \\
$\rho_\text{total}$ & 1.973 & 1.670 --- 2.276 & {\bf $<$0.001} & 1.624 & 1.391 --- 1.858 & {\bf $<$0.001} \\
\midrule\addlinespace[2.5pt]
\multicolumn{7}{l}{Asians vs Hispanics*} \\[2.5pt]
\midrule\addlinespace[2.5pt]
$\rho_{R \rightarrow M_1 \leadsto Y}$ & 1.230 & 1.138 --- 1.322 & {\bf $<$0.001} & 1.306 & 1.219 --- 1.393 & {\bf $<$0.001} \\
$\rho_{R \rightarrow M_2 \leadsto Y}$ & 1.255 & 1.205 --- 1.305 & {\bf $<$0.001} & 1.244 & 1.199 --- 1.289 & {\bf $<$0.001} \\
$\rho_{R \rightarrow M_3 \leadsto Y}$ & 1.018 & 1.003 --- 1.033 & {\bf 0.017} & 0.998 & 0.981 --- 1.016 & 0.834 \\
$\rho_{R \rightarrow M_4 \leadsto Y}$ & 0.867 & 0.816 --- 0.918 & {\bf $<$0.001} & 0.848 & 0.793 --- 0.903 & {\bf $<$0.001} \\
$\rho_{R \rightarrow Y}$ & 1.017 & 0.941 --- 1.092 & 0.666 & 1.097 & 1.013 --- 1.180 & {\bf 0.023} \\
$\rho_\text{total}$ & 1.471 & 1.193 --- 1.749 & {\bf 0.001} & 1.410 & 1.139 --- 1.680 & {\bf 0.003} \\
\bottomrule
\end{tabular}
\begin{tablenotes}
\item  \textsuperscript{*}Reference group; $M_1$: SES, $M_2$: Insurance, $M_3$: Health behaviors, $M_4$: Health status.
\end{tablenotes}
\label{tab:PSEs}
\end{table}
\end{center}

% summary
\subsection{Summary of findings} 

Our analysis shows persistent racial disparities in healthcare expenditures across both years, with Whites generally having higher expenditures than marginalized racial groups and the largest total disparities consistently observed for Whites vs. Hispanics, although this gap declined from 3.705 in 2009 to 3.371 in 2016. Across comparisons, SES and health status were the most consistent contributors to expenditure disparities, with SES-attributed disparity ratios ranging from 1.010 to 1.230 in 2009 and from 1.015 to 1.306 in 2016, and health status-attributed disparities corresponding to relative changes of 13.3\% to 35.5\% in 2009 and 15.2\% to 39.4\% in 2016. Insurance also played a critical role, particularly in shaping outcomes for Hispanics, whereas health behaviors had consistently small effects, with estimates generally close to the null. Outcome-attributed disparities were most pronounced in comparisons between Whites and other racial minority groups, but were often small or not statistically significant in non-White pairwise comparisons. Therefore, our analyses suggest a consistent ranking of mediator importance, with SES as the dominant contributor, followed by health status and insurance, and with health behaviors contributing minimally.

Because these decomposition results are defined with respect to the assumed causal ordering of mediators, their interpretation depends on this specification. To assess robustness, we conducted a sensitivity analysis under an alternative ordering of health status and health behaviors (Appendix Section~\ref{app:order}). The resulting estimates were broadly similar, with consistent patterns in the relative importance of mediators, suggesting that our substantive conclusions are not driven by the particular ordering assumed in the main analysis.

As discussed in Section~\ref{subsec:method_def}, mediator-attributable disparities do not sum additively to the total disparity. To compare the relative contributions of different components and identify dominant pathways, we compute cumulative disparity measures using a sequential decomposition in Appendix Section~\ref{app:pse_decomp} and report the findings in Appendix Section~\ref{app:meps_sequential}.

%####################################################
\section{Discussion}
\label{sec:discussion}
%####################################################

% summary 
Our findings highlight the central role of structural determinants, particularly socioeconomic status and insurance access, in shaping racial disparities in healthcare expenditures. By decomposing disparities into mediator-specific components, our framework provides insight into how these structural inequities translate into differences in healthcare utilization.

% policy implications
The mediator-attributable disparities offer valuable insights for policy development. The substantial contribution of socioeconomic status suggests that investments in education, job training, and income support for disadvantaged populations may play a central role in reducing disparities in healthcare utilization. Persistent gaps mediated by insurance access further highlight the importance of targeted coverage expansions, particularly for populations with high uninsurance rates such as Hispanic individuals. More broadly, these findings underscore that healthcare disparities are shaped by interconnected structural and systemic factors. In addition to improving economic and insurance access, efforts to reduce inequities may benefit from interventions within the healthcare system itself, including provider training to address implicit bias and initiatives that improve patient engagement and trust. By quantifying the relative contribution of each mediating factor, our analysis offers a data-driven basis for prioritizing interventions that target the root causes of disparities rather than their downstream consequences.
 
% link to fairness 
Beyond policy, our findings also speak to the design of predictive algorithms in healthcare. Cost data are often used to allocate resources or identify high-risk patients, yet they may reflect underlying racial disparities. Prior research has shown that algorithms trained solely on cost data may underestimate the healthcare needs of marginalized groups, particularly Black patients relative to White patients \citep{obermeyer_dissecting_2019}. This underscores the need for fairness-aware adjustments. Causal and distributional reasoning tools, including path-specific decompositions, can help identify whether observed disparities and unfair treatments are mediated by actionable variables \citep{nabi_fair_2018, nabi_learning_2018, kusner_counterfactual_2017}. If predictive models ignore these disparities, they risk perpetuating structural bias. Fairness-aware algorithms could explicitly constrain disparities along specific pathways, such as those mediated by SES, to align predictions with equity goals \citep{nabi2024statistical, nabi2024fair}. 

% limiations 
Despite its strengths, this study has several limitations. 
\textit{First,} reliance on self‐reported data introduces potential reporting bias, since participants may misclassify diagnoses or utilization due to recall errors or social desirability. Although validation with clinical records could help mitigate this issue, such data are often inaccessible. 
\textit{Second,} although MEPS uses a complex survey design with weighting and oversampling to produce nationally representative estimates of the U.S. civilian noninstitutionalized population, it excludes individuals in institutional settings (e.g., long-term care and correctional facilities). This represents a limitation in population coverage rather than sampling bias within the target population. As a result, our findings may not generalize to these groups, who may have systematically different healthcare needs and utilization patterns; future research should assess the extent to which these results extend to institutionalized populations using data sources that explicitly capture these settings.
\textit{Third,} because race is a social construct and not a manipulable treatment, causal interpretations must be approached with care. Our analysis focuses on disparities defined by distributional differences and emphasizes modifiable mediators, rather than positing counterfactual interventions on race. While this strategy yields meaningful insights into mechanisms, the resulting estimands are inherently more complex to interpret than conventional causal contrasts. 
\textit{Fourth,} healthcare utilization is inherently dynamic and may depend on prior healthcare use, illness recognition, and patient experiences with the healthcare system. Factors such as patient satisfaction and prior interactions with providers can influence subsequent care-seeking behavior and expenditures. Because our analysis is based on cross-sectional measures and does not explicitly model prior utilization or feedback processes, these pathways are not captured in the current decomposition. Future work incorporating longitudinal data and dynamic mediation structures may help to better characterize these mechanisms. As an exploratory analysis, we also examined healthcare satisfaction as a potential mediator in a restricted subpopulation with at least one healthcare visit (Appendix Section~\ref{app:satisfaction}); however, because this analysis conditions on utilization, it is not directly comparable to the main results.
Fifth, our analysis uses a complete case approach, excluding observations with missing data on variables included in the analysis. This approach may introduce bias if individuals with incomplete data differ systematically from those retained in the analytic sample. Its validity relies on a missing completely at random assumption, which may not hold in practice. As a result, estimates of mediator contributions and disparities should be interpreted with this limitation in mind. Future work could examine the robustness of findings using alternative approaches such as multiple imputation or weighting-based methods.

% future directions 
Future work should expand the set of mediators to isolate specific mechanisms, for example, distinguishing the effects of education and income separately, or examining neighborhood-level exposures. Sensitivity analyses assessing the impact of unmeasured confounding are essential to validate the robustness of the decomposition. In addition, extending the framework to dynamic settings, such as repeated measures or time-varying exposures, could offer a richer understanding of how disparities evolve. Incorporating alternative health outcomes may also provide a more complete picture of healthcare equity.

%####################################################
\section{Conclusion}
\label{sec:conclusion}
%####################################################

This study develops a mediator distribution-shift framework for decomposing disparities in healthcare expenditures. By integrating influence-function-based asymptotically linear estimators, flexible machine learning, and a two-part model tailored to zero-inflated and right-skewed expenditures, the proposed approach provides a robust and practical tool for estimating specific disparity components in complex observational settings. The results indicate that racial disparities in healthcare expenditures remain substantial across both study years, with the largest gaps observed between Whites and Hispanics. Our findings show that socioeconomic status and health status are the primary contributors to these disparities, with insurance access playing an additional important role, particularly for Hispanic populations, while health behaviors contribute relatively little. 

From a policy perspective, these results suggest that efforts to reduce disparities in healthcare utilization should prioritize interventions targeting socioeconomic inequalities and barriers to insurance coverage. At the same time, the persistence of residual disparities points to the influence of broader structural factors, including systemic racism and unequal access to care, that are not fully captured by observed mediators. This study is subject to several limitations, including reliance on self-reported data, potential selection bias due to complete case analysis, and the possibility of unmeasured mediators or confounding. Future work should extend this framework to incorporate more granular mediators, conduct sensitivity analyses, and examine dynamic and longitudinal processes underlying healthcare disparities. Overall, advancing healthcare equity will require both rigorous methodological tools to better understand disparity mechanisms and sustained policy efforts addressing the structural determinants of health.

%\bibliography{Fairness}
\putbib[Fairness]  % Use only citations from this part of the documen
\end{bibunit}

%####################################################
%####################################################
%####################################################
%%  appendix

\newpage
\appendix
\newgeometry{
    left=1.15in,      % Left margin
    right=1.15in,     % Right margin
    top=1in,         % Top margin
    bottom=1in       % Bottom margin
} %
\captionsetup{margin=0in}
\doublespacing

\begin{bibunit}[apalike]
\pagenumbering{arabic}      % Optional: reset page numbers if you want
\setcounter{page}{1}        % Start appendix from page 1
\setcounter{equation}{0} % Restart equation numbering
\renewcommand{\theequation}{S\arabic{equation}} % Prefix equations with "S"
\setcounter{section}{0} % Restart section numbering
\renewcommand{\thesection}{S\arabic{section}} % Prefix section numbers with "S"
\setcounter{figure}{0} % Restart section numbering
\renewcommand{\thefigure}{S\arabic{figure}} % Prefix section numbers with "S"
\setcounter{table}{0} % Restart table numbering
\renewcommand{\thetable}{S\arabic{table}} % Prefix table numbers with "S"

\mbox{}
%\vspace{0.5cm}
\begin{center}
{\large\bf SUPPLEMENTARY MATERIALS}
\end{center}

\vspace{0.25cm}
\begin{description}

  \item[GitHub repository:] \ The GitHub repository \href{https://github.com/xxou/Racial-Disparities-Healthcare-Expenditures}{\blue{xxou/Racial-Disparities-Healthcare-Expenditures}} contains the 2009 and 2016 MEPS data used in this study, along with documentation (\texttt{MEPSinfo.txt}), code, and results for MEPS data analysis and simulation studies.

  \vspace{0.15cm}
  \item[R package:] \ The R package \href{https://github.com/xxou/flexPaths}{\textcolor{blue}{flexPaths}} (available on GitHub at \href{https://github.com/xxou/flexPaths}{xxou/flexPaths}) provides code for robust and flexible estimation of causal path-specific effects using one‐step corrected plug‐in estimators.

  \vspace{0.15cm}
  \item[Appendix (PDF):] \ The Appendix is organized as follows. 

  \textbf{Appendix~\ref{app:causal_pse}} formalizes the connection between disparity estimands and causal path-specific effects, detailing their definitions, identification conditions, and decomposition strategies. 

  \textbf{Appendix~\ref{app:proofs}} presents proofs for all theoretical developments, including identification results, influence-function derivations, and asymptotic inference procedures. 
  
  \textbf{Appendix~\ref{app:meps_details}} provides additional details on the MEPS data and analytic sample, including variable definitions and sample construction. It further presents analyses of cumulative disparity components under a sequential decomposition, discusses the scale and interpretation of reported disparities in the presence of zero-inflated outcomes, and includes sensitivity analyses related to mediator specification, including alternative mediator choices and ordering.
  
  \textbf{Appendix~\ref{app:simulation}} details the data-generation process for the first simulation (mimicking real-world healthcare-expenditure complexities) and presents results from a second simulation assessing estimator robustness under various model misspecifications. 
  
\end{description}

%  ---  ---  ---  ---  ---  ---  ---  ---  ---  ---  ---  ---  --- -
\section{Connections to causal path-specific effects and decomposition strategies}
\label{app:causal_pse}
%  ---  ---  ---  ---  ---  ---  ---  ---  ---  ---  ---  ---  --- -

%  ---  ---  ---  ---  ---  ---  ---  ---  ---  ---  ---  ---  --- -
\subsection{Causal path-specific effects: Definition}
\label{app:pse_def}
%  ---  ---  ---  ---  ---  ---  ---  ---  ---  ---  ---  ---  --- - 

In this appendix, we formalize the connection between the disparity estimands defined in the main manuscript and causal path-specific effects (PSEs). To do so, we consider the DAG shown in Figure~\ref{fig:race_dag}, where $R$ denotes a binary treatment or exposure that may influence the outcome $Y$ either directly or indirectly through four sequential mediators, $M_1, \ldots, M_4$.

We define PSEs as population-level contrasts between counterfactual outcomes under two treatment scenarios. In the baseline scenario, treatment is set to a reference level ($R = 0$), allowing its influence to propagate naturally through all downstream variables. In the comparison scenario, treatment is set to the non-reference level ($R = 1$), but only along selected causal pathways---specifically, certain mediators are set to the values they would take under $R = 1$, while the remaining mediators are held at their values under $R = 0$. This follows the path intervention framework of \cite{shpitser2016causal} and ensures edge consistency, avoiding the \textit{recanting witness} problem associated with parameter non-identifiability. 

We consider five PSEs: the direct effect, corresponding to the direct pathway $\{R \rightarrow Y\}$, and four mediated effects, each capturing the impact of treatment through a distinct mediator $M_k$ ($k = 1, \dots, 4$). A mediated effect includes all paths from $R$ to $Y$ passing through $M_k$, represented as $\{R \rightarrow M_k \rightarrow Y, R \rightarrow M_k \rightarrow \ldots \rightarrow Y\}$, or more compactly, $\{R \rightarrow M_k \leadsto Y\}$. 

To formalize this, let $(r_0, \mathbf{r})$ denote the vector of treatment values along the five specified pathways, where $r_0 \in \{0, 1\}$ and $\mathbf{r} \coloneqq (r_1, r_2, r_3, r_4) \in \{0,1\}^4$. The setting $\mathbf{r} = \mathbf{0}$ reflects a scenario where all mediators take values under the reference treatment level. For a mediated effect through $M_k$, we set $\mathbf{r}$ to $\mathbf{1}_k$, an indicator vector with the $k$-th element set to $1$, meaning treatment is set to the non-reference level only along pathways involving $R \rightarrow M_k$.

We define the potential outcome:  
\begin{align}
   Y(r_0, \mathbf{r}) \coloneqq Y 
   \Big( 
            r_0,  
            \underbrace{M_1(r_1)}_{\coloneqq M^c_1}, 
            \underbrace{M_2\big(r_2,M^c_1\big)}_{\coloneqq M^c_2}, 
            \underbrace{M_3\big(r_3,M^c_1, M^c_2\big)}_{\coloneqq M^c_3}, 
            M_4\big(r_4, M^c_1, M^c_2, M^c_3 \big) 
   \Big) \ , 
   \label{eq:nested_counterfactual}
\end{align}

where mediators are recursively defined as follows: $M_1(r_1)$ (shorthand: $M^c_1$) is the counterfactual $M_1$ if $R=r_1$, $M_2(r_2, M^c_1)$ (shorthand: $M^c_2$) is the counterfactual $M_2$ if $R=r_2$ and $M_1 = M^c_1$. This recursive structure continues for all four mediators. Using this notation, we define the expected potential outcomes: 
\begin{align}
    \gamma_{R \rightarrow Y} \coloneqq \E[Y(1,\mathbf{0})] \ , \quad 
    \gamma_{R \rightarrow M_k \leadsto Y} \coloneqq \E[Y(0,\mathbf{1}_k)] \ , \quad
    \gamma_\text{ref} \coloneqq \E[Y(0, \mathbf{0})] \ . 
    \label{eq:pse_counterfactual_means}
\end{align}% 

\noindent The corresponding path-specific effects are defined as: 
\begin{align}
    \rho_{R \rightarrow Y} \coloneqq  \gamma_{R \rightarrow Y} - \gamma_\text{ref}  \ , \qquad 
    \rho_{R \rightarrow M_k \leadsto Y} \coloneqq   \gamma_{R \rightarrow M_k \leadsto Y} - \gamma_\text{ref}  \ . 
    \label{eq:pse_effs}
\end{align}%

In the definitions above, we adopt a \textit{reference-zero} potential outcome, i.e., $Y(0, \mathbf{0})$. This approach sets treatment to $R=1$ (the ``active'' value) along the pathways of interest while holding it at $R=0$ (the ``inactive'' value) elsewhere, and compares the resulting outcome to the baseline $Y(0, \mathbf{0})$. The resulting contrasts are often referred to as \textit{natural path-specific effects} \citep{zhou_semiparametric_2022}. In the main manuscript, the quantity $\gamma_\text{dis}$ serves as the reference-zero benchmark against which all disparity components are evaluated. 

Importantly, the natural PSEs are not mutually exclusive and do not decompose the total effect additively. Rather than partitioning the total effect across mediators, natural PSEs focus on the individual contribution of each pathway in isolation. One could also consider a sequential decomposition where the total effect is broken down cumulatively across mediators \citep{daniel_causal_2015, steen_flexible_2017}, detailed in Appendix~\ref{app:pse_decomp}.

%  ---  ---  ---  ---  ---  ---  ---  ---  ---  ---  ---  ---  --- -
\subsection{Causal path-specific effects: Identification}
\label{app:pse_id}
%  ---  ---  ---  ---  ---  ---  ---  ---  ---  ---  ---  ---  --- - 

Let $\overline{M}_k =(M_1, \cdots, M_k)$ and  $\overline{m}_k$ be a realization of $\overline{M}_k$ (for $k = 1, \ldots, 4$), with  $\overline{M}_0$ and $\overline{m}_0$ assumed to be the empty sets. 
We rely on the following assumptions to identify the counterfactual parameters defined in \eqref{eq:pse_effs}: 

\begin{itemize}
    \item[(A1)] \label{id:potential_consistency} \ Consistency, which indicates that observed outcome and mediators match their counterfactuals when treatment and mediator values are set at observed values; i.e., $Y(r, \overline{m}_4) = Y$ if $R=r$ and $\overline{M}_4 = \overline{m}_4$, and $M_k(r, \overline{m}_{k-1}) = M_k$ if $R=r$ and $\overline{M}_{k-1} = \overline{m}_{k-1}$.
    
    \item[(A2)]  \label{id:potential_positivity} \ Positivity, which  declares that 
    $P(R=1 \mid X=x) > 0$ when $P(X=x) > 0$, and $P(R=1 \mid \overline{M}_k=\overline{m}_k, X=x) > 0$ when $P(\overline{M}_k=\overline{m}_k,X=x) > 0$. 
    
    \item[(A3)] \label{id:potential_ignorability} \ Ignorability, which states that treatment is independent of all counterfactuals given $X$, and any mediator counterfactual is independent of future mediator and outcome counterfactuals given the observed past; i.e., for any $\overline{m}_k, r, r_0, r_k$,   
    $Y(r_0, \overline{m}_4), \underline{M}_4(r_4, \overline{m}_{3}) \perp R \mid X$  
    and 
    $Y(r_0, \overline{m}_4), \underline{M}_{k+1}(r_{k+1}, \overline{m}_{k}) \perp M_k(r, \overline{m}_{k-1}) \mid \overline{M}_{k-1}, R, X$, where  $\underline{M}_5(r_5, \overline{m}_{4})$ is an empty set and $\underline{M}_k(r_k, \overline{m}_{k-1})$ is defined as $(M_k(r_k, \overline{m}_{k-1}), \ldots, M_4(r_4, \overline{m}_{3}))$. Explicitly: 
    \begin{align}
        Y(r_0, \overline{m}_4), M_4(r_4, \overline{m}_{3}), M_3(r_3, \overline{m}_{2}), M_2(r_2, m_{1}), M_1(r_1) \perp R &\mid X \ , 
        \tag{A3.1}
        \label{id:identify_1} 
        \\ 
        Y(r_0, \overline{m}_4), M_4(r_4, \overline{m}_{3}), M_3(r_3, \overline{m}_{2}), M_2(r_2, m_{1}) \perp M_1(r) &\mid R, X \ , 
        \tag{A3.2}
        \label{id:identify_2} 
        \\
        Y(r_0, \overline{m}_4), M_4(r_4, \overline{m}_{3}), M_3(r_3, \overline{m}_{2}) \perp M_2(r, m_{1}) &\mid M_1, R, X \ , 
        \tag{A3.3}
        \label{id:identify_3} 
        \\
        Y(r_0, \overline{m}_4), M_4(r_4, \overline{m}_{3}) \perp M_3(r, \overline{m}_{2}) &\mid \overline{M}_2, R, X \ , 
        \tag{A3.4}
        \label{id:identify_4} 
        \\
        Y(r_0, \overline{m}_4) \perp M_4(r, \overline{m}_{3}) &\mid \overline{M}_3, R, X 
        \tag{A3.5}
        \label{id:identify_5} \ . 
    \end{align}
\end{itemize}

Assumption (A1) indicates that one's observed outcome under the actual value of a target variable equals the outcome that would be observed upon intervening to set the target variable to that value. 
Assumption (A2) indicates that there is sufficient overlap in the distribution of covariates across levels of treatment and mediators. 
Assumption (A3) implies: (i) The effects of treatment on $M_1$ through $M_4$ and $Y$ are unconfounded given covariates; (ii) The effects of $M_1$ on $M_2$ through $M_4$ and $Y$ are unconfounded given treatment and  covariates; (iii) The effects of $M_2$ on $M_3, M_4$, and $Y$ are unconfounded given $M_1$, treatment, and  covariates; (iv) The effects of $M_3$ on $M_4$ and $Y$ are unconfounded given $M_1, M_2$, treatment, and  covariates; (v) The effect of $M_4$ on $Y$ is unconfounded given $M_1$ through $M_3$, treatment, and  covariates. 

Assumptions (A1) and (A2) are standard in the causal inference literature. Assumption (A3) involves ``cross-world'' independencies, which hold under nonparametric structural equation models with independent errors, as described by Pearl \citep{pearl_causality_2009}. In this framework, each variable is generated by an unrestricted structural equation---a nonparametric function of its direct causes (parents in a DAG) and an exogenous error term---where error terms are assumed to be mutually independent. The cross-world assumptions in (A3) are subject to debate, as they govern interdependencies between race, mediators, and outcomes across hypothetical scenarios that may not co-occur in observable reality. Alternative mediation effect definitions, such as \textit{separable effects} or \textit{stochastic interventions} \citep{diaz2020causal, stensrud2022separable, miles2023causal, diaz2024non}, provide different perspectives on mediation estimands and cross-world identification assumptions. While these approaches offer useful insights, we do not pursue them here.  

Under these assumptions, the counterfactual means $\gamma_\text{ref}$, $\gamma_{R \rightarrow Y}$, and $\gamma_{R \rightarrow M_k \leadsto Y}$, for $k = 1, 2, 3, 4$, defined in \eqref{eq:pse_counterfactual_means}, can be identified using the \textit{edge g-formula}, as described by \cite{shpitser13cogsci} and \cite{shpitser15hierarchy}.

\begin{theorem} 
\label{thm:id_results}
Given Assumptions (A1), (A2), and (A3), the counterfactual means defined in \eqref{eq:pse_counterfactual_means}, are identified as follows:  
{\small
\begin{align}
    &\gamma_\text{ref} = \int y dP(y \mid R=0,x) dP(x)  \ ,  \notag \\
    &\gamma_{R \rightarrow Y} = \int y dP(y \mid \overline{m}_4, R=1,x) \prod_{k=1}^4 dP(m_k \mid \overline{m}_{k-1},R=0,x)  dP(x) \ , 
     \label{app:eq:pse_effs_ID_2}  \\
    &\gamma_{R \rightarrow M_k \leadsto Y} \! = \!\! \int \! y dP(y \mid \overline{m}_4, R=0,x) dP(m_k \mid \overline{m}_{k-1}, R=1,x) \!\! \prod_{\substack{j=1, \\ j \neq k}}^4 \!\! dP(m_j \mid \overline{m}_{j-1}, R=0,x) dP(x) \ .  \notag 
\end{align}%
}
\end{theorem} 
Given the identification functionals in Theorem~\ref{thm:id_results}, the effects defined in \eqref{eq:pse_effs} are simply identified by contrasts of identification functionals for $\gamma_{R \rightarrow Y}$ and $\gamma_{R \rightarrow M_k \leadsto Y}$ against $\gamma_\text{ref}$.  

See a proof in Appendix~\ref{app:proofs_ID}.  

%  ---  ---  ---  ---  ---  ---  ---  ---  ---  ---  ---  ---  --- -
\subsection{Decomposition strategies}
\label{app:pse_decomp}
%  ---  ---  ---  ---  ---  ---  ---  ---  ---  ---  ---  ---  --- - 

There are various ways to define path-specific effects when dealing with multiple ordered mediators, as discussed by \cite{daniel_causal_2015} and \cite{steen_flexible_2017}. Assume there are $K$ ordered mediators, $M_1, \ldots, M_K$. We can generalize \eqref{eq:nested_counterfactual} to incorporate $K$ mediators.

Let $(r_0, \mathbf{r})$ denote the vector of values for binary treatment $R$  along the $K+1$ specified pathways, where $r_0 \in \{0, 1\}$ and $\mathbf{r} \coloneqq (r_1, \ldots, r_K) \in \{0,1\}^K$. We define the potential outcome:   
\begin{align}
   Y(r_0, \mathbf{r}) \coloneqq Y 
   \Big( 
        r_0,  
        \underbrace{M_1(r_1)}_{\coloneqq M^c_1}, 
        \underbrace{M_2\big(r_2,M^c_1\big)}_{\coloneqq M^c_2}, 
        \ldots,
        M_K\big(r_K, M^c_1, M^c_2, \ldots, M^c_{K-1} \big) 
   \Big) \ , 
   \label{eq:nested_counterfactual_K}
\end{align}%
where mediators are recursively defined as follows: $M_1(r_1)$ (shorthand: $M^c_1$) is the counterfactual $M_1$ if $R=r_1$, $M_2(r_2, M^c_1)$ (shorthand: $M^c_2$) is the counterfactual $M_2$ if $R=r_2$ and $M_1 = M^c_1$. This recursive structure continues for all mediators. Using this notation, the effect through $M_k$ ($k= 1,\ldots, K$) can be defined as a contrast of the form:
\begin{align*}
    \tilde{\rho}_{R \rightarrow M_k \leadsto Y} 
    = \E[Y(r_0, (r_1, \ldots, r_k=1, \ldots, r_K))] - \E[Y(r_0, (r_1, \ldots, r_k=0, \ldots, r_K))] \ . 
\end{align*}
Given the possible value combinations for $r_0$ and the vector $\bf r$ (with the $k$-th element fixed), there are $2^K$ potential contrasts. This also holds for the direct effect, defined as 
\begin{align*}
    \tilde{\rho}_{R \rightarrow Y} = \E[Y(1,{\bf r})] -\E[Y(0,{\bf r})] \ . 
\end{align*}

This flexibility allows for nuanced interpretations of how distinct pathways contribute to the overall effect, and two common approaches to decomposing PSEs are the \textit{sequential} and \textit{reference-zero} decompositions. To illustrate, consider a setting with two mediators, shown in Figure~\ref{fig:2medDAG}. Let $Y(r_0,r_1,r_2) = Y(r_0,M_1(r_1),M_2(r_2,M_1(r_1)))$  represent the potential outcome if $R$ were set to $r_0$, $M_1$ to its natural value under $R=r_1$, and $M_2$  to its natural value under $R=r_2$ and $M_1(r_1)$. Below, we give examples of these two decompositions.

\begin{figure}[t] 
    \begin{center}
    \scalebox{0.85}{
    \begin{tikzpicture}[>=stealth, node distance=1.5cm]
        \tikzstyle{format} = [thick, circle, minimum size=1.0mm, inner sep=2pt]
        \tikzstyle{square} = [draw, thick, minimum size=4.5mm, inner sep=2pt]
    
		\path[->, line width=1pt]
		
		node[] (a) {$R$}
		node[right of=a, xshift=0.75cm] (m1) {$M_1$}
        node[right of=m1, xshift=0.75cm] (m2) {$M_2$}
		node[above of=m1, xshift=1.2cm,yshift=-0.5cm] (x) {$X$}
		node[right of=m2, xshift=0.75cm] (y) {$Y$}

		(x) edge[black] (a) 
		(x) edge[black] (m1) 
        (x) edge[black] (m2) 
		(x) edge[black, bend right=0] (y) 
        (a) edge[black, bend right=30] (y)
		(a) edge[black] (m1) 
        (a) edge[black, bend right=20] (m2)
		(m1) edge[black] (m2) 
        (m1) edge[black,bend right=20] (y)
        (m2) edge[black] (y)  ; 
	\end{tikzpicture}
	}
        \vspace{-0.2cm}
	\caption{\centering A DAG with two ordered mediators.} 
	\label{fig:2medDAG}
    \end{center}
\end{figure}

%  ---  ---  ---  ---  ---  ---  ---  ---  ---  ---  --- 
\noindent \textit{(1) Sequential decomposition:} In this approach, specific pathways are ``deactivated'' in a fixed order. For the two-mediator setup shown in Figure~\ref{fig:2medDAG}, the PSEs can be defined as:
%  ---  ---  ---  ---  ---  ---  ---  ---  ---  ---  --- 
\begin{align}
    \tilde{\rho}_{R \rightarrow M_1 \leadsto Y} 
    &= \E[Y(1,1,1)] - \E[Y(1,0,1)] 
      \ , \label{eq:seq_m1} \\
    \tilde{\rho}_{R \rightarrow M_2 \rightarrow Y} 
    &= \E[Y(1,0,1)] - \E[Y(1,0,0)] \ ,  \label{eq:seq_m2} \\
   \tilde{\rho}_{R \rightarrow  Y} 
    &= \E[Y(1,0,0)] - \E[Y(0,0,0)] \ . \label{eq:seq_direct}
\end{align}
These effects are referred to as \textit{cumulative path-specific effects}  \citep{zhou_semiparametric_2022}. The total effect is partitioned into $K+1$ components, with each component representing  the cumulative contribution of a specific mediator to the total effect. This decomposition is particularly valuable in applications where investigators aim to quantify the proportion of the overall effect attributable to each component.  

We derive the PSEs using a saturated model without confounders as an illustrative example. Consider the following expression for the mean of the nested potential outcome: 
\begin{align} 
   \E[Y(r_0,r_1,r_2)]  
    =&  \beta_1 r_1 + \beta_{12}r_1 r_2 + \beta_{01} r_0 r_1 + \beta_{012} r_0 r_1 r_2 
    + \beta_2 r_2 + \beta_{02}  r_0 r_2 
    + \beta_0 r_0 
    + \theta  \ . \label{eq:a_saturated}
\end{align}
Thus, based on \eqref{eq:seq_m1} -- \eqref{eq:seq_direct}, the PSEs are given by: 
\begin{align*}
    \tilde{\rho}_{R \rightarrow M_1 \leadsto Y} = \beta_1+ \beta_{12}+ \beta_{01} + \beta_{012} \ , \quad 
    \tilde{\rho}_{R \rightarrow M_2 \rightarrow Y}  = \beta_2+ \beta_{02} \ , \quad 
   \tilde{\rho}_{R \rightarrow Y} = \beta_0 \ . 
\end{align*}
Notably, $\tilde{\rho}_{R \rightarrow M_1 \leadsto Y}$ includes the main effect of $r_ 
1$ ($\beta_1$) but also all interaction terms involving $r_1$ ($\beta_{12}, \beta_{01}, \beta_{012}$). Similarly, $\tilde{\rho}_{R \rightarrow M_2 \rightarrow Y}$ captures the main effect of $r_2$ ($\beta_2$) and the interaction terms involving $r_2$ that does not relate to $r_1$ ($\beta_{02}$). The direct effect, $\tilde{\rho}_{R \rightarrow Y}$, does not include any interaction terms.

\vspace{0.4cm}
%  ---  ---  ---  ---  ---  ---  ---  ---  ---  ---  ---  ---  ---  ---  --- 
\noindent \textit{(2) Reference-zero decomposition: } This method focuses on specific pathways of interest, treating variables as if the treatment is set to the ``active value'' ($R=1$) along the pathways of interest, while along other pathways, variables behave as if the treatment variable is set to the ``baseline value'' ($R=0$). For the two-mediator setup shown in Figure~\ref{fig:2medDAG}, the PSEs can be defined differently, as:
%  ---  ---  ---  ---  ---  ---  ---  ---  ---  ---  ---  ---  ---  ---  --- -
\begin{align}
   \tilde{\rho}_{R \rightarrow M_1 \leadsto Y} &= \E[Y(0,1,0)] - \E[Y(0,0,0)] \ , \\
   \tilde{\rho}_{R \rightarrow M_2 \rightarrow Y} &= \E[Y(0,0,1)] - \E[Y(0,0,0)] \ ,   \\
   \tilde{\rho}_{R \rightarrow Y}  &= \E[Y(1,0,0)] - \E[Y(0,0,0)] \ . 
\end{align}
These effects are referred to as \textit{natural path-specific effects} \citep{daniel_causal_2015}. Cumulative PSEs and natural PSEs share the same representation for the direct effect but differ in how they represent effects through specific mediators. Natural PSEs offer a more intuitive interpretation, such as the average change in $Y$ if the controlled group's  mediator is set to levels observed for the treatment group. 

The natural PSEs derived using the model in \ref{eq:a_saturated} are given by:
\begin{align*}
    \tilde{\rho}_{R \rightarrow M_1 \leadsto Y} = \beta_1 \ , \quad 
    \tilde{\rho}_{R \rightarrow M_2 \rightarrow Y} = \beta_2 \ , \quad 
    \tilde{\rho}_{R \rightarrow Y} = \beta_0 \ . 
\end{align*}
Natural PSEs capture only the main terms $\beta_1, \beta_2, \beta_0$ (for effects through $M_1$, $M_2$, and the direct effect, respectively), excluding any interaction terms. When there are no interactions among ($r_0, \ldots, r_K$), natural PSEs and cumulative PSEs coincide; otherwise, they can diverge---except for the direct effect, which remains the same under both definitions. Additionally, natural PSEs cannot simply be summed to obtain the total effect,  nor do their proportions match the ``proportion mediated'' often reported in mediation analysis. We have already elaborated on natural PSEs in the main text from the perspective of mediator intervention; in Appendix~\ref{app:meps_sequential}, we extend this framework to cumulative PSEs within the same investigation of racial disparities, utilizing the MEPS data. Beyond above decomposition,  \cite{tai2022conventional} proposed decomposing fully mediated interaction from the average causal effect, thereby offering further insight into how complex mediator interactions shape exposure-outcome relationships.

% \clearpage
%  ---  ---  ---  ---  ---  ---  ---  ---  ---  ---  ---  ---  --- -
\section{Proofs}
\label{app:proofs}
%  ---  ---  ---  ---  ---  ---  ---  ---  ---  ---  ---  ---  --- -

\subsection{Identification claims}
\label{app:proofs_ID}

Under identification assumptions \eqref{id:identify_1}---\eqref{id:identify_5}, the counterfactual mean $\E (Y(r_0,r_1,r_2,r_3,r_4))$ is identified as follows: 
{\footnotesize
\begin{align*}
&\hspace{-0.5cm}\E (Y(r_0,r_1,r_2,r_3,r_4))  \\
%  ---  ---  ---  ---  ---  ---  ---  ---  ---  ---  ---  --- -
=& \int \E \Big[Y(r_0,\overline{m}_4) \mid M_1(r_1)=m_1,M_2(r_2, m_1)=m_2, M_3(r_3,\overline{m}_2)=m_3, M_4(r_4,\overline{m}_3)=m_4, x\Big]\\
&\hspace{0.5cm} dP \big(M_4(r_4,\overline{m}_3)=m_4 \mid M_1(r_1)=m_1,M_2(r_2, m_1)=m_2, M_3(r_3,\overline{m}_2)=m_3, x\big)\\
&\hspace{0.5cm} dP\big(M_3(r_3,\overline{m}_2)=m_3 \mid M_1(r_1)=m_1,M_2(r_2, m_1)=m_2,  x\big)\\
&\hspace{0.5cm} dP\big(M_2(r_2, m_1)=m_2 \mid M_1(r_1)=m_1,  x\big)
dP\big(M_1(r_1)=m_1 \mid x\big)dP(x)\\
%  ---  ---  ---  ---  ---  ---  ---  ---  ---  ---  ---  --- -
\stackrel{\ref{id:identify_1}}{=} 
& \int \E \Big[Y(r_0,\overline{m}_4) \mid M_1(r_1)=m_1,M_2(r_2, m_1)=m_2, M_3(r_3,\overline{m}_2)=m_3,M_4(r_4,\overline{m}_3)=m_4, {R=r_0},x\Big]\\
&\hspace{0.5cm} dP\big(M_4(r_4,\overline{m}_3)=m_4 \mid M_1(r_1)=m_1,M_2(r_2, m_1)=m_2, M_3(r_3,\overline{m}_2)=m_3, {R=r_4}, x\big)\\
&\hspace{0.5cm}  dP\big(M_3(r_3,\overline{m}_2)=m_3 \mid M_1(r_1)=m_1,M_2(r_2, m_1)=m_2, {R=r_3}, x\big)\\
&\hspace{0.5cm} dP\big(M_2(r_2, m_1)=m_2 \mid M_1(r_1)=m_1, {R=r_2}, x\big)
dP\big( M_1(r_1)=m_1 \mid {R=r_1},x\big)dP(x)\\
%  ---  ---  ---  ---  ---  ---  ---  ---  ---  ---  ---  --- -
\stackrel{\ref{id:identify_5}}{=} 
& \int \E \Big[Y(r_0,\overline{m}_4) \mid M_1(r_1)=m_1,M_2(r_2, m_1)=m_2, M_3(r_3,\overline{m}_2)=m_3, {R=r_0},x\Big]\\
&\hspace{0.5cm}  dP\big(M_4(r_4,\overline{m}_3)=m_4 \mid M_1(r_1)=m_1,M_2(r_2, m_1)=m_2, M_3(r_3,\overline{m}_2)=m_3, {R=r_4}, x\big)\\
&\hspace{0.5cm}  dP\big(M_3(r_3,\overline{m}_2)=m_3 \mid M_1(r_1)=m_1,M_2(r_2, m_1)=m_2, {R=r_3}, x\big)\\
&\hspace{0.5cm}  dP\big(M_2(r_2, m_1)=m_2 \mid M_1(r_1)=m_1, {R=r_2}, x\big)
dP\big( M_1(r_1)=m_1 \mid {R=r_1},x\big)dP(x)\\
%  ---  ---  ---  ---  ---  ---  ---  ---  ---  ---  ---  --- -
\stackrel{\ref{id:identify_4}}{=} 
& \int \E \Big[Y(r_0,\overline{m}_4) \mid M_1(r_1)=m_1,M_2(r_2, m_1)=m_2,  {R=r_0},x\Big]\\
&\hspace{0.5cm}  dP\big(M_4(r_4,\overline{m}_3)=m_4 \mid M_1(r_1)=m_1,M_2(r_2, m_1)=m_2, {R=r_4}, x\big)\\
&\hspace{0.5cm}  dP\big(M_3(r_3,\overline{m}_2)=m_3 \mid M_1(r_1)=m_1,M_2(r_2, m_1)=m_2, {R=r_3}, x\big)\\
&\hspace{0.5cm}  dP\big(M_2(r_2, m_1)=m_2 \mid M_1(r_1)=m_1, {R=r_2}, x\big)
dP\big( M_1(r_1)=m_1 \mid {R=r_1},x\big)dP(x)\\
%  ---  ---  ---  ---  ---  ---  ---  ---  ---  ---  ---  --- -
\stackrel{\ref{id:identify_3}}{=} 
& \int \E \Big[Y(r_0,\overline{m}_4) \mid M_1(r_1)=m_1,  {R=r_0},x\Big] 
dP\big(M_4(r_4,\overline{m}_3)=m_4 \mid M_1(r_1)=m_1, {R=r_4}, x\big) \\
&\hspace{0.5cm} dP\big(M_3(r_3,\overline{m}_2)=m_3 \mid M_1(r_1)=m_1, {R=r_3}, x\big) 
dP\big(M_2(r_2, m_1)=m_2 \mid M_1(r_1)=m_1, {R=r_2}, x\big) \\
&\hspace{0.5cm} dP\big( M_1(r_1)=m_1 \mid {R=r_1},x\big)dP(x)\\
%  ---  ---  ---  ---  ---  ---  ---  ---  ---  ---  ---  --- -
\stackrel{\ref{id:identify_2}}{=} 
& \int \E \Big[Y(r_0,\overline{m}_4) \mid   {R=r_0},x\Big]dP\big(M_4(r_4,\overline{m}_3)=m_4 \mid  {R=r_4}, x\big)
dP\big(M_3(r_3,\overline{m}_2)=m_3 \mid  {R=r_3}, x\big)\\
&\hspace{0.5cm} dP\big(M_2(r_2, m_1)=m_2 \mid  {R=r_2}, x\big)dP\big( M_1(r_1)=m_1 \mid {R=r_1},x\big)dP(x)\\
%  ---  ---  ---  ---  ---  ---  ---  ---  ---  ---  ---  --- -
\stackrel{\ref{id:identify_2} \& \hyperref[id:potential_consistency]{A1}}{=} 
& \int \E \Big[Y(r_0,\overline{m}_4) \mid M_1=m_1,  {R=r_0},x\Big] 
dP\big(M_4(r_4,\overline{m}_3)=m_4 \mid M_1=m_1, {R=r_4}, x\big)\\
&\hspace{0.5cm} dP\big(M_3(r_3,\overline{m}_2)=m_3 \mid  M_1=m_1, {R=r_3}, x\big)
dP\big(M_2(r_2, m_1)=m_2 \mid  M_1=m_1, {R=r_2}, x\big)\\
&\hspace{0.5cm} dP\big( M_1(r_1)=m_1 \mid {R=r_1},x\big)dP(x)\\
%  ---  ---  ---  ---  ---  ---  ---  ---  ---  ---  ---  --- -
\stackrel{\ref{id:identify_3} \& \hyperref[id:potential_consistency]{A1}}{=}
& \int \E \Big[Y(r_0,\overline{m}_4) \mid M_1=m_1, M_2 = m_2, {R=r_0},x\Big]
dP\big(M_4(r_4,\overline{m}_3)=m_4 \mid M_1=m_1, M_2 = m_2,{R=r_4}, x\big)\\
&\hspace{0.5cm} dP\big(M_3(r_3,\overline{m}_2)=m_3 \mid  M_1=m_1, M_2 = m_2,{R=r_3}, x\big)
dP\big(M_2(r_2, m_1)=m_2 \mid  M_1=m_1,{R=r_2}, x\big)\\
&\hspace{0.5cm} dP\big( M_1(r_1)=m_1 \mid {R=r_1},x\big)dP(x)\\
%  ---  ---  ---  ---  ---  ---  ---  ---  ---  ---  ---  --- -
\stackrel{\ref{id:identify_4} \& \hyperref[id:potential_consistency]{A1}}{=} 
& \int \E \Big[Y(r_0,\overline{m}_4) \mid M_1=m_1, M_2 = m_2, M_3 = m_3,{R=r_0},x\Big]\\
&\hspace{0.5cm} dP\big(M_4(r_4,\overline{m}_3)=m_4 \mid M_1=m_1, M_2 = m_2,M_3 = m_3,{R=r_4}, x\big)\\
&\hspace{0.5cm} dP\big(M_3(r_3,\overline{m}_2)=m_3 \mid  M_1=m_1, M_2 = m_2,{R=r_3}, x\big)\\
&\hspace{0.5cm} dP\big(M_2(r_2, m_1)=m_2 \mid  M_1=m_1,{R=r_2}, x\big)\\
&\hspace{0.5cm} dP\big( M_1(r_1)=m_1 \mid {R=r_1},x\big)dP(x)\\
%  ---  ---  ---  ---  ---  ---  ---  ---  ---  ---  ---  --- -
\stackrel{\ref{id:identify_5} \& \hyperref[id:potential_consistency]{A1}}{=} 
& \int \E \Big[Y(r_0,\overline{m}_4) \mid M_1=m_1, M_2 = m_2, M_3 = m_3,M_4 = m_4,{R=r_0},x\Big]\\
&\hspace{0.5cm} dP\big(M_4(r_4,\overline{m}_3)=m_4 \mid M_1=m_1, M_2 = m_2,M_3 = m_3,{R=r_4}, x\big)\\
&\hspace{0.5cm} dP\big(M_3(r_3,\overline{m}_2)=m_3 \mid  M_1=m_1, M_2 = m_2,{R=r_3}, x\big)\\
&\hspace{0.5cm} dP\big(M_2(r_2, m_1)=m_2 \mid  M_1=m_1,{R=r_2}, x\big)\\
&\hspace{0.5cm} dP\big( M_1(r_1)=m_1 \mid {R=r_1},x\big)dP(x)\\
%  ---  ---  ---  ---  ---  ---  ---  ---  ---  ---  ---  --- -
\stackrel{\hyperref[id:potential_consistency]{A1}}{=} & \int ydP(y \mid r_0,\overline{m}_4, x)dP(m_4 \mid r_4,\overline{m}_3, x)dP(m_3 \mid r_3,\overline{m}_2,x)dP(m_2 \mid r_2, m_1, x)dP(m_1 \mid r_1,x)dP(x)  \ . 
\end{align*}
}

These derivations yield the identification functionals for the estimands in Theorem \ref{thm:id_results}. 

% ****************************************
\subsection{Estimation claims}
\label{app:proofs_est} 
% ****************************************

Let $o = (x, r, \overline{m}_4, y)$ denote the vector values of $O = (X, R, \overline{M}_{4}, Y )$. 

First, note that by the Bayes' rule, we can write: 
\begin{equation}
\begin{aligned}
\frac{p(m_k \mid \overline{m}_{k-1}, R=1,x)}{p(m_k \mid \overline{m}_{k-1}, R=0,x)} 
& = \frac{p(R=1 \mid \overline{m}_k,x)p(m_k \mid \overline{m}_{k-1},x)/p(R=1 \mid \overline{m}_{k-1},x)}{p(R=0 \mid \overline{m}_k,x)p(m_k \mid \overline{m}_{k-1},x)/p(R=0 \mid \overline{m}_{k-1},x)} \\
& = \frac{g_k(\overline{m}_k,x)}{1-g_k(\overline{m}_k,x)}\frac{1-g_{k-1}(\overline{m}_{k-1},x)}{g_{k-1}(\overline{m}_{k-1},x)}  \ . 
\end{aligned}
\label{eq:bayes}
\end{equation}

\vspace{0.2cm}

\noindent $\bullet$ Efficient influence function (EIF) derivation for $\gamma_{R \rightarrow Y} $:
\begin{align}
&\frac{\partial}{\partial \varepsilon} \gamma_{R \rightarrow Y}(P_{\varepsilon})\bigg|_{\varepsilon=0}  \nonumber\\
& = \frac{\partial}{\partial \varepsilon}
\int y dP_{\varepsilon} (y \mid \overline{m}_4, R=1,x) dP_{\varepsilon} (\overline{m}_4 \mid R=0,x) dP_{\varepsilon} (x)\bigg|_{\varepsilon=0} \nonumber\\
& = \int y S(y \mid \overline{m}_4, R=1,x) dP(y \mid \overline{m}_4, R=1,x) dP(\overline{m}_4 \mid R=0,x)  dP(x)  \tag{1} \\
& + \int y S(\overline{m}_4 \mid R=0,x) dP(y \mid \overline{m}_4, R=1,x) dP(\overline{m}_4 \mid R=0,x)  dP(x)  \tag{2} \\
& + \int y S(x) dP(y \mid \overline{m}_4, R=1,x) dP(\overline{m}_4 \mid R=0,x)  dP(x) \ .   \tag{3} 
\end{align}

Line (1) simplifies to:
\begin{equation*}
\begin{aligned}
& \int y S(y \mid \overline{m}_4, R=1,x) dP(y \mid \overline{m}_4, R=1,x) dP(\overline{m}_4 \mid R=0,x)  dP(x) \\
& = \int \frac{\mathbb{I}(R=1)}{p(R=1 \mid x)} \frac{p(\overline{m}_4 \mid R=0,x)}{p(\overline{m}_4 \mid R=1,x)}y S(y \mid \overline{m}_4, R,x) dP(y, \overline{m}_4, R,x) \\
&  \overset{\ref{eq:bayes}}{=} \int \frac{\mathbb{I}(R=1)}{1-\pi(x)} \frac{1-g_4(\overline{m}_4 ,x)}{g_4(\overline{m}_4 ,x)}\big(y - \mu_4(\overline{m}_4, R=1,x)\big) S(o) dP(o) \ .  
\end{aligned}
\end{equation*}

Line (2) simplifies to:
\begin{equation*}
\begin{aligned}
& \int y S(\overline{m}_4 \mid R=0,x) dP(y \mid \overline{m}_4, R=1,x) dP(\overline{m}_4 \mid R=0,x)  dP(x)  \\
& = \int \frac{\mathbb{I}(R=0)}{p(R=0 \mid x)}\mu_4(\overline{m}_4, R=1,x) S(\overline{m}_4 \mid R,x)  dP(\overline{m}_4, R,x)  \\
& = \int \frac{\mathbb{I}(R=0)}{1-\pi(x)}\big(\mu_4(\overline{m}_4, R=1,x) - \mathscr{C}_{\mu_4}(R=0,x)\big)S(o)dP(o) \ .   
\end{aligned}
\end{equation*}

Line (3) simplifies to:
\begin{equation*}
\begin{aligned}
& \int y S(x) dP(y \mid \overline{m}_4, R=1,x) dP(\overline{m}_4 \mid R=0,x)dP(x) \\
& = \int \mathscr{C}_{\mu_4}(R=0,x)S(x)dP(o)\\
& = \int \big(\mathscr{C}_{\mu_4}(R=0,x) - \gamma_{R \rightarrow Y}\big)S(o)dP(o) \ . 
\end{aligned}
\end{equation*}

Therefore, the EIF  for $\gamma_{R\rightarrow Y}$, denoted by $\Phi_{\gamma_{R\rightarrow Y}}(Q)$, is given as follows: 
{\small
\begin{equation}
\begin{aligned}
\Phi_{\gamma_{R\rightarrow Y}}(Q)(O)
 =& \frac{R}{1-\pi(X)} \frac{1-g_4(\overline{M}_4 ,X)}{g_4(\overline{M}_4 ,X)}\left\{Y - \mu_4(\overline{M}_4, R=1,X)\right\} \\
& + \frac{1-R}{1-\pi(X)}\left\{\mu_4(\overline{M}_4, R=1,X) - \mathscr{C}_{\mu_4}(R=0,X)\right\} + \mathscr{C}_{\mu_4}(R=0,X) - \gamma_{R \rightarrow Y} \ . 
\end{aligned}
\end{equation}
}

\vspace{0.5cm}
\noindent $\bullet$ EIF derivation for $\gamma_{R \rightarrow M_k \leadsto Y}$, $k=2,3,4$, where:   
\begin{equation*}
    \gamma_{R\rightarrow M_k \leadsto Y} 
    = \int y dP(y \mid \overline{m}_k, R=0,x) dP(m_k \mid \overline{m}_{k-1}, R=1,x) dP(\overline{m}_{k-1} \mid  R=0,x)  dP(x) \ .   
\end{equation*}

{\footnotesize
\begin{align}
&\frac{\partial}{\partial \varepsilon} \gamma_{R\rightarrow M_k \leadsto Y}(P_{\varepsilon})\bigg|_{\varepsilon=0}  \nonumber\\
& = \frac{\partial}{\partial \varepsilon} \int y dP_{\varepsilon}(y \mid \overline{m}_k, R=0,x) dP_{\varepsilon}(m_k \mid \overline{m}_{k-1}, R=1,x) dP_{\varepsilon}(\overline{m}_{k-1} \mid  R=0,x)  dP_{\varepsilon}(x) \bigg|_{\varepsilon=0} \notag \\
& = \int y S(y \mid \overline{m}_k, R=0,x)dP(y \mid \overline{m}_k, R=0,x) dP(m_k \mid \overline{m}_{k-1}, R=1,x) dP(\overline{m}_{k-1} \mid  R=0,x)  dP(x) \tag{1} \\
& + \int y S(m_k \mid \overline{m}_{k-1}, R=1,x)dP(y \mid \overline{m}_k, R=0,x) dP(m_k \mid \overline{m}_{k-1}, R=1,x) dP(\overline{m}_{k-1} \mid  R=0,x)  dP(x) \tag{2} \\
& + \int y S(\overline{m}_{k-1} \mid  R=0,x)dP(y \mid \overline{m}_k, R=0,x) dP(m_k \mid \overline{m}_{k-1}, R=1,x) dP(\overline{m}_{k-1} \mid  R=0,x)  dP(x) \tag{3} \\
& + \int y S(x)dP(y \mid \overline{m}_k, R=0,x) dP(m_k \mid \overline{m}_{k-1}, R=1,x) dP(\overline{m}_{k-1} \mid  R=0,x)  dP(x) \ .  \tag{4} 
\end{align}
}

Line (1) simplifies to:
{\small
$$
\begin{aligned}
& \int y S(y \mid \overline{m}_k, R=0,x)dP(y \mid \overline{m}_k, R=0,x) dP(m_k \mid \overline{m}_{k-1}, R=1,x) dP(\overline{m}_{k-1} \mid  R=0,x)  dP(x) \\
& = \int \frac{\mathbb{I}(R=0)}{1-\pi(x)} \frac{p(m_k \mid \overline{m}_{k-1}, R=1,x)}{p(m_k \mid \overline{m}_{k-1}, R=0,x)}y S(y \mid \overline{m}_k, R,x)dP(y, \overline{m}_k, R,x)  \\
& = \int \frac{\mathbb{I}(R=0)}{1-\pi(x)} \frac{p(m_k \mid \overline{m}_{k-1}, R=1,x)}{p(m_k \mid \overline{m}_{k-1}, R=0,x)}\big(y-\mu_k(\overline{m}_k, R=0,x)\big) S(o)dP(o)  \\
&  \overset{\ref{eq:bayes}}{=} \int \frac{\mathbb{I}(R=0)}{1-\pi(x)} \frac{g_k(\overline{m}_k,x)}{1-g_k(\overline{m}_k,x)}\frac{1-g_{k-1}(\overline{m}_{k-1},x)}{g_{k-1}(\overline{m}_{k-1},x)} \big(y - \mu_k( \overline{m}_k, R=0,x)\big) S(o) d P(o)  \ . 
\end{aligned}
$$}

Line (2) simplifies to:
{\small
$$
\begin{aligned}
& \int y S(m_k \mid \overline{m}_{k-1}, R=1,x)dP(y \mid \overline{m}_k, R=0,x) dP(m_k \mid \overline{m}_{k-1}, R=1,x) dP(\overline{m}_{k-1} \mid  R=0,x)  dP(x) \\
& = \int \frac{\mathbb{I}(R=1)}{p(R=1 \mid x)}  \frac{p(\overline{m}_{k-1} \mid  R=0,x)}{p(\overline{m}_{k-1} \mid  R=1,x)}\mu_k(\overline{m}_k,R=0,x)S(m_k \mid \overline{m}_{k-1}, R,x) dP(\overline{m}_{k}, R,x) \\
& \overset{\ref{eq:bayes}}{=} \int \frac{\mathbb{I}(R=1)}{1-\pi(x)}  \frac{1-g_{k-1}(\overline{m}_{k-1},x)}{g_{k-1}(\overline{m}_{k-1},x)}\big(\mu_k(\overline{m}_k,R=0,x)-\mathscr{B}_k(\overline{m}_{k-1}, R=1,x)\big)S(o) dP(o) \ . 
\end{aligned}
$$}

Line (3) simplifies to:
{\small
$$
\begin{aligned}
& \int y S(\overline{m}_{k-1} \mid  R=0,x)dP(y \mid \overline{m}_k, R=0,x) dP(m_k \mid \overline{m}_{k-1}, R=1,x) dP(\overline{m}_{k-1} \mid  R=0,x)  dP(x)  \\
& = \int \frac{\mathbb{I}(R=0)}{p(R=0 \mid x)} \mathscr{B}_k(\overline{m}_{k-1}, R=1,x) S(\overline{m}_{k-1} \mid R,x)dP(\overline{m}_{k-1},R,x)\\
& =   \int \frac{\mathbb{I}(R=0)}{1-\pi(x)}  \big(\mathscr{B}_k(\overline{m}_{k-1}, R=1,x) - 
\mathscr{C}_{\mathscr{B}_k}(R=0,x)
\big)S(o)dP(o) \ .  \\
\end{aligned}
$$}

Line (4) simplifies to:
$$
\begin{aligned}
& \int y S(x)dP(y \mid \overline{m}_k, R=0,x) dP(m_k \mid \overline{m}_{k-1}, R=1,x) dP(\overline{m}_{k-1} \mid  R=0,x)  dP(x) \\
& = \int\mathscr{C}_{\mathscr{B}_k}(R=0,x)S(x)dP(x) \\
& = \int \big(\mathscr{C}_{\mathscr{B}_k}(R=0,x) - \gamma_{R\rightarrow M_k \leadsto Y}\big)S(o)dP(o) \ . 
\end{aligned}
$$

Therefore, the EIF for $\gamma_{R\rightarrow M_k \leadsto Y}$, denoted by $\Phi_{\gamma_{R\rightarrow M_k \leadsto Y}}(Q)$, is given by:  
\begin{equation}
\begin{aligned}
 \Phi_{\gamma_{R\rightarrow M_k \leadsto Y}}(Q)(O) 
 &= \frac{1-R}{1-\pi(X)} \frac{g_k(\overline{M}_k,X)}{1-g_k(\overline{M}_k,X)}\frac{1-g_{k-1}(\overline{M}_{k-1},X)}{g_{k-1}(\overline{M}_{k-1},X)} \left\{Y - \mu_k(\overline{M}_{k}, R=0, X)\right\} \\
 & + \frac{R}{1-\pi(X)}  \frac{1-g_{k-1}(\overline{M}_{k-1},X)}{g_{k-1}(\overline{M}_{k-1},X)}\left\{\mu_k(\overline{M}_k,R=0,X)-\mathscr{B}_k(\overline{M}_{k-1}, R=1,X)\right\} \\
 & + \frac{1-R}{1-\pi(X)}  \left\{\mathscr{B}_k(\overline{M}_{k-1}, R=1,X) - \mathscr{C}_{\mathscr{B}_k}(R=0,X) \right\} \\
 &+ \mathscr{C}_{\mathscr{B}_k}(R=0,X) - \gamma_{R\rightarrow M_k \leadsto Y} \ . 
\end{aligned}
\end{equation}

\vspace{0.5cm}
\noindent $\bullet$ EIF derivation for $\gamma_{R\rightarrow M_1 \leadsto Y}$, where 
\begin{equation*}
    \gamma_{R\rightarrow M_1 \leadsto Y} 
    = \int y dP(y \mid m_1, R=0,x) dP(m_1 \mid  R=1,x) dP(x) \ . 
\end{equation*}
\begin{align}
&\frac{\partial}{\partial \varepsilon} \gamma_{R\rightarrow M_1 \leadsto Y}(P_{\varepsilon})\bigg|_{\varepsilon=0}  \nonumber\\
& = \frac{\partial}{\partial \varepsilon} \int y dP_{\varepsilon}(y \mid m_1, R=0,x) dP_{\varepsilon}(m_1 \mid  R=1,x) dP_{\varepsilon}(x)  \bigg|_{\varepsilon=0} \notag \\
& = \int y S(y \mid m_1, R=0,x)dP(y \mid m_1, R=0,x) dP(m_1 \mid  R=1,x) dP(x) \tag{1}\\
& + \int y S(m_1 \mid  R=1,x)dP(y \mid m_1, R=0,x) dP(m_1 \mid  R=1,x) dP(x) \tag{2}\\
& + \int y S(x)dP(y \mid m_1, R=0,x) dP(m_1 \mid  R=1,x) dP(x) \ . \tag{3} 
\end{align}

Line (1) simplifies to:
$$
\begin{aligned}
& \int y S(y \mid m_1, R=0,x)dP(y \mid m_1, R=0,x) dP(m_1 \mid  R=1,x) dP(x) \\
& = \int \frac{\mathbb{I}(R=0)}{p(R=0 \mid x)} \frac{p(m_1 \mid  R=1,x)}{p(m_1 \mid  R=0,x)}y S(y \mid m_1, R=0,x)dP(y , m_1, R=0,x) \\
& = \int \frac{\mathbb{I}(R=0)}{\pi(x)} \frac{p(m_1 \mid  R=1,x)}{p(m_1 \mid  R=0,x)}y S(y \mid m_1, R,x)dP(y , m_1, R,x) \\
& \overset{\ref{eq:bayes}}{=} \int \frac{\mathbb{I}(R=0)}{\pi(x)} \frac{g_1(m_1, x)}{1-g_1(m_1, x)}\big(y - \mu_1(m_1,R=0,x) \big)S(o)dP(o) \ . 
\end{aligned}
$$

Line (2) simplifies to:
$$
\begin{aligned}
& \int y S(m_1 \mid  R=1,x)dP(y \mid m_1, R=0,x) dP(m_1 \mid  R=1,x) dP(x)\\
& = \int \frac{\mathbb{I}(R=1)}{p(R=1 \mid x)} \mu_1(m_1,R=0,x) S(m_1 \mid  R,x) dP(m_1,  R,x)  \\
& = \int \frac{\mathbb{I}(R=1)}{\pi(x)} \big(\mu_1(m_1,R=0,x) - \mathscr{C}_{\mu_1}(R=1,x)\big) S(o) dP(o) \ .  
\end{aligned}
$$

Line (3) simplifies to:
$$
\begin{aligned}
& \int y S(x)dP(y \mid m_1, R=0,x) dP(m_1 \mid  R=1,x) dP(x)\\
& = \int \mathscr{C}_{\mu_1}(R=1,x) S(x)dP(x)  \\
& = \int \big(\mathscr{C}_{\mu_1}(R=1,x) - \gamma_{R\rightarrow M_1 \leadsto Y}  \big)S(x)dP(x) \ . 
\end{aligned}
$$

Therefore, the EIF for $\gamma_{R\rightarrow M_1 \leadsto Y}$, denoted by $\Phi_{\gamma_{R\rightarrow M_1 \leadsto Y}}(Q)$, is given by:
{\small
\begin{equation}
\begin{aligned}
 \Phi_{\gamma_{R\rightarrow M_1 \leadsto Y}}(Q)(O) 
 &= \frac{1-R}{\pi(X)} \frac{g_1(M_1,X)}{1-g_1(M_1,X)} \left\{Y - \mu_1(M_{1}, R=0, X)\right\} \\
 &\quad + \frac{R}{\pi(X)} \left\{\mu_1(M_1,R=0,X) - \mathscr{C}_{\mu_1}(R=1,X)\right\} +\mathscr{C}_{\mu_1}(R=1,X) - \gamma_{R\rightarrow M_1 \leadsto Y} \ . 
\end{aligned}
\end{equation}
}

\vspace{0.5cm}
Due to the identities $g_0(M_0, X) = \pi(X)$ and $\mathscr{B}_1(R=1, X) = \mathscr{C}_{\mathscr{B}_1}(R=1, X)$, the EIF for $\gamma_{R\rightarrow M_1 \leadsto Y}$ can be  incorporated into the expression for $\gamma_{R\rightarrow M_k \leadsto Y}$ for $k = 2, 3, 4$. 

\subsection{Inference claims}
\label{app:proofs_inf} 

In Theorem \ref{thm:asymp_lin} and Corollary \ref{cor:robustness}, certain regularity conditions are required for the empirical process term to be negligible, i.e., $(P_n - P) (\Phi(\hat{Q}) - \Phi(Q)) = o_P(n^{-1/2})$. These conditions are as follows: 

\begin{enumerate}
    \item $\Phi(\hat{Q}) - \Phi(Q)$ belongs to a P-Donsker class with probability tending to 1, and
    \item $\Phi(\hat{Q})$ is $L^2(P)$-consistent: $P\{ \Phi(\hat{Q}) - \Phi(Q) \}^2 = o_P(1)$. 
\end{enumerate}
The first condition can be relaxed using sample-splitting procedures \citep{double17chernozhukov}. Additionally, we require, for $\delta > 0$: $\delta < \hat{\pi} < 1 - \delta$ and  $\delta < \hat{g}_k < 1- \delta$, $k = 1,2,3,4$.

It remains to derive the remainder terms for $\gamma^{+}_{R \rightarrow Y}(\hat{Q})$ and $\gamma^{+}_{R \rightarrow M_k \leadsto Y}(\hat{Q})$, denoted by $R_{2, \gamma_{R \rightarrow Y}}(\hat{Q}, Q)$ and $R_{2, \gamma_{R \rightarrow M_k \leadsto Y}}(\hat{Q}, Q)$, respectively. Below, we show these remainder terms are:  ($\pi \equiv g_{0}$ and $\mathscr{B}_1 \equiv \mathscr{C}_{\mathscr{B}_1}$)
\begin{align}
    &R_{2, \gamma_{R \rightarrow Y}}(\hat{Q}, Q) 
    = P\left[ \frac{1}{1-\hat{\pi}} \frac{1}{\hat{g}_4} (\hat{g}_4 - g_4) (\hat{\mu}_4 - \mu_4) + \frac{1}{1-\hat{\pi}} (\pi - \hat{\pi}) (\hat{\mathscr{C}}_{\mu_4} - \mathscr{C}_{\mu_4}) \right],
    \\ 
    &R_{2, \gamma_{R \rightarrow M_k \leadsto Y}}(\hat{Q}, Q) 
    =  P\Big[ \frac{1}{1-\hat{\pi}} \frac{1}{\hat{g}_{k-1}} \Big\{ \frac{1-\hat{g}_{k-1}}{1-\hat{g}_k} (g_k - \hat{g}_k) (\hat{\mu}_k - \mu_k) 
    + (\hat{g}_{k-1} - g_{k-1}) (\hat{\mathscr{B}}_k - \mathscr{B}_k) \Big\}  \notag \\ 
    &\hspace{4cm} +  \frac{1}{1-\hat{\pi}}  (\pi - \hat{\pi}) (\hat{\mathscr{C}}_{\mathscr{B}_k} - \mathscr{C}_{\mathscr{B}_k}) \Big] \ ,  \ k=1,2,3,4 \ . 
\end{align}
Note that conditions for $R_2(\hat{Q}, Q) = o_P(n^{-1/2})$ are equivalent to each nuisance product term having an $L^2(P)$ convergence rate equal or faster than $o_P(n^{-1/2})$, with finite scaling factors. 

Let $h(Q)(O) = \Phi(Q)(O) + \gamma(Q)$, and thus $\gamma^{+}(\hat{Q}) = P_n[h(\hat{Q})] = \frac{1}{n} \sum_{i=1}^n h(\hat{Q})(O_i)$. We propose a special set of estimated nuisance parameters  $\widetilde{Q} = (\hat{\pi}, \hat{g}, \mathscr{C}, \mathscr{B}, \mu)$ where all the outcome and sequential regression nuisances are correctly estimated. Our first step is to prove that $P[h(\widetilde{Q})] = \gamma$, where $P[h(Q)] = \int h(Q)(o)dP(o).$  

\noindent $\bullet$ For $\gamma_{R \rightarrow Y}$: 
\begin{align}
    P\left[h_{\gamma_{R \rightarrow Y}}(\widetilde{Q})\right]  
    = & P\left[\frac{R}{1-\hat{\pi}} \frac{1-\hat{g}_4}{\hat{g}_4} \underbrace{E\left(Y-\mu_4 \mid \overline{M}_4, R=1, X\right)}_{=0}\right] \nonumber \\ 
    & + P\left[\frac{1-R}{1-\hat{\pi}} \underbrace{E\left(\mu_4-\mathscr{C}_{\mu_4} \mid R=0, X\right)}_{=0}\right] + P\left[ \mathscr{C}_{\mu_4} \right] \nonumber \\
    = & P\left[ \mathscr{C}_{\mu_4} \right] = \gamma_{R \rightarrow Y} \ .
\end{align}

\noindent $\bullet$ For $\gamma_{R \rightarrow M_k \leadsto Y}$: 
\begin{align}
    P[h_{\gamma_{R \rightarrow M_k \leadsto Y}}(\widetilde{Q})]
    = & P\left[\frac{1-R}{1-\hat{\pi}} \frac{\hat{g}_k}{1-\hat{g}_k} \frac{1-\hat{g}_{k-1}}{\hat{g}_{k-1}} \underbrace{E\left(Y-\mu_k \mid \overline{M}_k, R=0, X\right)}_{=0}\right] \nonumber \\
    & + P\left[\frac{R}{1-\hat{\pi}} \frac{1-\hat{g}_{k-1}}{\hat{g}_{k-1}} \underbrace{E\left(\mu_k-\mathscr{B}_k \mid \overline{M}_{k-1}, R=1, X\right)}_{=0}\right] \nonumber \\ 
    & + P\left[\frac{1-R}{1-\hat{\pi}} \underbrace{E\left(\mathscr{B}_k-\mathscr{C}_{\mathscr{B}_k} \mid R=0, X\right)}_{=0}\right]  
     + P\left[\mathscr{C}_{\mathscr{B}_k}\right] \nonumber \\ 
    = & P\left[\mathscr{C}_{\mathscr{B}_k}\right] = \gamma_{R \rightarrow M_k \leadsto Y} \ .
\end{align}

With $P[h(\widetilde{Q})] = \gamma(Q)$, the second-order remainder term can be re-written as $R_2(\hat{Q},Q) = P[h(\hat{Q})] - \gamma(Q) = P[h(\hat{Q}) - h(\widetilde{Q})]$. Using this fact, the second-order remainder terms can be derived as follows: 
\begin{align}
    R_{2, R \rightarrow Y}(\hat{Q}, Q) 
    = & P\left\{\frac{R}{1-\hat{\pi}} \frac{1-\hat{g}_4}{\hat{g}_4}\left[\left(Y-\hat{\mu}_4\right)-\left(Y-\mu_4\right)\right]\right\} \nonumber \\ 
    & + P\left\{\frac{1-R}{1-\hat{\pi}}\left[\left(\hat{\mu}_4-\hat{\mathscr{C}}_{\mu_4}\right)-\left(\mu_4-\mathscr{C}_{\mu_4}\right)\right]\right\} + P\left(\hat{\mathscr{C}}_{\mu_4}-\mathscr{C}_{\mu_4}\right) \nonumber \\
    = & -P\left[\frac{g_4}{1-\pi} \frac{1-\hat{g}_4}{\hat{g}_4}\left(\hat{\mu}_4-\mu_4\right)\right] + P\left[\frac{1-g_4}{1-\hat{\pi}}\left(\hat{\mu}_4-\mu_4\right)\right] \nonumber \\
    & - P\left[\frac{1-\pi}{1-\hat{\pi}}\left(\hat{\mathscr{C}}_{\mu_4}-\mathscr{C}_{\mu_4}\right)\right] + P\left[\hat{\mathscr{C}}_{\mu_4}-\mathscr{C}_{\mu_4}\right] \nonumber \\
    = & P\left[ \frac{1}{1-\hat{\pi}} \frac{1}{\hat{g}_4} (\hat{g}_4 - g_4) (\hat{\mu}_4 - \mu_4) \right] + P\left[ \frac{1}{1-\hat{\pi}} (\pi - \hat{\pi}) (\hat{\mathscr{C}}_{\mu_4} - \mathscr{C}_{\mu_4}) \right] \ . 
\end{align}

\begin{align}
    R_{2, R \rightarrow M_k \leadsto Y}(\hat{Q}, Q) 
    = & P\left\{\frac{1-R}{1-\hat{\pi}} \frac{\hat{g}_k}{1-\hat{g}_k} \frac{1-\hat{g}_{k-1}}{\hat{g}_{k-1}}\left[\left(Y-\hat{\mu}_k\right)-\left(Y-\mu_k\right)\right]\right\} \nonumber \\
    & + P\left\{\frac{R}{1-\hat{\pi}} \frac{1-\hat{g}_{k-1}}{\hat{g}_{k-1}}\left[\left(\hat{\mu}_k-\hat{\mathscr{B}}_k\right)-\left(\mu_k-\mathscr{B}_k\right)\right]\right\} \nonumber \\
    & + P\left\{\frac{1-R}{1-\hat{\pi}}\left[\left(\hat{\mathscr{B}}_k-\hat{\mathscr{C}}_{\mathscr{B}_k}\right)-\left(\mathscr{B}_k-\mathscr{C}_{\mathscr{B}_k}\right)\right]\right\} \nonumber \\
    & + P\left\{\hat{\mathscr{C}}_{\mathscr{B}_k}-\mathscr{C}_{\mathscr{B}_k}\right\} \nonumber \\
    = & -P\left[\frac{1-g_k}{1-\hat{\pi}} \frac{\hat{g}_k}{1-\hat{g}_k} \frac{1-\hat{g}_{k-1}}{\hat{g}_{k-1}}\left(\hat{\mu}_k-\mu_k\right)\right] + P\left[\frac{g_k}{1-\hat{\pi}} \frac{1-\hat{g}_{k-1}}{\hat{g}_{k-1}}\left(\hat{\mu}_k-\mu_k\right)\right] \nonumber \\
    & - P\left[\frac{g_{k-1}}{1-\hat{\pi}} \frac{1-\hat{g}_{k-1}}{\hat{g}_{k-1}}\left(\hat{\mathscr{B}}_k-\mathscr{B}_k\right)\right] + P\left[\frac{1-g_{k-1}}{1-\hat{\pi}}\left(\hat{\mathscr{B}}_k-\mathscr{B}_k\right)\right] \nonumber \\
    & - P\left[\frac{1-\pi}{1-\hat{\pi}}\left(\hat{\mathscr{C}}_{\mathscr{B}_k}-\mathscr{C}_{\mathscr{B}_k}\right)\right] + P\left[\hat{\mathscr{C}}_{\mathscr{B}_k}-\mathscr{C}_{\mathscr{B}_k}\right] \nonumber \\
    = & P\left[ \frac{1}{1-\hat{\pi}} \frac{1}{1-\hat{g}_k} \frac{1-\hat{g}_{k-1}}{\hat{g}_{k-1}} (g_k - \hat{g}_k) (\hat{\mu}_k - \mu_k) \right] \nonumber \\ 
    & + P\left[ \frac{1}{1-\hat{\pi}} \frac{1}{\hat{g}_{k-1}} (\hat{g}_{k-1} - g_{k-1}) (\hat{\mathscr{B}}_k - \mathscr{B}_k) \right] \nonumber \\ 
    & + P\left[ \frac{1}{1-\hat{\pi}} (\pi - \hat{\pi}) (\hat{\mathscr{C}}_{\mathscr{B}_k} - \mathscr{C}_{\mathscr{B}_k}) \right] \ , 
\end{align}%

for $k=1,2,3,4$. Note that when $k=1$, the $R_2$ term reduces to: 
\begin{align}
    R_{2, R \rightarrow M_1 \leadsto Y} (\hat{Q}, Q) 
    = & P\left[ \frac{1}{\hat{\pi}} \frac{1}{1-\hat{g}_1} (g_1 - \hat{g}_1) (\hat{\mu}_1 - \mu_1) \right] + P\left[ \frac{1}{\hat{\pi}} (\hat{\pi} - \pi) (\hat{\mathscr{B}}_1 - \mathscr{B}_1) \right].
\end{align}

With the second-order remainder terms expressed as a sum of cross-product terms, regularity conditions are required to ensure that these terms are negligible, i.e., $o_P(n^{-1/2})$. Specifically, all denominators must be bounded away from zero. Thus, the propensity score estimates $\hat{\pi}$ and $\hat{g}_k$ for $k=1,2,3,4$ must satisfy $0 <\hat{\pi} < 1$ and $0 < \hat{g}_k < 1$. Under this regularity assumption, the second-order remainder terms can be expressed as:
\begin{align}
    R_{2, R \rightarrow Y}(\hat{Q}, Q) =& P\left[m_1(\hat{\pi}, \hat{g}_4) \cdot (\hat{g}_4 - g_4) \cdot (\hat{\mu}_4 - \mu_4 )\right] + P\left[m_2(\hat{\pi}) \cdot (\pi - \hat{\pi}) \cdot (\hat{\mathscr{C}}_{\mu_4} - \mathscr{C}_{\mu_4}) \right] \ , 
    \\
    R_{2, R \rightarrow M_k \leadsto Y}(\hat{Q}, Q) =& P\left[m_3(\hat{\pi}, \hat{g}_k, \hat{g}_{k-1}) \cdot (g_k - \hat{g}_k) \cdot (\hat{\mu}_k - \mu_k) \right] \nonumber \\
    &+ P\left[m_1(\hat{\pi}, \hat{g}_{k-1}) \cdot (\hat{g}_{k-1} - g_{k-1}) \cdot (\hat{\mathscr{B}}_k - \mathscr{B}_k) \right] \nonumber \\ 
    &+ P\left[m_2(\hat{\pi}) \cdot (\pi - \hat{\pi}) \cdot (\hat{\mathscr{C}}_{\mathscr{B}_k} - \mathscr{C}_{\mathscr{B}_k} )\right] \ . 
\end{align}

Here, the functions $m_1$, $m_2$ and $m_3$ are bounded.  Consequently, the overall negligibility of the second-order remainder terms depends only on the $L^2(P)$ convergence rates of the nuisance estimates in combinations corresponding to the product terms. Specifically, as long as the combined $L^2(P)$ convergence rate of the two nuisance estimates in each product term is faster than $o_p(n^{-1/2})$,  the remainder term $R_2(\hat{Q},Q)$ would also be $o_p(n^{-1/2})$. This negligibility condition enables the discussion of the asymptotic linearity of the one-step corrected plug-in estimators. Given that $\gamma^{+}(\hat{Q}) - \gamma(Q) = P_n(\Phi(Q)) + o_p(n^{-1/2})$, the central limit theorem implies $\sqrt{n}(\gamma^{+}(\hat{Q}) - \gamma) \rightarrow^d N(0, \E[\Phi^2(Q)])$. This is formally presented in Theorem \ref{thm:asymp_lin}.

Regarding consistency, as long as at least one component of each nuisance product term is consistently estimated (i.e., the difference between the nuisance estimate and its true value is $o_p(1)$), the one-step corrected plug-in estimator will be consistent. This robustness property is discussed in detail in Corollary \ref{cor:robustness}.

% \newpage
%  ---  ---  ---  ---  ---  ---  ---  ---  ---  ---  ---  ---  --- -
\section{Details of the MEPS data}
\label{app:meps_details}
%  ---  ---  ---  ---  ---  ---  ---  ---  ---  ---  ---  ---  --- -  

\subsection{Description of the MEPS data}
\label{app:meps_description}

The Medical Expenditures Panel Survey (MEPS), co-sponsored by the Agency for Healthcare Research and Quality and the National Center for Health Statistics, is a large-scale survey that collects detailed data on healthcare costs, use, and insurance coverage from families, individuals, medical providers, and employers across the United States. MEPS is a crucial resource for health services research and policy analysis due to its comprehensive individual-level data. For our analysis, we used the MEPS household components of the 2009 and 2016. 

The sample exclusion process is shown in Figure~\ref{fig:exclusion}. We restricted the analysis to adults aged 18 years or older and individuals reporting a single race/ethnicity. Participants reporting more than one race were excluded to ensure mutually exclusive racial/ethnic categories. After applying these two exclusion criteria, the sample yielded a total of  $n=25{,}454$ for 2009 and  $n=24{,}475$ for 2016.

The MEPS data do not encode missing values as standard missing indicators such as NA. Instead, special codes are used, including $-1$ (inapplicable), $-7$ (refused), $-8$ (don't know), and $-9$ (not ascertained). We recoded all such values as missing and conducted a complete case analysis, excluding observations with missing data on any variables included in the analysis. We report the extent of missingness across all variables in both datasets in Table~\ref{tab:missing_summary}. Overall, missingness was low for most variables (generally below 1\%), with the exception of smoking and selected health status measures, including PCS and MCS, which exhibited higher levels of missingness.

The final analytic samples for 2009 included self-reported non-Hispanic Whites ($n=9{,}955$), non-Hispanic Blacks ($n=3{,}967$), non-Hispanic Asians ($n=1{,}468$), and Hispanics ($n=5{,}399$), yielding a total of $n=20{,}789$. For 2016, the final sample included non-Hispanic Whites ($n=8{,}763$), non-Hispanic Blacks ($n=3{,}576$), non-Hispanic Asians ($n=1{,}535$), and Hispanics ($n=5{,}634$), for a total of $n=19{,}508$. 

\begin{figure}[t]
\centering
\scalebox{0.8}{
\begin{tikzpicture}[
box/.style={draw, rectangle, rounded corners, minimum width=2cm, minimum height=1cm, align=center},
arrow/.style={->, thick},
node distance=3cm
]
% ---------- Row 1 : MEPS 2009 ----------
\node[box] (raw09) {MEPS 2009 \\ $n=36{,}855$};

\node[box, right=3cm of raw09] (age09) {$n=26{,}008$};
\node[box, right=4cm of age09] (race09) {$n=25{,}454$};
\node[box, right=3cm of race09] (final09) {$n=20{,}789$};

\draw[arrow] (raw09) -- node[above]{Age $\geq 18$} (age09);
\draw[arrow] (age09) -- node[above]{One race/ethnicity} (race09);
\draw[arrow] (race09) -- node[above]{\shortstack{Complete case}} (final09);

% ---------- Row 2 : MEPS 2016 ----------
\node[box, below=0.5cm of raw09] (raw16) {MEPS 2016 \\ $n=34{,}655$};

\node[box, right=3cm of raw16] (age16) {$n=25{,}200$};
\node[box, right=4cm of age16] (race16) {$n=24{,}475$};
\node[box, right=3cm of race16] (final16) {$n=19{,}508$};

\draw[arrow] (raw16) -- node[above]{Age $\geq 18$} (age16);
\draw[arrow] (age16) -- node[above]{One race/ethnicity} (race16);
\draw[arrow] (race16) -- node[above]{\shortstack{Complete case}} (final16);
\end{tikzpicture}
}
\caption{Sample exclusion process for MEPS 2009 and 2016. The analysis was restricted to adults aged 18 years or older and individuals reporting a single self-identified race/ethnicity; participants reporting multiple races were excluded to ensure mutually exclusive groups. Complete case analysis was used, excluding observations with missing or invalid values on any variables included in the analysis. In MEPS, such values include codes for inapplicable ($-1$), refused ($-7$), don't know ($-8$), and not ascertained ($-9$).}
\label{fig:exclusion}
\end{figure}

\begin{table}[t]
\caption{Number of missing observations by variable after restricting to adults with a single race/ethnicity (n = 25,454 in 2009 and n = 24,475 in 2016)}
\label{tab:missing_summary}
 % or \footnotesize for slightly larger font
\renewcommand{\arraystretch}{1.2} % Row spacing (default is 1)
\setlength{\tabcolsep}{12pt} 
\centering
\begin{tabular}{lcc}
\toprule
Variable & Missing in 2009 & Missing in 2016 \\ 
\midrule\addlinespace[2.5pt]
Education &   281 (1.1\%) &   311 (1.3\%) \\ 
Employment &   134 (0.5\%) &   133 (0.5\%) \\ 
Smoke & 3,253 (12.8\%) & 3,863 (15.8\%) \\ 
Exercise &   445 (1.7\%) &   495 (2.0\%) \\ 
BMI &   875 (3.4\%) &   771 (3.2\%) \\ 
Mental health &   153 (0.6\%) &   161 (0.7\%) \\ 
Health &   145 (0.6\%) &   157 (0.6\%) \\ 
PCS & 3,029 (11.9\%) & 3,693 (15.1\%) \\ 
MCS & 3,017 (11.9\%) & 3,681 (15.0\%) \\ 
Any limitation &   822 (3.2\%) &   557 (2.3\%) \\ 
Social limitation &    98 (0.4\%) &   113 (0.5\%) \\ 
Cognitive limitation &   209 (0.8\%) &   147 (0.6\%) \\ 
Diabetes &    54 (0.2\%) &    46 (0.2\%) \\ 
Asthma &    50 (0.2\%) &    47 (0.2\%) \\ 
High blood pressure &    69 (0.3\%) &    60 (0.2\%) \\ 
Coronary heart disease &    58 (0.2\%) &    55 (0.2\%) \\ 
Angina &    59 (0.2\%) &    54 (0.2\%) \\ 
Myocardial infarction &    51 (0.2\%) &    49 (0.2\%) \\ 
Stroke &    50 (0.2\%) &    47 (0.2\%) \\ 
Emphysema &    51 (0.2\%) &    48 (0.2\%) \\ 
Cholesterol &    72 (0.3\%) &    73 (0.3\%) \\ 
Arthritis &    61 (0.2\%) &    56 (0.2\%) \\ 
Cancer &    58 (0.2\%) &    47 (0.2\%) \\ 
\bottomrule
\end{tabular}
\end{table}

These samples collected information on individuals' baseline characteristics, family SES, SES, health insurance access, health behaviors, health status, and healthcare expenditures across different racial groups. A detailed breakdown of these variables is provided below. 

\textit{Family SES} includes family poverty.  Poverty level was computed by dividing family income by the applicable poverty line (based on family size and composition) and classified into one of five categories: negative or poor (less than 100\%), near poor (100\% to less than 125\%), low income (125\% to less than 200\%), middle income (200\% to less than 400\%), and high income (greater than or equal to 400\%  of the poverty line). 

\textit{Baseline characteristics} include demographic information such as age and sex, as well as geographic region. Age is recorded as the exact age of each individual as of December 31 of the survey year, with the sample ranging from 18 to 85 years old. Sex, which includes male and female,  was verified and corrected during each MEPS interview. Geographic region is categorized according to U.S. Census regions: Northeast, Midwest, South, and West. 

\textit{SES} was measured by income, education and employment status. Total income was defined as the sum of all person-level income components, excluding the person's sales income, to align as closely as possible with the \textit{Current Population Survey} definition of income. Education was categorized into four levels: less than high school, high school, college, and graduate education. Employment status was defined as a binary variable. Individuals were classified as employed if they reported being currently employed and identified a current main job, or if they reported having a job to which they expected to return.

For \textit{insurance access}, individuals were considered uninsured if they were not covered by one of the following sources in the survey year: TRICARE, Medicare, Medicaid, State Children's Health Insurance Program (SCHIP), or other public hospital/physician insurance, or private hospital/physician insurance. 

\textit{Health behaviors} were assessed using two variables: smoking and exercise. Smoking status indicated whether an individual was a current smoker, while exercise indicated whether a person had currently spent half hour or more in moderate to vigorous physical activity at least five times a week. 

\textit{Health status} was measured across several dimensions: (1) anthropometric measures, such as BMI ($\text{kg}/\text{m}^2$); (2) health perception, including perceived health status and perceived mental health status (both measured on a 5-point scale: excellent, very good, good, fair, and poor), as well as Physical Component Summary (PCS) and Mental Component Summary (MCS) scores; (3) functional status, assessed by  cognition limitations, social limitations (such as the use of assistive technology and recreation), and any limitations in daily living activities, functional, or sensory abilities; and (4) chronic conditions, including diabetes, asthma, high blood pressure, coronary heart disease, angina, myocardial infarction, stroke, emphysema, cholesterol, arthritis, and cancer. 

The \textit{outcome} of interest is annual total healthcare expenditures, defined as the sum of direct payments for care provided during the year, including out-of-pocket payments and payments by private insurance, Medicaid, Medicare, and other sources. Payments for over-the-counter drugs are not included in MEPS total expenditures. 

Table~\ref{tab:description_09_16} presents descriptive statistics on observed covariates, SES, insurance access, health behaviors, health status, and healthcare expenditures across the four racial groups in both 2009 and 2016. The variable types are reflected in Table~\ref{tab:description_09_16}, where continuous variables are summarized using medians, binary variables are presented as proportions, and categorical variables are displayed across levels. The racial composition was similar between 2009 and 2016, with non-Hispanic Whites comprising approximately half of the overall sample, while Asians accounted for the smallest proportion, around 7\%. Whites had the highest median healthcare expenditures at \$ 1,675  in 2009 and at \$ 2,093  in 2016 respectively, whereas Hispanics reported the lowest median expenditures during the same periods. The medians of healthcare expenditures increased across all racial groups from 2009 to 2016.  
To assess whether various factors differed significantly across the racial groups, categorical variables were compared across racial groups using the Chi-square test, while continuous variables were compared using Kruskal-Wallis rank sum test. Significant differences in SES, insurance access, health behaviors, and health status were observed across all racial groups within 2009 and 2016.  

Table~\ref{tab:expenditure09_16} shows the median healthcare expenditures in both 2009 and 2016 stratified by race and other characteristic levels. Overall, older adults and those living in northern and Midwest regions tended to have higher median expenditures. Females spent more in healthcare compared with males. Additionally, individuals with higher educational attainment and income levels, as well as those enrolled in insurance programs, had significantly higher healthcare expenditures --- nearly \$ 1,400  difference of median for the insured compared to the uninsured. Conversely, participants who engaged in regular exercise and reported better health status had lower healthcare expenditures. These expenditure trends were consistent across the four racial groups.

All variables were included in the estimation procedure according to their original measurement scale, as summarized in Table~\ref{tab:description_09_16}. Continuous variables (e.g., BMI, PCS, MCS) were entered as continuous measures; binary variables (e.g., smoking status, exercise, and specific conditions) were included as indicator variables; and categorical variables (e.g., self-reported health status) were represented using factor variables, with one category designated as the reference level. Variables corresponding to the same mediator domain were included jointly in the model. Estimation was conducted using super learner techniques, which flexibly accommodate mixed variable types without requiring restrictive parametric assumptions.

%######################################
% table 
%######################################
% table for description 

\begin{table}[t]
\caption{Characteristics across different racial groups.}
\centering 
\scriptsize
\setstretch{0.75}
\setlength{\tabcolsep}{4.2pt}
\begin{tabular}{lccccc|ccccc} % !{\vrule width 0.8pt}
\toprule
 & \multicolumn{5}{c}{\textbf{MEPS data in year 2009}} & \multicolumn{5}{c}{\textbf{MEPS data in year 2016}} \\ 
 \cmidrule(lr){2-6} \cmidrule(lr){7-11}
Characteristic &Overall & Asians & Blacks & Hispanics & \multicolumn{1}{c}{Whites}  &Overall & Asians  & Blacks  & Hispanics  & Whites \\ 
\midrule\addlinespace[2.5pt]
N & 20,789 & 1,468 & 3,967 & 5,399 & 9,955 & 19,508 & 1,535 & 3,576 & 5,634 & 8,763\\
Expenditure & 921.0 & 542.0 & 758.0 & 284.0 & 1,675.0 & 1,118.0 & 777.0 & 895.0 & 396.0 & 2,094.0  \\ 
Expenditure > 0 (\%) & 81.0\% & 80.4\% & 79.8\% & 67.3\% & 89.0\% & 82.0\% & 82.3\% & 78.5\% & 70.6\% & 90.6\% \\ 
 & &  &  &  &  &  & & & & \\
\textbf{\textit{Baseline characteristics}} & &  &  &  &  &  & & & & \\ 
Age (years) & 44.0 & 43.0 & 44.0 & 39.0 & 48.0 & 46.0 & 44.0 & 46.0 & 41.0 & 52.0 \\ 
Male & 45.6\% & 46.8\% & 40.2\% & 46.8\% & 46.9\% & 45.9\% & 47.4\% & 41.5\% & 45.9\% & 47.3\% \\ 
Region &  &  &  &  &  &  &  &  &  &  \\ 
   \quad North & 15.0\% & 14.9\% & 17.1\% & 13.4\% & 15.1\% & 16.1\% & 15.8\% & 16.7\% & 14.9\% & 16.8\% \\ 
  \quad  Midwest & 20.0\% & 10.8\% & 16.1\% & 10.2\% & 28.3\% & 19.4\% & 12.1\% & 16.4\% & 8.7\% & 28.7\% \\ 
   \quad South & 38.3\% & 17.2\% & 58.5\% & 34.4\% & 35.6\% & 38.4\% & 20.3\% & 57.7\% & 38.5\% & 33.7\% \\ 
  \quad  West & 26.6\% & 57.2\% & 8.3\% & 42.1\% & 21.0\% & 26.1\% & 51.9\% & 9.1\% & 37.9\% & 20.9\% \\ 
Marriage &  &  &  &  &  &  &  &  &  &  \\ 
  \quad  Married & 52.7\% & 64.2\% & 33.4\% & 52.9\% & 58.5\% & 48.9\% & 66.3\% & 30.5\% & 46.0\% & 55.3\% \\ 
   \quad Previously married & 20.4\% & 11.0\% & 26.6\% & 15.8\% & 21.7\% & 21.2\% & 10.3\% & 25.3\% & 17.9\% & 23.6\% \\ 
   \quad Never married & 27.0\% & 24.8\% & 40.1\% & 31.3\% & 19.7\% & 29.9\% & 23.5\% & 44.2\% & 36.1\% & 21.1\% \\ 
 & &  &  &  &  &  & & & & \\ 
\textbf{\textit{Family SES}} & &  &  &  &  &  & & & & \\ 
  Family poverty &  &  &  &  &  &  &  &  &  &  \\ 
   \quad Below poverty & 17.2\% & 9.9\% & 25.3\% & 24.0\% & 11.3\% & 17.4\% & 9.6\% & 26.0\% & 23.8\% & 11.0\% \\ 
   \quad Near poverty & 5.5\% & 2.9\% & 6.6\% & 7.5\% & 4.5\% & 5.4\% & 4.8\% & 6.6\% & 7.6\% & 3.6\% \\ 
   \quad Low & 16.3\% & 13.4\% & 18.4\% & 22.0\% & 12.8\% & 15.6\% & 11.7\% & 17.3\% & 20.9\% & 12.1\% \\ 
   \quad Middle & 31.1\% & 29.0\% & 30.2\% & 31.9\% & 31.4\% & 29.2\% & 23.5\% & 29.3\% & 31.2\% & 28.9\% \\ 
   \quad High & 29.9\% & 44.8\% & 19.4\% & 14.7\% & 40.1\% & 32.5\% & 50.4\% & 20.8\% & 16.5\% & 44.3\% \\ 
  & &  &  &  &  &  & & & &  \\ 
\textbf{\textit{SES}} & &  &  &  &  &  & & & & \\ 
Total income (\$) & 20,350 & 24,960 & 16,900 & 15,080 & 25,306 & 23,808 & 30,030 & 19,000 & 17,680 & 31,680 \\ 
Education &  &  &  &  &  &  &  &  &  &  \\ 
  \quad  < High school & 26.5\% & 14.4\% & 26.4\% & 49.4\% & 15.8\% & 23.6\% & 13.2\% & 22.0\% & 42.8\% & 13.8\% \\ 
  \quad  High school & 44.4\% & 30.9\% & 51.9\% & 37.2\% & 47.3\% & 42.8\% & 29.1\% & 53.9\% & 39.2\% & 43.1\% \\ 
  \quad  College & 14.7\% & 31.0\% & 10.2\% & 6.9\% & 18.3\% & 16.4\% & 30.2\% & 10.5\% & 9.3\% & 21.1\% \\ 
  \quad  Graduate & 14.5\% & 23.8\% & 11.6\% & 6.5\% & 18.5\% & 17.1\% & 27.5\% & 13.6\% & 8.7\% & 22.0\% \\ 
Employed & 60.2\% & 64.9\% & 55.6\% & 61.0\% & 60.9\% & 61.0\% & 65.2\% & 58.1\% & 63.1\% & 60.1\% \\
 & &  &  &  &  &  & & & & \\ 
\textbf{\textit{Insurance access}} & &  &  &  &  &  & & & & \\ 
Uninsured & 20.2\% & 14.0\% & 18.5\% & 38.4\% & 11.8\% & 12.0\% & 5.5\% & 10.0\% & 25.4\% & 5.3\% \\
 & &  &  &  &  &  & & & & \\ 
\textbf{\textit{Health behaviors}} & &  &  &  &  &  & & & & \\ 
Smoke & 18.1\% & 8.8\% & 21.5\% & 12.0\% & 21.4\% & 14.1\% & 7.3\% & 19.5\% & 8.9\% & 16.4\% \\ 
Exercise & 56.6\% & 58.9\% & 53.1\% & 52.5\% & 59.9\% & 49.6\% & 45.1\% & 50.8\% & 46.4\% & 51.9\% \\ 
 & &  &  &  &  &  & & & & \\ 
\textbf{\textit{Health status}} & &  &  &  &  &  & & & & \\ 
BMI & 27.1 & 23.7 & 28.3 & 27.5 & 26.6 & 27.4 & 24.1 & 29.0 & 28.2 & 27.1 \\ 
Mental health &  &  &  &  &  &  &  &  &  &  \\ 
  \quad  Excellent & 36.2\% & 42.3\% & 37.4\% & 34.7\% & 35.7\% & 35.2\% & 40.1\% & 38.2\% & 36.1\% & 32.6\% \\ 
   \quad Very good & 29.4\% & 29.6\% & 25.5\% & 28.2\% & 31.6\% & 28.7\% & 30.3\% & 25.1\% & 24.4\% & 32.8\% \\ 
   \quad Good & 26.5\% & 23.4\% & 27.5\% & 29.6\% & 24.9\% & 26.9\% & 23.1\% & 27.1\% & 30.2\% & 25.4\% \\ 
   \quad Fair & 6.4\% & 3.3\% & 7.8\% & 6.6\% & 6.1\% & 7.4\% & 5.3\% & 7.8\% & 8.1\% & 7.1\% \\ 
  \quad  Poor & 1.5\% & 1.4\% & 1.8\% & 0.9\% & 1.7\% & 1.8\% & 1.2\% & 1.8\% & 1.2\% & 2.2\% \\ 
Health &  &  &  &  &  &  &  &  &  &  \\ 
   \quad Excellent & 23.4\% & 26.8\% & 21.5\% & 21.2\% & 24.8\% & 23.1\% & 27.9\% & 22.3\% & 23.9\% & 22.1\% \\ 
   \quad Very good & 31.5\% & 34.3\% & 28.4\% & 28.1\% & 34.1\% & 31.9\% & 36.2\% & 28.3\% & 26.0\% & 36.3\% \\ 
   \quad Good & 30.1\% & 29.0\% & 31.8\% & 33.7\% & 27.7\% & 29.5\% & 27.3\% & 31.3\% & 32.3\% & 27.4\% \\ 
   \quad Fair & 11.5\% & 7.7\% & 14.5\% & 14.1\% & 9.6\% & 12.3\% & 6.5\% & 14.6\% & 15.1\% & 10.5\% \\ 
   \quad Poor & 3.5\% & 2.2\% & 3.8\% & 2.9\% & 3.8\% & 3.2\% & 2.1\% & 3.5\% & 2.7\% & 3.7\% \\ 
PCS & 53.2 & 54.2 & 52.1 & 53.7 & 52.9 & 53.5 & 54.8 & 52.6 & 53.8 & 53.2 \\ 
MCS & 53.0 & 54.0 & 53.3 & 51.7 & 53.7 & 54.4 & 54.9 & 54.8 & 54.5 & 54.2 \\ 
Any limitation & 25.6\% & 12.8\% & 28.0\% & 16.4\% & 31.5\% & 25.8\% & 12.1\% & 29.6\% & 17.5\% & 32.0\% \\ 
Social limitation & 4.3\% & 1.5\% & 5.6\% & 2.3\% & 5.4\% & 6.3\% & 2.8\% & 7.2\% & 3.6\% & 8.2\% \\ 
Cognition limitation & 4.4\% & 2.2\% & 5.9\% & 3.1\% & 4.8\% & 6.3\% & 3.8\% & 7.9\% & 4.5\% & 7.2\% \\ 
Diabetes & 9.4\% & 7.5\% & 12.4\% & 9.4\% & 8.6\% & 11.6\% & 9.5\% & 14.8\% & 11.8\% & 10.4\% \\ 
Asthma & 8.8\% & 5.3\% & 10.2\% & 6.5\% & 10.0\% & 9.3\% & 5.5\% & 11.8\% & 7.5\% & 10.0\% \\ 
High blood pressure & 32.8\% & 25.7\% & 43.1\% & 24.2\% & 34.4\% & 34.7\% & 25.4\% & 45.1\% & 26.7\% & 37.2\% \\ 
Coronary heart disease & 5.6\% & 2.5\% & 5.3\% & 3.7\% & 7.2\% & 5.3\% & 2.7\% & 4.8\% & 4.3\% & 6.6\% \\ 
Angina & 2.7\% & 1.2\% & 2.3\% & 1.8\% & 3.6\% & 2.3\% & 1.3\% & 1.7\% & 1.4\% & 3.3\% \\ 
Myocardial infarction & 3.6\% & 1.2\% & 3.6\% & 1.9\% & 4.9\% & 3.8\% & 1.6\% & 3.8\% & 2.4\% & 5.1\% \\ 
Stroke & 3.5\% & 1.4\% & 5.0\% & 1.9\% & 4.1\% & 4.3\% & 2.1\% & 6.3\% & 2.4\% & 5.0\% \\ 
Emphysema & 2.1\% & 0.4\% & 1.6\% & 0.6\% & 3.3\% & 1.9\% & 0.6\% & 1.4\% & 0.6\% & 3.1\% \\ 
Cholesterol & 30.3\% & 28.0\% & 28.7\% & 24.7\% & 34.3\% & 31.6\% & 28.0\% & 29.1\% & 27.0\% & 36.1\% \\ 
Arthritis & 24.0\% & 12.5\% & 27.2\% & 13.8\% & 30.0\% & 26.4\% & 13.6\% & 28.1\% & 16.0\% & 34.6\% \\ 
Cancer & 8.4\% & 2.7\% & 5.0\% & 3.2\% & 13.4\% & 9.5\% & 2.7\% & 6.0\% & 4.5\% & 15.3\% \\ 
\bottomrule
\end{tabular}

\begin{tablenotes}
\item Descriptive statistics stratified by racial groups (Asians, Blacks, Hispanics, and Whites). The table displays key baseline characteristics, family SES, SES, insurance access, health behaviors, health status, and healthcare expenditures. Continuous variables are presented as \textit{median} and categorical variables are presented as \textit{percentage \%}. The Chi-square test was used to compare categorical variables, and the Kruskal-Wallis rank sum test was used to compare continuous variables across racial groups. All comparisons are significant ($p < 0.001$).
\end{tablenotes}
\label{tab:description_09_16}
\end{table}

% table for expenditure 
% \newpage
\begin{table}[t]
\caption{Median healthcare expenditures stratified by race and characteristics.}
% Healthcare expenditures are presented as \textit{median}. 
\centering 
\scriptsize
\setstretch{0.55}
\setlength{\tabcolsep}{2pt}
\begin{tabular}{llccccc|ccccc}
\toprule
 & &\multicolumn{5}{c}{\textbf{Expenditures in year 2009}} &  \multicolumn{5}{c}{\textbf{Expenditures in year 2016}} \\ 
 \cmidrule(lr){3-7} \cmidrule(lr){8-12}
Characteristic& &Overall & Asians & Blacks & Hispanics & \multicolumn{1}{c}{Whites} &Overall &  Asians  & Blacks  & Hispanics  & Whites \\ 
\midrule
\multicolumn{2}{l}{\textbf{\textit{Baseline characteristics}}} &  & & &  &  & & & & & \\ 
Age (years) & > 45 & 2,167 & 1,149 & 1,796 & 923 & 2,901 & 2,516 & 1,816 & 2,296 & 1,193 & 3,399 \\ 
 & <= 45 & 364 & 323 & 283 & 121 & 729 & 390 & 360 & 267 & 180 & 867 \\ 
Male & No & 1,326 & 778 & 1,108 & 542 & 2,235 & 1,576 & 1,037 & 1,276 & 662 & 2,697 \\ 
 & Yes & 529 & 349 & 327 & 87 & 1,147 & 681 & 486 & 421 & 181 & 1,470 \\ 
Region & North & 1,251 & 687 & 964 & 516 & 1,926 & 1,459 & 723 & 762 & 770 & 2,477 \\ 
 & Midwest & 1,173 & 371 & 941 & 305 & 1,585 & 1,447 & 535 & 1,118 & 406 & 1,914 \\ 
 & South & 857 & 342 & 680 & 250 & 1,656 & 929 & 499 & 849 & 290 & 2,041 \\ 
 & West & 691 & 634 & 732 & 245 & 1,696 & 993 & 1,020 & 986 & 418 & 2,191 \\ 
Marriage & Married & 1,099 & 696 & 872 & 360 & 1,841 & 1,342 & 903 & 1,194 & 494 & 2,218 \\ 
 & Previously married & 2,000 & 1,252 & 1,774 & 889 & 2,722 & 2,603 & 2,185 & 2,383 & 1,161 & 3,585 \\ 
 & Never married & 297 & 192 & 279 & 84 & 660 & 387 & 359 & 361 & 181 & 911 \\ 
 & &  & & &  &  & & & & & \\ 
 \multicolumn{2}{l}{\textbf{\textit{Family SES}}} &  & & &  &  & & & & & \\
Family poverty & Below poverty & 554 & 376 & 572 & 179 & 1,492 & 884 & 1,335 & 897 & 386 & 2,175 \\ 
 & Near poverty & 708 & 342 & 862 & 230 & 1,668 & 774 & 242 & 1,124 & 280 & 2,487 \\ 
 & Low & 566 & 477 & 561 & 199 & 1,297 & 754 & 589 & 779 & 300 & 1,919 \\ 
 & Middle & 819 & 352 & 842 & 277 & 1,408 & 923 & 832 & 689 & 377 & 1,728 \\ 
 & High & 1,527 & 725 & 1,036 & 779 & 2,028 & 1,692 & 803 & 1,219 & 808 & 2,340 \\ 
 & &  & & &  &  & & & & & \\
\multicolumn{2}{l}{\textbf{\textit{SES}}} &  & & &  &  & & & & & \\ 
Total income  (\$) & <= 22000 & 667 & 346 & 563 & 230 & 1,585 & 942 & 773 & 893 & 377 & 2,125 \\ 
 &  > 22000 & 1,174 & 741 & 987 & 381 & 1,760 & 1,281 & 785 & 898 & 408 & 2,086 \\ 
Education & < High school & 493 & 280 & 746 & 210 & 1,415 & 696 & 881 & 1,007 & 370 & 1,902 \\ 
 & High school & 842 & 349 & 616 & 273 & 1,536 & 956 & 720 & 673 & 299 & 1,933 \\ 
 & College & 1,320 & 690 & 1,277 & 710 & 1,766 & 1,532 & 839 & 1,265 & 679 & 2,098 \\ 
 & Graduate & 1,577 & 883 & 1,149 & 657 & 2,116 & 1,809 & 775 & 1,408 & 1,129 & 2,573 \\ 
Employed & No & 1,792 & 797 & 1,264 & 558 & 3,213 & 2,580 & 1,497 & 2,191 & 869 & 4,128 \\ 
 & Yes & 644 & 458 & 529 & 190 & 1,111 & 685 & 566 & 517 & 269 & 1,288 \\ 
 & &  & & &  &  & & & & & \\ 
\multicolumn{2}{l}{\textbf{\textit{Insurance access}}} &  & & &  &  & & & & & \\ 
Uninsured & No & 1,428 & 706 & 1,098 & 701 & 2,052 & 1,444 & 875 & 1,126 & 694 & 2,292 \\ 
 & Yes & 40 & 40 & 68 & 0 & 150 & 0 & 0 & 0 & 0 & 150 \\ 
& &  & & &  &  & & & & & \\ 
\multicolumn{2}{l}{\textbf{\textit{Health behaviors}}} &  & & &  &  & & & & & \\ 
Smoke & No & 985 & 590 & 851 & 290 & 1,842 & 1,152 & 848 & 920 & 384 & 2,252 \\ 
 & Yes & 619 & 240 & 385 & 204 & 1,017 & 931 & 332 & 767 & 547 & 1,285 \\ 
Exercise & No & 1,214 & 490 & 1,147 & 301 & 2,541 & 1,482 & 832 & 1,460 & 466 & 2,864 \\ 
 & Yes & 757 & 577 & 482 & 260 & 1,261 & 857 & 747 & 528 & 320 & 1,568 \\ 
& &  & & &  &  & & & & & \\ 
\multicolumn{2}{l}{\textbf{\textit{Health status}}} &  & & &  &  & & & & & \\ 
BMI & < 18.5 & 617 & 335 & 469 & 170 & 1,246 & 1,066 & 1,028 & 666 & 206 & 2,058 \\ 
 & 18.5-24.9 & 728 & 497 & 340 & 199 & 1,306 & 942 & 733 & 421 & 273 & 1,768 \\ 
 & >24.9 & 1,049 & 642 & 921 & 326 & 1,906 & 1,231 & 913 & 1,044 & 448 & 2,307 \\ 
Mental health & Excellent & 615 & 386 & 428 & 154 & 1,156 & 642 & 520 & 427 & 228 & 1,419 \\ 
 & Very good & 915 & 739 & 594 & 283 & 1,574 & 1,106 & 735 & 825 & 369 & 1,830 \\ 
 & Good & 1,118 & 605 & 1,149 & 347 & 2,272 & 1,475 & 1,234 & 1,303 & 503 & 3,097 \\ 
 & Fair & 3,095 & 2,084 & 2,808 & 1,588 & 4,720 & 3,410 & 3,357 & 3,451 & 2,216 & 4,757 \\ 
 & Poor & 6,090 & 1,785 & 4,132 & 5,905 & 7,050 & 7,123 & 5,201 & 7,160 & 8,329 & 6,856 \\ 
Health & Excellent & 383 & 300 & 184 & 51 & 823 & 409 & 396 & 190 & 114 & 1,046 \\ 
 & Very good & 793 & 467 & 512 & 203 & 1,436 & 949 & 609 & 569 & 333 & 1,689 \\ 
 & Good & 1,078 & 697 & 1,024 & 313 & 2,236 & 1,441 & 1,254 & 1,303 & 484 & 3,045 \\ 
 & Fair & 2,905 & 2,044 & 2,663 & 1,229 & 5,581 & 3,315 & 2,187 & 3,386 & 1,435 & 6,382 \\ 
 & Poor & 8,513 & 2,756 & 11,078 & 6,138 & 9,785 & 11,324 & 7,190 & 8,147 & 10,895 & 13,009 \\ 
PCS & > 50 & 480 & 360 & 328 & 120 & 946 & 549 & 492 & 347 & 196 & 1,173 \\ 
 & <= 50 & 2,715 & 1,196 & 2,193 & 1,170 & 4,092 & 3,571 & 2,238 & 3,057 & 1,891 & 5,243 \\ 
MCS &  > 50 & 751 & 500 & 559 & 160 & 1,427 & 861 & 659 & 610 & 272 & 1,750 \\ 
 &  <= 50 & 1,251 & 595 & 1,173 & 542 & 2,212 & 1,867 & 1,054 & 1,845 & 845 & 3,039 \\ 
Any limitation & No & 540 & 406 & 389 & 173 & 1,079 & 620 & 597 & 396 & 250 & 1,255 \\ 
 & Yes & 3,717 & 2,322 & 3,130 & 2,770 & 4,244 & 5,236 & 4,162 & 4,567 & 3,775 & 6,144 \\ 
Social limitation & No & 829 & 518 & 645 & 256 & 1,536 & 968 & 735 & 740 & 357 & 1,839 \\ 
 & Yes & 8,852 & 7,775 & 8,852 & 9,997 & 8,503 & 9,092 & 9,005 & 9,097 & 9,148 & 9,093 \\ 
Cognition limitation & No & 835 & 506 & 645 & 250 & 1,556 & 981 & 725 & 732 & 353 & 1,908 \\ 
 & Yes & 7,539 & 4,338 & 6,407 & 6,770 & 8,142 & 7,962 & 8,590 & 7,709 & 8,196 & 7,821 \\ 
Diabetes & No & 741 & 466 & 517 & 201 & 1,430 & 878 & 623 & 596 & 292 & 1,750 \\ 
 & Yes & 4,745 & 3,599 & 5,396 & 2,693 & 6,051 & 5,890 & 3,142 & 5,423 & 3,691 & 7,624 \\ 
Asthma & No & 830 & 490 & 673 & 246 & 1,557 & 992 & 729 & 767 & 346 & 1,933 \\ 
 & Yes & 2,503 & 1,480 & 2,071 & 1,256 & 3,395 & 3,117 & 2,613 & 2,555 & 2,207 & 3,937 \\ 
High blood pressure & No & 470 & 339 & 245 & 128 & 994 & 558 & 461 & 268 & 223 & 1,234 \\ 
 & Yes & 2,707 & 1,896 & 2,229 & 1,548 & 3,634 & 3,191 & 2,567 & 2,674 & 1,825 & 4,307 \\ 
Coronary heart disease & No & 800 & 501 & 645 & 250 & 1,477 & 995 & 744 & 807 & 356 & 1,883 \\ 
 & Yes & 6,223 & 4,220 & 7,982 & 3,650 & 6,799 & 7,394 & 6,656 & 7,569 & 4,526 & 7,984 \\ 
Angina & No & 865 & 511 & 723 & 266 & 1,579 & 1,059 & 772 & 859 & 382 & 1,973 \\ 
 & Yes & 6,113 & 7,285 & 6,324 & 5,523 & 6,219 & 8,239 & 2,465 & 7,422 & 7,351 & 9,362 \\ 
Myocardial infarction & No & 846 & 516 & 693 & 263 & 1,550 & 1,022 & 766 & 820 & 372 & 1,931 \\ 
 & Yes & 6,317 & 4,796 & 8,095 & 4,615 & 6,828 & 6,937 & 4,803 & 6,775 & 7,116 & 7,100 \\ 
Stroke & No & 847 & 508 & 660 & 265 & 1,564 & 1,017 & 748 & 779 & 372 & 1,933 \\ 
 & Yes & 6,353 & 4,352 & 6,318 & 3,092 & 7,182 & 7,259 & 6,014 & 7,865 & 4,446 & 7,462 \\ 
Emphysema & No & 877 & 537 & 731 & 276 & 1,586 & 1,071 & 777 & 869 & 391 & 1,983 \\ 
 & Yes & 6,386 & 1,665 & 6,810 & 6,648 & 6,599 & 8,119 & 903 & 5,570 & 7,869 & 9,330 \\ 
Cholesterol & No & 493 & 342 & 365 & 135 & 974 & 570 & 441 & 398 & 212 & 1,196 \\ 
 & Yes & 2,715 & 1,636 & 2,682 & 1,314 & 3,588 & 3,295 & 2,381 & 3,346 & 1,713 & 4,365 \\ 
Arthritis & No & 549 & 400 & 381 & 184 & 1,028 & 605 & 569 & 441 & 256 & 1,208 \\ 
 & Yes & 3,614 & 3,299 & 2,826 & 2,468 & 4,367 & 4,442 & 3,803 & 3,590 & 3,185 & 5,076 \\ 
Cancer & No & 757 & 498 & 680 & 252 & 1,355 & 916 & 741 & 799 & 357 & 1,690 \\ 
 & Yes & 4,912 & 5,806 & 3,713 & 4,489 & 5,088 & 5,703 & 5,246 & 5,393 & 4,109 & 5,931 \\ 
\bottomrule
\end{tabular}
\begin{tablenotes}
\item Healthcare expenditures are presented as \textit{median}
\end{tablenotes}

\label{tab:expenditure09_16}
\end{table}

%  ---  ---  ---  ---  ---  ---  ---  ---  ---  ---  ---  ---  --- -
\subsection{Analysis of the cumulative disparity components in MEPS data}
\label{app:meps_sequential}
%  ---  ---  ---  ---  ---  ---  ---  ---  ---  ---  ---  ---  --- - 

In this appendix, we report cumulative disparity components across racial group comparisons in the MEPS data, using a sequential decomposition framework in which components sum to the total disparity in healthcare expenditures.

\subsubsection{Cumulative disparity components as measures of disparity}
\label{app:sequential_PSEs_id}

We define cumulative (or sequential) disparity components by decomposing the total disparity in healthcare expenditures into a sequence of contributions from ordered mediators. This decomposition follows the causal path-specific effect framework described in Appendix~\ref{app:pse_decomp}, with the key distinction that mediator pathways are sequentially ``deactivated'' one at a time.

Let $R$ denote race, $Y$ the outcome, and $(M_1, M_2, M_3, M_4)$ the ordered mediators (SES, insurance access, health behaviors, health status). Define the following covariate-standardized outcome means under modified mediator distributions: 
{\small 
\begin{align*}
&\gamma_\text{dis} = \int y \, dP(y \mid R=0,x) \, dP(x)  \ ,  \\
&\gamma_\text{adv} = \int y \, dP(y \mid R=1,x) \, dP(x)  \ ,   \\
&\gamma_{R \rightarrow Y} = \int y \, dP(y \mid \overline{m}_4, R=1,x) \prod_{j=1}^4 dP(m_j \mid \overline{m}_{j-1},R=0,x) \, dP(x) \ , \numberthis  \\
&\gamma^*_{R \rightarrow M_k \leadsto Y} = \int  y \, dP(y \mid \overline{m}_4, R=1,x)  \prod_{ \substack{j= k+1, \\ k \neq 4} }^4 dP(m_j \mid \overline{m}_{j-1}, R=1,x)  
\prod_{i=1}^{k}  dP(m_i \mid \overline{m}_{i-1}, R=0,x) \, dP(x) \ .  
\end{align*}
}

Here, $\gamma_\text{adv}$ and $\gamma_\text{dis}$ represent the covariate-standardized mean outcomes for the advantaged and disadvantaged groups, respectively, and their difference defines the total disparity. The intermediate quantities $\gamma^*_{R \rightarrow M_k \leadsto Y}$ correspond to scenarios in which the first $k$ mediators are drawn from the disadvantaged group and the remaining from the advantaged group, allowing for sequential attribution of the disparity. We note that $\gamma^*_{R \rightarrow M_4 \leadsto Y}$ is equivalent to $\gamma_{R \rightarrow Y}$. 

We define the sequential disparity components, each corresponding to a mediator or outcome disparity, as follows:
\begin{equation}
\begin{aligned}
    \rho^*_{R \rightarrow M_1 \leadsto Y} & \coloneqq \gamma_\text{adv} - \gamma^*_{R \rightarrow M_1 \leadsto Y}  \ ,   \\
    \rho^*_{R \rightarrow M_2 \leadsto Y} & \coloneqq \gamma^*_{R \rightarrow M_1 \leadsto Y} - \gamma^*_{R \rightarrow M_2 \leadsto Y}  \ ,   \\
    \rho^*_{R \rightarrow M_3 \leadsto Y} & \coloneqq \gamma^*_{R \rightarrow M_2 \leadsto Y} - \gamma^*_{R \rightarrow M_3 \leadsto Y}  \ ,  \\
    \rho^*_{R \rightarrow M_4 \leadsto Y} & \coloneqq \gamma^*_{R \rightarrow M_3 \leadsto Y} - \gamma^*_{R \rightarrow M_4 \leadsto Y}  \ ,  \\
    \rho^*_{R \rightarrow Y} & \coloneqq \gamma^*_{R \rightarrow M_4 \leadsto Y} -  \gamma_\text{dis}\ .  
    \label{eq:sequential_disparity}
\end{aligned}
\end{equation}%
By construction, these components satisfy the identity: $\rho^*_\text{total} = \rho^*_{R \rightarrow Y} + \sum_{k = 1}^4 \rho^*_{R \rightarrow M_k \leadsto Y}$. 

Each component $\rho^*_{R \rightarrow M_k \leadsto Y}$ captures the reduction in disparity achieved by replacing the advantaged group's distribution of mediator $M_k$ with that of the disadvantaged group, while holding all earlier mediators at their disadvantaged distributions and allowing later mediators and the outcome to respond as they would under the advantaged group. The final term, $\rho^*_{R \rightarrow Y}$, represents the residual disparity that remains after all mediators have been set to follow the disadvantaged group, isolating effects not captured by the specified mediating pathways. 

$\rho^*_{R \rightarrow M_1 \leadsto Y}$: This represents the portion of the total disparity in healthcare expenditures attributable to differences in the distribution of $M_1$ (socioeconomic status) between racial groups, assuming that all downstream mediators ($M_2, M_3, M_4$) and the outcome evolve as they would for the advantaged group ($R = 1$). It quantifies the reduction in disparity that would occur if, within each covariate stratum $X$, the advantaged group had the same distribution of $M_1$ as the disadvantaged group, while retaining their own levels of downstream mediators and outcome. 

$\rho^*_{R \rightarrow M_2 \leadsto Y}$: This represents the portion of the total disparity in healthcare expenditures attributable to differences in the distribution of $M_2$ (insurance access) across racial groups, after accounting for differences in $M_1$ (socioeconomic status). It assumes that downstream mediators ($M_3$, $M_4$) and the outcome evolve as they would for the advantaged group ($R = 1$), while $M_1$ is already aligned to the disadvantaged group ($R = 0$). This component quantifies the additional disparity reduction achieved by equalizing the distribution of $M_2$ across groups, conditional on already having equalized $M_1$.

$\rho^*_{R \rightarrow M_3 \leadsto Y}$: This represents the portion of the total disparity in healthcare expenditures attributable to differences in the distribution of $M_3$ (health-related behaviors) across racial groups, after accounting for differences in $M_1$ and $M_2$. It assumes that the downstream mediator ($M_4$) and the outcome evolve as they would for the advantaged group ($R = 1$), while $M_1$ and $M_2$ are already aligned to the disadvantaged group ($R = 0$). This component quantifies the additional disparity reduction achieved by equalizing the distribution of $M_3$, given that disparities in the first two mediators have already been addressed.

$\rho^*_{R \rightarrow M_4 \leadsto Y}$: This represents the portion of the total disparity in healthcare expenditures attributable to differences in the distribution of $M_4$ (e.g., health status) between racial groups, assuming that $M_1$, $M_2$, and $M_3$ follow the disadvantaged group's distribution ($R = 0$), and that the outcome responds as it would for the advantaged group ($R = 1$). It quantifies the disparity reduction achieved by replacing the advantaged group's distribution of $M_4$ with that of the disadvantaged group, holding all prior mediators at their disadvantaged levels and allowing only the outcome to reflect advantaged conditions. 

$\rho^*_{R \rightarrow Y}$: This is structurally equivalent to the outcome-attributed disparity defined in the main manuscript. Both quantify the portion of the total disparity that remains after replacing all mediators with their distributions under the disadvantaged group, while allowing the outcome to respond as it would under the advantaged group.

\subsubsection{Empirical results}
\label{app:sequential_PSEs}

We derived one-step corrected plug-in estimators for the disparity components in \eqref{eq:sequential_disparity} using nonparametric influence functions, following a process similar to that outlined in Section~\ref{subsec:method_est} and Appendix~\ref{app:proofs_est}, and incorporated the same super learning estimation techniques as those described in Section~\ref{subsec:implement} for nuisance estimations. 

Table~\ref{tab:PSEs_seq} reports the total disparity and cumulative disparities as ratios of scaled geometric means. By construction, the product of these disparity components equals the total disparity. A cumulative disparity farther from 1 signifies a greater contribution of mediator to racial disparities in healthcare expenditures.

Consistent with the mediator-attributable disparities reported in Table~\ref{tab:PSEs} of the main manuscript, unexplained disparities in 2009 were statistically significant only when comparing Whites vs. marginalized racial groups, but not between two marginalized racial groups. Moreover, these unexplained disparities emerged as the dominant drivers of disparities between Whites and marginalized racial populations, as reflected in their geometric‐mean ratios. 

% Among the four mediators, the contribution from SES component was dominant in 2009 for disparities between Whites and Blacks and between Asians and Hispanics; the health-insurance disparity component drove the White vs. Hispanic and Black vs. Hispanic comparisons; and health status component was most influential in the White vs. Asian and Black vs. Asian disparities. These patterns reinforce the main manuscript's conclusions, particularly highlighting SES and insurance coverage as critical levers for improving healthcare utilization among Hispanic individuals. In 2016, these patterns largely persisted, except that health status replaced SES as the primary mediator of the White-Black disparity, a shift that may reflect rising chronic-disease prevalence potentially driven by changes in economic conditions, dietary habits, and other lifestyle factors \citep{ansah2023projecting}. 

Among the four mediators, SES was the dominant contributor to disparities in 2009 for all pairwise comparisons except Black versus Asian, and it remained the important contributor in 2016 for all comparisons except White versus Asian. The health insurance component also played a major role, driving disparities in all 2009 comparisons except White versus Black; however, its contribution attenuated to nearly zero by 2016 in the White versus Asian and Black versus Asian comparisons. Health status was an influential mediator across all comparisons in both 2009 and 2016. These patterns reinforce the conclusions of the main manuscript and further highlight SES and insurance coverage as critical levers for improving healthcare utilization among Hispanic individuals.

% Overall, natural and sequential decompositions yield largely consistent results with a few notable exceptions. Differences in statistical significance emerge for (1) health‐status disparity components in the White vs. Black comparison, (2) SES and health‐behavior disparity components in the White vs. Asian comparison, and (3) SES disparity components in the Black vs. Hispanic comparison. Most strikingly, the health‐behavior disparity components in the White vs. Hispanic comparison reversed direction between the two decompositions, an indication of underlying interaction effects (see Appendix~\ref{app:pse_decomp}).
Overall, natural and sequential decompositions yield similar patterns, with several notable exceptions. In particular, the disparity component attributable to health behaviors is nearly uniformly non-significant under the natural decomposition, but is significant in most pairwise comparisons in both 2009 and 2016 under the sequential decomposition. Moreover, for the Asian versus Hispanic comparison, the health-behavior component reverses direction across the two decompositions. These discrepancies suggest the presence of underlying interaction effects (see Appendix~\ref{app:pse_decomp}).

% table for PSEs 

\begin{table}[t]
\caption{Cumulative disparity components across racial group comparisons, reported on the scaled geometric mean ratios. }
\footnotesize
\renewcommand{\arraystretch}{0.9}
\setlength{\tabcolsep}{12.5pt} 
\centering
\begin{tabular}{lcccccc}
\toprule
& \multicolumn{3}{c}{\textbf{MEPS data in year 2009}} & \multicolumn{3}{c}{\textbf{MEPS data in year 2016}} \\
\cmidrule(lr){2-4} \cmidrule(lr){5-7}
Disparity & Value & 95\% CI & p-value & Value & 95\% CI & p-value \\ 
\midrule\addlinespace[2.5pt]
\multicolumn{7}{l}{Whites vs Blacks*} \\[2.5pt]
\midrule\addlinespace[2.5pt]
$\rho^*_{R \rightarrow M_1 \leadsto Y}$ & 1.037 & 1.017 --- 1.058 & {\bf $<$0.001} & 1.089 & 1.062 --- 1.116 & {\bf $<$0.001} \\
$\rho^*_{R \rightarrow M_2 \leadsto Y}$ & 0.997 & 0.972 --- 1.023 & 0.838 & 1.001 & 0.980 --- 1.022 & 0.899 \\
$\rho^*_{R \rightarrow M_3 \leadsto Y}$ & 0.983 & 0.969 --- 0.997 & {\bf 0.019} & 0.984 & 0.969 --- 0.999 & {\bf 0.034} \\
$\rho^*_{R \rightarrow M_4 \leadsto Y}$ & 1.055 & 1.012 --- 1.097 & {\bf 0.011} & 1.099 & 1.048 --- 1.151 & {\bf $<$0.001} \\
$\rho^*_{R \rightarrow Y}$ & 1.772 & 1.614 --- 1.930 & {\bf $<$0.001} & 1.768 & 1.607 --- 1.929 & {\bf $<$0.001} \\
$\rho^*_\text{total}$ & 1.901 & 1.680 --- 2.122 & {\bf $<$0.001} & 2.084 & 1.840 --- 2.329 & {\bf $<$0.001} \\
\midrule\addlinespace[2.5pt]
\multicolumn{7}{l}{Whites vs Asians*} \\[2.5pt]
\midrule\addlinespace[2.5pt]
$\rho^*_{R \rightarrow M_1 \leadsto Y}$ & 0.948 & 0.919 --- 0.977 & {\bf $<$0.001} & 0.972 & 0.938 --- 1.006 & 0.102 \\
$\rho^*_{R \rightarrow M_2 \leadsto Y}$ & 1.068 & 1.024 --- 1.112 & {\bf 0.002} & 1.021 & 0.993 --- 1.049 & 0.143 \\
$\rho^*_{R \rightarrow M_3 \leadsto Y}$ & 0.967 & 0.941 --- 0.992 & {\bf 0.011} & 0.939 & 0.918 --- 0.960 & {\bf $<$0.001} \\
$\rho^*_{R \rightarrow M_4 \leadsto Y}$ & 1.276 & 1.199 --- 1.353 & {\bf $<$0.001} & 1.355 & 1.275 --- 1.436 & {\bf $<$0.001} \\
$\rho^*_{R \rightarrow Y}$ & 2.316 & 1.999 --- 2.632 & {\bf $<$0.001} & 2.028 & 1.770 --- 2.287 & {\bf $<$0.001} \\
$\rho^*_\text{total}$ & 2.893 & 2.408 --- 3.378 & {\bf $<$0.001} & 2.561 & 2.184 --- 2.937 & {\bf $<$0.001} \\
\midrule\addlinespace[2.5pt]
\multicolumn{7}{l}{Whites vs Hispanics*} \\[2.5pt]
\midrule\addlinespace[2.5pt]
$\rho^*_{R \rightarrow M_1 \leadsto Y}$ & 1.196 & 1.146 --- 1.245 & {\bf $<$0.001} & 1.234 & 1.184 --- 1.285 & {\bf $<$0.001} \\
$\rho^*_{R \rightarrow M_2 \leadsto Y}$ & 1.299 & 1.257 --- 1.341 & {\bf $<$0.001} & 1.250 & 1.214 --- 1.286 & {\bf $<$0.001} \\
$\rho^*_{R \rightarrow M_3 \leadsto Y}$ & 0.923 & 0.893 --- 0.954 & {\bf $<$0.001} & 0.915 & 0.888 --- 0.941 & {\bf $<$0.001} \\
$\rho^*_{R \rightarrow M_4 \leadsto Y}$ & 1.248 & 1.192 --- 1.303 & {\bf $<$0.001} & 1.305 & 1.246 --- 1.365 & {\bf $<$0.001} \\
$\rho^*_{R \rightarrow Y}$ & 2.071 & 1.901 --- 2.240 & {\bf $<$0.001} & 1.830 & 1.676 --- 1.984 & {\bf $<$0.001} \\
$\rho^*_\text{total}$ & 3.705 & 3.318 --- 4.093 & {\bf $<$0.001} & 3.371 & 3.005 --- 3.737 & {\bf $<$0.001} \\
\midrule\addlinespace[2.5pt]
\multicolumn{7}{l}{Blacks vs Asians*} \\[2.5pt]
\midrule\addlinespace[2.5pt]
$\rho^*_{R \rightarrow M_1 \leadsto Y}$ & 1.002 & 0.934 --- 1.070 & 0.954 & 0.901 & 0.838 --- 0.963 & {\bf 0.002} \\
$\rho^*_{R \rightarrow M_2 \leadsto Y}$ & 1.092 & 1.020 --- 1.163 & {\bf 0.012} & 1.009 & 0.946 --- 1.071 & 0.791 \\
$\rho^*_{R \rightarrow M_3 \leadsto Y}$ & 1.001 & 0.971 --- 1.032 & 0.933 & 0.990 & 0.958 --- 1.023 & 0.550 \\
$\rho^*_{R \rightarrow M_4 \leadsto Y}$ & 1.351 & 1.213 --- 1.488 & {\bf $<$0.001} & 1.461 & 1.297 --- 1.625 & {\bf $<$0.001} \\
$\rho^*_{R \rightarrow Y}$ & 0.991 & 0.811 --- 1.171 & 0.922 & 0.921 & 0.777 --- 1.065 & 0.281 \\
$\rho^*_\text{total}$ & 1.466 & 1.115 --- 1.817 & {\bf 0.009} & 1.210 & 0.913 --- 1.506 & 0.165 \\
\midrule\addlinespace[2.5pt]
\multicolumn{7}{l}{Blacks vs Hispanics*} \\[2.5pt]
\midrule\addlinespace[2.5pt]
$\rho^*_{R \rightarrow M_1 \leadsto Y}$ & 1.188 & 1.095 --- 1.281 & {\bf $<$0.001} & 1.175 & 1.093 --- 1.258 & {\bf $<$0.001} \\
$\rho^*_{R \rightarrow M_2 \leadsto Y}$ & 1.346 & 1.294 --- 1.398 & {\bf $<$0.001} & 1.334 & 1.272 --- 1.396 & {\bf $<$0.001} \\
$\rho^*_{R \rightarrow M_3 \leadsto Y}$ & 0.978 & 0.952 --- 1.004 & 0.100 & 0.971 & 0.943 --- 0.998 & {\bf 0.038} \\
$\rho^*_{R \rightarrow M_4 \leadsto Y}$ & 1.224 & 1.161 --- 1.286 & {\bf $<$0.001} & 1.191 & 1.127 --- 1.256 & {\bf $<$0.001} \\
$\rho^*_{R \rightarrow Y}$ & 1.031 & 0.952 --- 1.109 & 0.442 & 0.896 & 0.822 --- 0.969 & {\bf 0.005} \\
$\rho^*_\text{total}$ & 1.973 & 1.670 --- 2.276 & {\bf $<$0.001} & 1.624 & 1.391 --- 1.858 & {\bf $<$0.001} \\
\midrule\addlinespace[2.5pt]
\multicolumn{7}{l}{Asians vs Hispanics*} \\[2.5pt]
\midrule\addlinespace[2.5pt]
$\rho^*_{R \rightarrow M_1 \leadsto Y}$ & 1.438 & 1.275 --- 1.600 & {\bf $<$0.001} & 1.246 & 1.108 --- 1.384 & {\bf $<$0.001} \\
$\rho^*_{R \rightarrow M_2 \leadsto Y}$ & 1.213 & 1.159 --- 1.266 & {\bf $<$0.001} & 1.223 & 1.178 --- 1.268 & {\bf $<$0.001} \\
$\rho^*_{R \rightarrow M_3 \leadsto Y}$ & 0.982 & 0.967 --- 0.997 & {\bf 0.020} & 1.010 & 0.993 --- 1.027 & 0.237 \\
$\rho^*_{R \rightarrow M_4 \leadsto Y}$ & 0.845 & 0.793 --- 0.897 & {\bf $<$0.001} & 0.835 & 0.783 --- 0.888 & {\bf $<$0.001} \\
$\rho^*_{R \rightarrow Y}$ & 1.017 & 0.941 --- 1.092 & 0.666 & 1.097 & 1.013 --- 1.180 & {\bf 0.023} \\
$\rho^*_\text{total}$ & 1.471 & 1.193 --- 1.749 & {\bf 0.001} & 1.410 & 1.139 --- 1.680 & {\bf 0.003} \\ 
\bottomrule
\end{tabular}
\begin{tablenotes}
\item  \textsuperscript{*}Reference group; $M_1$: SES, $M_2$: Insurance, $M_3$: Health behaviors, $M_4$: Health status.
\end{tablenotes}
\label{tab:PSEs_seq}
\end{table}

%  ---  ---  ---  ---  ---  ---  ---  ---  ---  ---  ---  ---  --- -
\subsection{Scale of reported disparities in the MEPS data}
\label{app:meps_response}
%  ---  ---  ---  ---  ---  ---  ---  ---  ---  ---  ---  ---  --- -

This appendix explores alternative summary measures for disparity measures beyond the arithmetic mean, particularly in settings where the outcome distribution is skewed. To convey these ideas, we rely on the potential outcomes framework introduced in \eqref{eq:nested_counterfactual_K}. 

\subsubsection{Geometric mean interpretation} 
\label{app:meps_geometric_mean}

% ======== new version:  ===========
\noindent {\bf Positive responses.} \ Assume responses are all positive. By the \textit{law of large numbers}, we can write: 
\begin{align*}
\frac{1}{n} \sum_{i=1}^n [\log Y_i(r_0, {\bf r}) - \log Y_i(0, {\bf 0})] 
\rightarrow^{\text{as}}
\E[\log Y(r_0, {\bf r}) - \log Y(0, {\bf 0})] \ . 
\end{align*}

To interpret the above estimand on a scale meaningful for healthcare expenditures, we apply the exponential function. By the \textit{continuous mapping theorem}:
\begin{align*}
\frac{ G_n\big(Y(r_0, {\bf r})\big)}{G_n\big(Y(0, {\bf 0})\big) } 
= \frac{ \Big\{ \prod_{i=1}^n Y_i(r_0, {\bf r})   \Big\}^{1/n} }{\Big\{ \prod_{i=1}^n Y_i(0, {\bf 0}) \Big\}^{1/n} } 
& = \exp \Big(\frac{1}{n}  \sum_{i=1}^n [\log Y_i(r_0, {\bf r}) - \log Y_i(0, {\bf 0})]  \Big)\\
&  \rightarrow^{\text{as}}
\exp \Big(\E[\log Y(r_0, {\bf r}) - \log Y(0, {\bf 0})] \Big)
\ , 
\end{align*}
where $G_n(f)$ denotes the geometric mean of $f$, i.e., $G_n(f) = \{\prod_{i=1}^n f_i \}^{1/n}.$

We note that identification and estimation arguments for $\E[\log Y(r_0, {\bf r}) - \log Y(0, {\bf 0})]$ remain the same by simply defining the outcome as log of healthcare expenditures. The identification functionals are given by:  
{\footnotesize 
\begin{align*}
\label{app:eq:pse_effs_ID} 
    \E[\log Y(0, {\bf 0})] 
    % &\coloneqq \E\left[ \log Y(0, {\bf 0})  \right] 
    &= \int \log y \, dP(y \mid R=0,x) \, dP(x)  \ ,   \\
    \E[\log Y(1, {\bf 0})]  
    % &\coloneqq \E\left[ \log Y(1, {\bf 0})  \right] 
    &= \int \log y \, dP(y \mid \overline{m}_4, R=1,x) \, \prod_{j=1}^4 dP(m_j \mid \overline{m}_{j-1},R=0,x) \,  dP(x) \ , \numberthis
       \\
    \E[\log Y(0, {\bf 1}_k)] 
    &=\int \log y \, dP(y \mid \overline{m}_4, R=0,x) \, dP(m_k \mid \overline{m}_{k-1}, R=1,x) \, \prod_{\substack{j=1 \\ j \neq k}}^4 dP(m_j \mid \overline{m}_{j-1}, R=0,x) \, dP(x) \ . 
\end{align*}
}

Table~\ref{tab:positive_Y} displays the decomposition for the subpopulation with positive expenditures ($Y|Y>0$) in geometric mean ratios.

%  ---  ---  ---  ---  ---  ---  ---  ---  ---  ---  ---  ---  ---  ---  ---  --- --
% ======== new version:  ===========

\noindent {\bf Positive and zero responses.} \ In our setting, we have both positive and zero responses. Let  $Y_\text{pos}(r_0, {\bf r})$ denote the positive counterfactual responses. By the \textit{law of large numbers}: 
\begin{align*}
& \hat{P}( Y(r_0, {\bf r})>0) \times \frac{1}{n} \sum_{i=1}^n \log Y_{i, \text{pos}}(r_0, {\bf r}) - \hat{P}( Y(0, {\bf 0})>0) \times \frac{1}{n} \sum_{i=1}^n \log Y_{i, \text{pos}}(0, {\bf 0}) \\
& \hspace{0.5cm} =  \frac{1}{n} \sum_{i=1}^n [ \mathbb{I}( Y_i(r_0, {\bf r}) >0) \log Y_i(r_0, {\bf r}) -  \mathbb{I}( Y_i(0, {\bf 0}) >0) \log Y_i(0, {\bf 0})] \\
& \hspace{0.5cm} \rightarrow^{\text{as}} \E[ \mathbb{I}( Y(r_0, {\bf r}) >0) \log Y (r_0, {\bf r}) -  \mathbb{I}( Y(0, {\bf 0}) >0) \log Y (0, {\bf 0})] \ . 
\end{align*}

To interpret the above estimand on a scale meaningful for healthcare expenditures, we apply the exponential function. By the \textit{continuous mapping theorem}: 
{\small
\begin{equation}
\begin{aligned}
\frac{G_n\big( Y_{\text{pos}}(r_0, {\bf r}) \big)^{\hat{P}( Y(r_0, {\bf r})>0)}}{G_n\big( Y_{\text{pos}}(0, {\bf 0}) \big)^{\hat{P}( Y(0, {\bf 0})>0)}}  
& =  \frac{\Big\{ \prod_{i=1}^n  Y_{i, \text{pos}}(r_0, {\bf r}) \Big\}^{\hat{P}( Y(r_0, {\bf r})>0) / n} }{ \Big\{ \prod_{i=1}^n  Y_{i, \text{pos}}(0, {\bf 0}) \Big\}^{\hat{P}( Y(0, {\bf 0})>0) / n}}  \\
& = \frac{\exp \Big\{ \frac{1}{n} \sum_{i=1}^n \mathbb{I}( Y_i(r_0, {\bf r}) >0) \log Y_i(r_0, {\bf r}) \Big\} }{\exp \Big\{ \frac{1}{n} \sum_{i=1}^n \mathbb{I}( Y_i(0, {\bf 0}) >0) \log Y_i(0, {\bf 0}) \Big\} }  \\
& = \exp \Big ( \frac{1}{n} \sum_{i=1}^n [ \mathbb{I}( Y_i(r_0, {\bf r}) >0) \log Y_i(r_0, {\bf r}) -  \mathbb{I}( Y_i(0, {\bf 0}) >0) \log Y_i(0, {\bf 0})] \Big )  \\
& \rightarrow^{\text{as}} 
\exp \Big ( \E[ \mathbb{I}( Y(r_0, {\bf r}) >0) \log Y (r_0, {\bf r}) -  \mathbb{I}( Y(0, {\bf 0}) >0) \log Y (0, {\bf 0})]  \Big )
\ ,
\label{app:eq:effect_interpretation} 
\end{aligned}
\end{equation}
}
where $G_n( Y_\text{pos}(r_0, {\bf r}))$ and $G_n( Y_\text{pos}(0, {\bf 0}))$ denote the geometric mean of positive counterfactual responses $ Y_\text{pos}(r_0, {\bf r})$ and $G_n( Y_\text{pos}(0, {\bf 0}))$, respectively. Therefore, the effect can be interpreted as ratio of  scaled geometric means. 

We note that identification and estimation arguments for $\E[ \mathbb{I}( Y(r_0, {\bf r}) >0) \log Y (r_0, {\bf r}) -  \mathbb{I}( Y(0, {\bf 0}) >0) \log Y (0, {\bf 0})]$ remain the same by simply defining the outcome as zero if expenditure is zero, and log of expenditure otherwise. The identification functionals are given by:  
{\small 
\begin{align*}
  % ----
    \E[\mathbb{I}(Y(0, {\bf 0})>0) \log Y(0, {\bf 0})] 
    % &\coloneqq \E\left[ \log  Y(0, {\bf 0})  \right] 
    &= \int \mathbb{I}(y>0) \, \log y \, dP(y \mid R=0,x) \, dP(x)  \ ,  
      \\
    % -----
    \E[\mathbb{I}(Y(1, {\bf 0})>0)\log  Y(1, {\bf 0})]  
    % &\coloneqq \E\left[ \log  Y(1, {\bf 0})  \right] 
    &= \int \mathbb{I}(y>0) \, \log y \, dP(y \mid \overline{m}_4, R=1,x) \, \prod_{j=1}^4 dP(m_j \mid \overline{m}_{j-1},R=0,x) \,  dP(x) \ , 
       \\
    % ----
    \E[\mathbb{I}(Y(0, {\bf 1}_k)>0) \log  Y(0, {\bf 1}_k)]
    % &\coloneqq \E\left[ \log  Y(0, {\bf 1}_k)  \right] 
    &=\int \bigg\{ \mathbb{I}(y>0) \, \log y \, dP(y \mid \overline{m}_4, R=0,x) \, dP(m_k \mid \overline{m}_{k-1}, R=1,x) \,  \\
    &\hspace{2cm} \times \prod_{\substack{j=1 \\ j \neq k}}^4 dP(m_j \mid \overline{m}_{j-1}, R=0,x) \, dP(x) \bigg\} \ . \numberthis
    \label{app:eq:pse_effs_ID_pos_zero}  
\end{align*}
}

%  ---  ---  ---  ---  ---  ---  ---  ---  ---  --- 

\begin{remark}[{\bf Asymptotic variance}] \ 
By delta method, we can write: 
\vspace{-0.3cm}
\begin{align*}
   &\sqrt{n}(\exp(\rho^{+}_{R \rightarrow Y}(\hat{Q})) - \exp(\rho_{R \rightarrow Y}(Q))) \\ 
    &\hspace{3cm} \rightarrow^d \mathcal{N}\left( 0, \exp(\rho_{R \rightarrow  Y}(Q))^2 \times \E[ \big(\Phi_{\gamma_{R \rightarrow Y}}(Q) - \Phi_{\gamma_{\text{dis}}}(Q)\big)^2 ] \right) \ , 
\end{align*} 
and 
\vspace{-0.3cm}
\begin{align*}
  &\sqrt{n}(\exp(\rho^{+}_{R \rightarrow M_k \leadsto Y}(\hat{Q})) - \exp(\rho_{R \rightarrow M_k \leadsto Y}(Q))) \\ 
    &\hspace{3cm} \rightarrow^d \mathcal{N}\left( 0, \exp(\rho_{R \rightarrow M_k \leadsto Y}(Q))^2 \times \E[ \big(\Phi_{\gamma_{R \rightarrow M_k \leadsto Y}}(Q) - \Phi_{\gamma_{\text{dis}}}(Q)\big)^2 ] \right) \ .  
\end{align*}
\end{remark}

%  ---  ---  ---  ---  ---  ---  ---  ---  ---  --- 
\begin{remark}[{\bf Probability of positive counterfactual responses}] \ 
In addition to reporting effects with the interpretations outlined in \eqref{app:eq:effect_interpretation}, we also report effects based on a binary indicator for zero or positive responses in Table~\ref{tab:prob_pos}, i.e., $P(Y(r_0, {\bf r})>0) - P(Y(0, {\bf 0})>0)$. The identification and estimation arguments remain unchanged, with the outcome simply redefined as $\mathbb{I}(Y >0)$. 
\end{remark}

%  ---  ---  ---  ---  ---  ---  ---  ---  ---  --- 
\begin{remark}[{\bf Smearing transformation}] \
% \color{red}
The smearing transformation is often applied to adjust for the bias introduced  when taking the difference after exponentiating $\E[\log Y(r_0, {\bf r}) ]$ and exponentiating $\E[ \log Y(0, {\bf 0})]$ to estimate the arithmetic mean of the differences, $\E[Y(r_0, {\bf r}) - Y(0, {\bf 0})]$, rather than the geometric mean. As an example, assume:  
\begin{align*}
    \log Y(r_0, {\bf r}) &\sim \mathcal{N}(\E[ \log Y(r_0, {\bf r})], \sigma^2) 
    \\
     \log Y_i(r_0, {\bf r})  &= \E[\log Y(r_0, {\bf r}) ] + \epsilon_i, \quad \epsilon_i \overset{\text{iid}}{\sim} \mathcal{N}(0, \sigma^2) \ . 
\end{align*}
Therefore: 
$Y_i(r_0, {\bf r})  = \exp \left( \E[\log Y(r_0, {\bf r})] + \epsilon_i \right)$, 
and 
\begin{align*}
    \E[Y(r_0, {\bf r})] 
   &= \E\left[ \exp \left(\E[\log Y(r_0, {\bf r}) ] + \epsilon \right)  \right] \\
   &= \exp \left( \E[\log Y(r_0, {\bf r})] \right) \times \E\left[ \exp \left( \epsilon \right)  \right] 
   \\
   &= \exp \left( \E[\log Y(r_0, {\bf r})] \right) \times \exp \left( \sigma^2/2 \right) \ . 
\end{align*}

The last equality holds by the moment-generating function of a Normal distribution. Here, $\sigma^2$ is the variance of $\log Y(r_0, {\bf r})$. If the assumption of a normally distributed error term is violated, the empirical mean can be used to estimate $\E\left[ \exp \left( \epsilon \right)  \right]$, specifically as $\frac{1}{n} \sum_{i = 1}^n \exp(\epsilon_i)$, where $\epsilon _i = \log Y_i(r_0, {\bf r})  - \hat{\E}[ \log Y(r_0, {\bf r}) ].$ Similarly, $\E[Y(0, {\bf 0})]$ can be obtained using the smearing transformation, and the arithmetic mean difference can then be estimated straightforwardly.
\end{remark}

\noindent 
{\bf Empirical results comparison.} Across the three specifications, positive and zero responses (Table~\ref{tab:PSEs}), probability of positive responses (Table~\ref{tab:prob_pos}), and positive responses only in the subpopulation (Table~\ref{tab:positive_Y}), the results are broadly consistent while highlighting distinct dimensions of disparities. For instance, in the White vs. Black comparison in 2016, SES plays a more pronounced role in the binary model, suggesting that it primarily affects the likelihood of healthcare utilization rather than expenditure levels conditional on use. The two-part model, which combines both the probability of any healthcare use and the level of expenditures, provides a comprehensive summary. The binary model isolates disparities in access to care, while the positive-expenditure model focuses on a selected subpopulation with nonzero spending and captures disparities in expenditure intensity conditional on use. Together, these results indicate that observed disparities arise through both differences in access to care and differences in spending intensity among users, and that the relative importance of mediators differs across these components.

\subsubsection{Two-stage super learner}
\label{app:meps_two_stage_SL}

Let $Y(r_0, {\bf r})$ be defined as the original healthcare expenditures, which include both positive and zero responses. The effects, as defined in \ref{eq:pse_effs}, are interpreted as differences in arithmetic means. To obtain the one-step estimates, outlined in \ref{equ:rho_define},  the  function $\mu_k(\overline{M}_{k}, r_0,X)$  was estimated using the two-stage super learner,  as demonstrated in an example here \href{https://github.com/wuziyueemory/Two-stage-SuperLearner/blob/main/MEPS%20data%20analysis/MEPS.R}{[link]}. 
The two-stage super learner library comprises all pairwise combinations of two constituent algorithms: one for estimating $P(Y > 0 \! \mid \! \overline{M}_{k}, r_0,X)$ and another for $\E[Y \! \mid \! Y > 0, \overline{M}_{k}, r_0,X]$. 
Using a two-stage super learner is expected to improve predictions for each individual outcome. 

Table~\ref{tab:PSEs_difference} presents the results of PSEs calculated as differences in arithmetic means. These findings differ notably from those in Table~\ref{tab:PSEs} and Table~\ref{tab:prob_pos}, where results in the latter two tables are mostly aligned. 
For instance, the disparities attributed to SES ($R \rightarrow M_1 \leadsto Y$) in the White vs. Black comparison in 2016 are positive and statistically significant in both Table~\ref{tab:PSEs} and Table~\ref{tab:prob_pos}, whereas they are negative, although not statistically significant, in Table~\ref{tab:PSEs_difference}. Similarly, for total disparities, Asians have higher expenditures than Hispanics in Table~\ref{tab:PSEs} and Table~\ref{tab:prob_pos}; however, the direction of the effects is reversed in Table~\ref{tab:PSEs_difference}. These differences suggest that relying solely on arithmetic means may obscure both the total disparities and the role of SES in explaining disparities in healthcare expenditures.
More broadly, the discrepancies highlight the limitations of using arithmetic means in the analysis of highly skewed data, which may lead to potential misinterpretations of the results.

%  ---  ---  --- - table
\begin{table}[t]
\caption{Disparity components across racial group comparisons, with healthcare expenditures binarized as zero or positive, reported on the difference scale.}
 % or \footnotesize for slightly larger font
% \setstretch{1}
\footnotesize
\renewcommand{\arraystretch}{0.9}
\setlength{\tabcolsep}{12.5pt} 
\centering
\begin{tabular}{lcccccc}
\toprule
& \multicolumn{3}{c}{\textbf{MEPS data in year 2009}} & \multicolumn{3}{c}{\textbf{MEPS data in year 2016}} \\
\cmidrule(lr){2-4} \cmidrule(lr){5-7}
Disparity & Value & 95\% CI & p value & Value & 95\% CI & p value \\ 
\midrule\addlinespace[2.5pt]
\multicolumn{7}{l}{Whites vs Blacks*} \\[2.5pt]
\midrule\addlinespace[2.5pt]
$\rho_{R \rightarrow M_1 \leadsto Y}$ & 0.004 & -0.001 --- 0.009 & 0.091 & 0.007 & 0.001 --- 0.012 & {\bfseries 0.013} \\
$\rho_{R \rightarrow M_2 \leadsto Y}$ & 0.001 & -0.003 --- 0.004 & 0.762 & 0.001 & -0.002 --- 0.003 & 0.723 \\
$\rho_{R \rightarrow M_3 \leadsto Y}$ & -0.002 & -0.005 --- 0.000 & 0.072 & -0.002 & -0.004 --- 0.000 & 0.096 \\
$\rho_{R \rightarrow M_4 \leadsto Y}$ & -0.001 & -0.008 --- 0.006 & 0.819 & 0.007 & 0.001 --- 0.014 & {\bfseries 0.023} \\
$\rho_{R \rightarrow Y}$ & 0.050 & 0.039 --- 0.061 & {\bfseries $<$0.001} & 0.051 & 0.039 --- 0.063 & {\bfseries $<$0.001} \\
$\rho_\text{total}$ & 0.056 & 0.042 --- 0.070 & {\bfseries $<$0.001} & 0.069 & 0.055 --- 0.083 & {\bfseries $<$0.001} \\
\midrule\addlinespace[2.5pt]
\multicolumn{7}{l}{Whites vs Asians*} \\[2.5pt]
\midrule\addlinespace[2.5pt]
$\rho_{R \rightarrow M_1 \leadsto Y}$ & -0.000 & -0.014 --- 0.013 & 0.956 & -0.004 & -0.015 --- 0.006 & 0.420 \\
$\rho_{R \rightarrow M_2 \leadsto Y}$ & 0.007 & -0.001 --- 0.015 & 0.086 & 0.001 & -0.003 --- 0.005 & 0.657 \\
$\rho_{R \rightarrow M_3 \leadsto Y}$ & -0.000 & -0.005 --- 0.005 & 0.975 & -0.004 & -0.008 --- -0.000 & {\bfseries 0.038} \\
$\rho_{R \rightarrow M_4 \leadsto Y}$ & 0.022 & 0.008 --- 0.036 & {\bfseries 0.002} & 0.023 & 0.013 --- 0.033 & {\bfseries $<$0.001} \\
$\rho_{R \rightarrow Y}$ & 0.060 & 0.042 --- 0.077 & {\bfseries $<$0.001} & 0.054 & 0.038 --- 0.070 & {\bfseries $<$0.001} \\
$\rho_\text{total}$ & 0.073 & 0.051 --- 0.095 & {\bfseries $<$0.001} & 0.067 & 0.049 --- 0.085 & {\bfseries $<$0.001} \\
\midrule\addlinespace[2.5pt]
\multicolumn{7}{l}{Whites vs Hispanics*} \\[2.5pt]
\midrule\addlinespace[2.5pt]
$\rho_{R \rightarrow M_1 \leadsto Y}$ & 0.021 & 0.014 --- 0.027 & {\bfseries $<$0.001} & 0.022 & 0.015 --- 0.028 & {\bfseries $<$0.001} \\
$\rho_{R \rightarrow M_2 \leadsto Y}$ & 0.028 & 0.024 --- 0.033 & {\bfseries $<$0.001} & 0.028 & 0.023 --- 0.032 & {\bfseries $<$0.001} \\
$\rho_{R \rightarrow M_3 \leadsto Y}$ & 0.005 & -0.000 --- 0.010 & 0.059 & 0.003 & -0.001 --- 0.008 & 0.175 \\
$\rho_{R \rightarrow M_4 \leadsto Y}$ & 0.022 & 0.015 --- 0.029 & {\bfseries $<$0.001} & 0.037 & 0.030 --- 0.043 & {\bfseries $<$0.001} \\
$\rho_{R \rightarrow Y}$ & 0.081 & 0.070 --- 0.092 & {\bfseries $<$0.001} & 0.058 & 0.047 --- 0.068 & {\bfseries $<$0.001} \\
$\rho_\text{total}$ & 0.136 & 0.123 --- 0.149 & {\bfseries $<$0.001} & 0.121 & 0.108 --- 0.134 & {\bfseries $<$0.001} \\
\midrule\addlinespace[2.5pt]
\multicolumn{7}{l}{Blacks vs Asians*} \\[2.5pt]
\midrule\addlinespace[2.5pt]
$\rho_{R \rightarrow M_1 \leadsto Y}$ & 0.006 & -0.011 --- 0.024 & 0.492 & -0.001 & -0.021 --- 0.019 & 0.944 \\
$\rho_{R \rightarrow M_2 \leadsto Y}$ & 0.011 & 0.000 --- 0.021 & {\bfseries 0.042} & -0.003 & -0.016 --- 0.011 & 0.673 \\
$\rho_{R \rightarrow M_3 \leadsto Y}$ & -0.009 & -0.019 --- 0.002 & 0.101 & 0.000 & -0.007 --- 0.007 & 0.958 \\
$\rho_{R \rightarrow M_4 \leadsto Y}$ & 0.016 & -0.006 --- 0.038 & 0.162 & 0.019 & -0.001 --- 0.039 & 0.061 \\
$\rho_{R \rightarrow Y}$ & -0.025 & -0.051 --- 0.001 & 0.060 & -0.023 & -0.047 --- 0.002 & 0.067 \\
$\rho_\text{total}$ & 0.004 & -0.029 --- 0.038 & 0.805 & -0.005 & -0.040 --- 0.029 & 0.756 \\
\midrule\addlinespace[2.5pt]
\multicolumn{7}{l}{Blacks vs Hispanics*} \\[2.5pt]
\midrule\addlinespace[2.5pt]
$\rho_{R \rightarrow M_1 \leadsto Y}$ & 0.020 & 0.014 --- 0.026 & {\bfseries $<$0.001} & 0.016 & 0.010 --- 0.022 & {\bfseries $<$0.001} \\
$\rho_{R \rightarrow M_2 \leadsto Y}$ & 0.041 & 0.035 --- 0.047 & {\bfseries $<$0.001} & 0.034 & 0.028 --- 0.040 & {\bfseries $<$0.001} \\
$\rho_{R \rightarrow M_3 \leadsto Y}$ & 0.003 & -0.001 --- 0.007 & 0.107 & 0.001 & -0.004 --- 0.006 & 0.625 \\
$\rho_{R \rightarrow M_4 \leadsto Y}$ & 0.021 & 0.014 --- 0.028 & {\bfseries $<$0.001} & 0.020 & 0.013 --- 0.027 & {\bfseries $<$0.001} \\
$\rho_{R \rightarrow Y}$ & 0.012 & 0.001 --- 0.022 & {\bfseries 0.028} & -0.014 & -0.025 --- -0.002 & {\bfseries 0.026} \\
$\rho_\text{total}$ & 0.086 & 0.065 --- 0.106 & {\bfseries $<$0.001} & 0.056 & 0.036 --- 0.075 & {\bfseries $<$0.001} \\
\midrule\addlinespace[2.5pt]
\multicolumn{7}{l}{Asians vs Hispanics*} \\[2.5pt]
\midrule\addlinespace[2.5pt]
$\rho_{R \rightarrow M_1 \leadsto Y}$ & 0.028 & 0.018 --- 0.039 & {\bfseries $<$0.001} & 0.027 & 0.018 --- 0.035 & {\bfseries $<$0.001} \\
$\rho_{R \rightarrow M_2 \leadsto Y}$ & 0.028 & 0.022 --- 0.033 & {\bfseries $<$0.001} & 0.027 & 0.022 --- 0.032 & {\bfseries $<$0.001} \\
$\rho_{R \rightarrow M_3 \leadsto Y}$ & 0.004 & 0.002 --- 0.006 & {\bfseries $<$0.001} & -0.001 & -0.003 --- 0.001 & 0.306 \\
$\rho_{R \rightarrow M_4 \leadsto Y}$ & -0.009 & -0.016 --- -0.001 & {\bfseries 0.028} & -0.008 & -0.016 --- -0.000 & {\bfseries 0.050} \\
$\rho_{R \rightarrow Y}$ & 0.026 & 0.016 --- 0.036 & {\bfseries $<$0.001} & 0.024 & 0.013 --- 0.034 & {\bfseries $<$0.001} \\
$\rho_\text{total}$ & 0.085 & 0.059 --- 0.111 & {\bfseries $<$0.001} & 0.065 & 0.038 --- 0.092 & {\bfseries $<$0.001} \\
\bottomrule
\end{tabular}
\begin{tablenotes}
\item \textsuperscript{*} Reference group; $M_1$: SES, $M_2$: Insurance access, $M_3$: Health behaviors, $M_4$: Health status.
\end{tablenotes}
\label{tab:prob_pos}
\end{table}

\begin{table}[t]
\caption{Disparity components across racial group comparisons, with only positive healthcare expenditures, reported on the geometric mean ratios.}
\footnotesize
\renewcommand{\arraystretch}{0.9}
\setlength{\tabcolsep}{12.5pt} 
\centering
\begin{tabular}{lcccccc}
\toprule
& \multicolumn{3}{c}{\textbf{MEPS data in year 2009}} & \multicolumn{3}{c}{\textbf{MEPS data in year 2016}} \\
\cmidrule(lr){2-4} \cmidrule(lr){5-7}
Disparity & Value & 95\% CI & p value & Value & 95\% CI & p value \\ 
\midrule\addlinespace[2.5pt]
\multicolumn{7}{l}{Whites vs Blacks*} \\[2.5pt]
\midrule\addlinespace[2.5pt]
$\rho_{R \rightarrow M_1 \leadsto Y}$ & 1.017 & 0.992 --- 1.041 & 0.184 & 1.027 & 0.994 --- 1.060 & 0.107 \\
$\rho_{R \rightarrow M_2 \leadsto Y}$ & 1.002 & 0.990 --- 1.013 & 0.782 & 0.989 & 0.980 --- 0.998 & {\bf 0.023} \\
$\rho_{R \rightarrow M_3 \leadsto Y}$ & 0.995 & 0.979 --- 1.010 & 0.495 & 1.007 & 0.990 --- 1.023 & 0.419 \\
$\rho_{R \rightarrow M_4 \leadsto Y}$ & 1.023 & 0.980 --- 1.066 & 0.300 & 1.016 & 0.975 --- 1.056 & 0.447 \\
$\rho_{R \rightarrow Y}$ & 1.328 & 1.251 --- 1.405 & {\bf $<$0.001} & 1.316 & 1.240 --- 1.392 & {\bf $<$0.001} \\
$\rho_\text{total}$ & 1.346 & 1.245 --- 1.447 & {\bf $<$0.001} & 1.338 & 1.237 --- 1.440 & {\bf $<$0.001} \\
\midrule\addlinespace[2.5pt]
\multicolumn{7}{l}{Whites vs Asians*} \\[2.5pt]
\midrule\addlinespace[2.5pt]
$\rho_{R \rightarrow M_1 \leadsto Y}$ & 0.997 & 0.950 --- 1.044 & 0.889 & 1.033 & 0.971 --- 1.095 & 0.297 \\
$\rho_{R \rightarrow M_2 \leadsto Y}$ & 1.020 & 0.984 --- 1.055 & 0.279 & 0.994 & 0.976 --- 1.012 & 0.518 \\
$\rho_{R \rightarrow M_3 \leadsto Y}$ & 1.012 & 1.000 --- 1.024 & 0.055 & 1.009 & 0.995 --- 1.023 & 0.219 \\
$\rho_{R \rightarrow M_4 \leadsto Y}$ & 1.195 & 1.126 --- 1.264 & {\bf $<$0.001} & 1.223 & 1.152 --- 1.293 & {\bf $<$0.001} \\
$\rho_{R \rightarrow Y}$ & 1.680 & 1.556 --- 1.804 & {\bf $<$0.001} & 1.462 & 1.351 --- 1.573 & {\bf $<$0.001} \\
$\rho_\text{total}$ & 1.894 & 1.724 --- 2.063 & {\bf $<$0.001} & 1.700 & 1.542 --- 1.858 & {\bf $<$0.001} \\
\midrule\addlinespace[2.5pt]
\multicolumn{7}{l}{Whites vs Hispanics*} \\[2.5pt]
\midrule\addlinespace[2.5pt]
$\rho_{R \rightarrow M_1 \leadsto Y}$ & 1.065 & 1.031 --- 1.099 & {\bf $<$0.001} & 1.138 & 1.096 --- 1.180 & {\bf $<$0.001} \\
$\rho_{R \rightarrow M_2 \leadsto Y}$ & 1.071 & 1.052 --- 1.089 & {\bf $<$0.001} & 1.060 & 1.043 --- 1.077 & {\bf $<$0.001} \\
$\rho_{R \rightarrow M_3 \leadsto Y}$ & 1.008 & 0.988 --- 1.027 & 0.438 & 1.005 & 0.988 --- 1.023 & 0.544 \\
$\rho_{R \rightarrow M_4 \leadsto Y}$ & 1.144 & 1.105 --- 1.183 & {\bf $<$0.001} & 1.128 & 1.089 --- 1.168 & {\bf $<$0.001} \\
$\rho_{R \rightarrow Y}$ & 1.320 & 1.250 --- 1.389 & {\bf $<$0.001} & 1.306 & 1.238 --- 1.374 & {\bf $<$0.001} \\
$\rho_\text{total}$ & 1.596 & 1.493 --- 1.700 & {\bf $<$0.001} & 1.576 & 1.468 --- 1.684 & {\bf $<$0.001} \\
\midrule\addlinespace[2.5pt]
\multicolumn{7}{l}{Blacks vs Asians*} \\[2.5pt]
\midrule\addlinespace[2.5pt]
$\rho_{R \rightarrow M_1 \leadsto Y}$ & 1.018 & 0.966 --- 1.071 & 0.494 & 1.058 & 0.978 --- 1.139 & 0.154 \\
$\rho_{R \rightarrow M_2 \leadsto Y}$ & 1.060 & 0.998 --- 1.122 & 0.057 & 1.010 & 0.985 --- 1.035 & 0.430 \\
$\rho_{R \rightarrow M_3 \leadsto Y}$ & 1.013 & 0.994 --- 1.033 & 0.185 & 1.017 & 0.995 --- 1.039 & 0.122 \\
$\rho_{R \rightarrow M_4 \leadsto Y}$ & 1.261 & 1.156 --- 1.367 & {\bf $<$0.001} & 1.256 & 1.159 --- 1.353 & {\bf $<$0.001} \\
$\rho_{R \rightarrow Y}$ & 1.194 & 1.074 --- 1.313 & {\bf 0.002} & 1.116 & 1.005 --- 1.228 & {\bf 0.040} \\
$\rho_\text{total}$ & 1.516 & 1.309 --- 1.723 & {\bf $<$0.001} & 1.331 & 1.154 --- 1.507 & {\bf $<$0.001} \\
\midrule\addlinespace[2.5pt]
\multicolumn{7}{l}{Blacks vs Hispanics*} \\[2.5pt]
\midrule\addlinespace[2.5pt]
$\rho_{R \rightarrow M_1 \leadsto Y}$ & 1.084 & 1.050 --- 1.118 & {\bf $<$0.001} & 1.125 & 1.091 --- 1.159 & {\bf $<$0.001} \\
$\rho_{R \rightarrow M_2 \leadsto Y}$ & 1.112 & 1.089 --- 1.136 & {\bf $<$0.001} & 1.090 & 1.069 --- 1.111 & {\bf $<$0.001} \\
$\rho_{R \rightarrow M_3 \leadsto Y}$ & 1.004 & 0.989 --- 1.018 & 0.645 & 0.997 & 0.983 --- 1.012 & 0.726 \\
$\rho_{R \rightarrow M_4 \leadsto Y}$ & 1.086 & 1.050 --- 1.122 & {\bf $<$0.001} & 1.071 & 1.035 --- 1.108 & {\bf $<$0.001} \\
$\rho_{R \rightarrow Y}$ & 0.947 & 0.897 --- 0.997 & {\bf 0.036} & 0.949 & 0.899 --- 1.000 & {\bf 0.048} \\
$\rho_\text{total}$ & 1.180 & 1.072 --- 1.288 & {\bf 0.001} & 1.168 & 1.068 --- 1.269 & {\bf 0.001} \\
\midrule\addlinespace[2.5pt]
\multicolumn{7}{l}{Asians vs Hispanics*} \\[2.5pt]
\midrule\addlinespace[2.5pt]
$\rho_{R \rightarrow M_1 \leadsto Y}$ & 1.028 & 0.980 --- 1.075 & 0.255 & 1.158 & 1.111 --- 1.206 & {\bf $<$0.001} \\
$\rho_{R \rightarrow M_2 \leadsto Y}$ & 1.059 & 1.041 --- 1.078 & {\bf $<$0.001} & 1.054 & 1.039 --- 1.070 & {\bf $<$0.001} \\
$\rho_{R \rightarrow M_3 \leadsto Y}$ & 1.005 & 0.995 --- 1.015 & 0.340 & 1.007 & 0.996 --- 1.017 & 0.204 \\
$\rho_{R \rightarrow M_4 \leadsto Y}$ & 0.864 & 0.827 --- 0.901 & {\bf $<$0.001} & 0.871 & 0.838 --- 0.904 & {\bf $<$0.001} \\
$\rho_{R \rightarrow Y}$ & 0.838 & 0.795 --- 0.882 & {\bf $<$0.001} & 0.950 & 0.908 --- 0.992 & {\bf 0.019} \\
$\rho_\text{total}$ & 0.810 & 0.722 --- 0.899 & {\bf $<$0.001} & 0.912 & 0.810 --- 1.014 & 0.092 \\
\bottomrule
\end{tabular}
\begin{tablenotes}
\item \textsuperscript{*} Reference group; $M_1$: SES, $M_2$: Insurance access, $M_3$: Health behaviors, $M_4$: Health status.
\end{tablenotes}
\label{tab:positive_Y}
\end{table}

\begin{table}[t]
\caption{Disparity components across racial group comparisons estimated using a two-stage super learner, reported on the difference scale (arithmetic mean). }
\footnotesize
\renewcommand{\arraystretch}{0.9}
\setlength{\tabcolsep}{12.5pt} 
\centering
\begin{tabular}{lcccccc}
\toprule
& \multicolumn{3}{c}{\textbf{MEPS data in year 2009}} & \multicolumn{3}{c}{\textbf{MEPS data in year 2016}} \\
\cmidrule(lr){2-4} \cmidrule(lr){5-7}
Disparity & Value & 95\% CI & p-value & Value & 95\% CI & p-value \\ 
\midrule\addlinespace[2.5pt]
\multicolumn{7}{l}{Whites vs Blacks*} \\[2.5pt]
\midrule\addlinespace[2.5pt]
$\rho_{R \rightarrow M_1 \leadsto Y}$ & -94.9 & -220.1 --- 30.3 & 0.137 & -22.1 & -155.7 --- 111.5 & 0.746 \\
$\rho_{R \rightarrow M_2 \leadsto Y}$ & 131.3 & 83.5 --- 179.0 & {\bfseries $<$0.001} & 74.4 & 33.2 --- 115.7 & {\bfseries $<$0.001} \\
$\rho_{R \rightarrow M_3 \leadsto Y}$ & -84.5 & -146.1 --- -22.9 & {\bfseries 0.007} & -3.3 & -59.6 --- 52.9 & 0.907 \\
$\rho_{R \rightarrow M_4 \leadsto Y}$ & 139.2 & -145.5 --- 423.9 & 0.338 & 325.8 & -15.4 --- 666.9 & 0.061 \\
$\rho_{R \rightarrow Y}$ & 1099.6 & 654.3 --- 1545.0 & {\bfseries $<$0.001} & 1735.5 & 1189.9 --- 2281.1 & {\bfseries $<$0.001} \\
$\rho_\text{total}$ & 337.0 & -111.2 --- 785.3 & 0.141 & 1036.1 & 461.0 --- 1611.1 & {\bfseries $<$0.001} \\
\midrule\addlinespace[2.5pt]
\multicolumn{7}{l}{Whites vs Asians*} \\[2.5pt]
\midrule\addlinespace[2.5pt]
$\rho_{R \rightarrow M_1 \leadsto Y}$ & 51.2 & -55.4 --- 157.8 & 0.346 & -11.4 & -147.5 --- 124.6 & 0.869 \\
$\rho_{R \rightarrow M_2 \leadsto Y}$ & 48.4 & -11.7 --- 108.6 & 0.115 & 1.8 & -40.2 --- 43.8 & 0.933 \\
$\rho_{R \rightarrow M_3 \leadsto Y}$ & -0.9 & -72.4 --- 70.6 & 0.981 & -16.3 & -84.6 --- 52.1 & 0.641 \\
$\rho_{R \rightarrow M_4 \leadsto Y}$ & 934.4 & 659.0 --- 1209.8 & {\bfseries $<$0.001} & 1808.7 & 1476.6 --- 2140.8 & {\bfseries $<$0.001} \\
$\rho_{R \rightarrow Y}$ & 1561.4 & 1123.0 --- 1999.9 & {\bfseries $<$0.001} & 1921.0 & 1369.2 --- 2472.8 & {\bfseries $<$0.001} \\
$\rho_\text{total}$ & 2374.3 & 1924.8 --- 2823.8 & {\bfseries $<$0.001} & 2979.9 & 2439.4 --- 3520.3 & {\bfseries $<$0.001} \\
\midrule\addlinespace[2.5pt]
\multicolumn{7}{l}{Whites vs Hispanics*} \\[2.5pt]
\midrule\addlinespace[2.5pt]
$\rho_{R \rightarrow M_1 \leadsto Y}$ & 150.1 & 61.8 --- 238.5 & {\bfseries $<$0.001} & 142.4 & 20.8 --- 263.9 & {\bfseries 0.022} \\
$\rho_{R \rightarrow M_2 \leadsto Y}$ & 503.0 & 423.6 --- 582.4 & {\bfseries $<$0.001} & 400.2 & 335.1 --- 465.3 & {\bfseries $<$0.001} \\
$\rho_{R \rightarrow M_3 \leadsto Y}$ & 9.3 & -66.7 --- 85.4 & 0.810 & 33.5 & -50.1 --- 117.0 & 0.432 \\
$\rho_{R \rightarrow M_4 \leadsto Y}$ & 740.2 & 547.0 --- 933.4 & {\bfseries $<$0.001} & 1069.8 & 807.2 --- 1332.5 & {\bfseries $<$0.001} \\
$\rho_{R \rightarrow Y}$ & 1150.5 & 840.5 --- 1460.5 & {\bfseries $<$0.001} & 1770.5 & 1364.4 --- 2176.7 & {\bfseries $<$0.001} \\
$\rho_\text{total}$ & 1490.1 & 1136.6 --- 1843.7 & {\bfseries $<$0.001} & 2147.5 & 1673.0 --- 2622.0 & {\bfseries $<$0.001} \\
\midrule\addlinespace[2.5pt]
\multicolumn{7}{l}{Blacks vs Asians*} \\[2.5pt]
\midrule\addlinespace[2.5pt]
$\rho_{R \rightarrow M_1 \leadsto Y}$ & 12.9 & -90.6 --- 116.3 & 0.807 & 57.8 & -88.4 --- 203.9 & 0.438 \\
$\rho_{R \rightarrow M_2 \leadsto Y}$ & 3.7 & -50.6 --- 58.0 & 0.894 & -17.3 & -63.5 --- 28.9 & 0.463 \\
$\rho_{R \rightarrow M_3 \leadsto Y}$ & 82.4 & -0.7 --- 165.6 & 0.052 & 82.4 & 0.4 --- 164.4 & {\bfseries 0.049} \\
$\rho_{R \rightarrow M_4 \leadsto Y}$ & 1806.8 & 1457.4 --- 2156.2 & {\bfseries $<$0.001} & 2198.3 & 1738.6 --- 2657.9 & {\bfseries $<$0.001} \\
$\rho_{R \rightarrow Y}$ & 634.7 & 211.9 --- 1057.6 & {\bfseries 0.003} & 432.0 & -114.2 --- 978.2 & 0.121 \\
$\rho_\text{total}$ & 1880.8 & 1386.1 --- 2375.6 & {\bfseries $<$0.001} & 1689.9 & 1107.2 --- 2272.5 & {\bfseries $<$0.001} \\
\midrule\addlinespace[2.5pt]
\multicolumn{7}{l}{Blacks vs Hispanics*} \\[2.5pt]
\midrule\addlinespace[2.5pt]
$\rho_{R \rightarrow M_1 \leadsto Y}$ & 222.2 & 165.7 --- 278.6 & {\bfseries $<$0.001} & 313.0 & 217.9 --- 408.2 & {\bfseries $<$0.001} \\
$\rho_{R \rightarrow M_2 \leadsto Y}$ & 335.7 & 260.4 --- 411.1 & {\bfseries $<$0.001} & 296.7 & 240.1 --- 353.3 & {\bfseries $<$0.001} \\
$\rho_{R \rightarrow M_3 \leadsto Y}$ & 41.3 & -18.3 --- 100.9 & 0.174 & -5.7 & -78.3 --- 66.8 & 0.877 \\
$\rho_{R \rightarrow M_4 \leadsto Y}$ & 801.8 & 527.2 --- 1076.5 & {\bfseries $<$0.001} & 817.6 & 501.8 --- 1133.4 & {\bfseries $<$0.001} \\
$\rho_{R \rightarrow Y}$ & 116.1 & -202.3 --- 434.5 & 0.475 & 91.4 & -266.2 --- 449.1 & 0.616 \\
$\rho_\text{total}$ & 992.8 & 563.3 --- 1422.3 & {\bfseries $<$0.001} & 957.9 & 449.5 --- 1466.3 & {\bfseries $<$0.001} \\
\midrule\addlinespace[2.5pt]
\multicolumn{7}{l}{Asians vs Hispanics*} \\[2.5pt]
\midrule\addlinespace[2.5pt]
$\rho_{R \rightarrow M_1 \leadsto Y}$ & 259.2 & 150.2 --- 368.2 & {\bfseries $<$0.001} & 205.6 & 38.0 --- 373.3 & {\bfseries 0.016} \\
$\rho_{R \rightarrow M_2 \leadsto Y}$ & 364.5 & 286.4 --- 442.6 & {\bfseries $<$0.001} & 401.1 & 331.9 --- 470.2 & {\bfseries $<$0.001} \\
$\rho_{R \rightarrow M_3 \leadsto Y}$ & -73.8 & -116.8 --- -30.8 & {\bfseries $<$0.001} & -46.2 & -125.1 --- 32.6 & 0.251 \\
$\rho_{R \rightarrow M_4 \leadsto Y}$ & -576.6 & -832.6 --- -320.7 & {\bfseries $<$0.001} & -433.3 & -846.2 --- -20.4 & {\bfseries 0.040} \\
$\rho_{R \rightarrow Y}$ & 29.1 & -280.0 --- 338.3 & 0.854 & 445.0 & 87.1 --- 802.8 & {\bfseries 0.015} \\
$\rho_\text{total}$ & -643.1 & -1034.3 --- -251.9 & {\bfseries 0.001} & -501.7 & -968.4 --- -35.0 & {\bfseries 0.035} \\
\bottomrule
\end{tabular}

\begin{tablenotes}
\item \textsuperscript{*} Reference group; $M_1$: SES, $M_2$: Insurance access, $M_3$: Health behaviors, $M_4$: Health status.
\end{tablenotes}
\label{tab:PSEs_difference}
\end{table}

%  ---  ---  ---  ---  ---  ---  ---  ---  ---  ---  ---  ---  --- -
\subsection{Assessment of mediators choice}
\label{app:mediator and order}
%  ---  ---  ---  ---  ---  ---  ---  ---  ---  ---  ---  ---  --- -

\subsubsection{Healthcare satisfaction as a mediator} 
\label{app:satisfaction}

We conducted an additional analysis to examine the role of patient satisfaction and healthcare experiences as a potential mediator. Patient satisfaction reflects the quality of patient-provider interactions, which may influence subsequent healthcare utilization and expenditures. We used the variable measuring the ``rating of healthcare from all doctors and other health providers,'' scored from 0 (worst healthcare possible) to 10 (best healthcare possible). Because this variable is recorded only for individuals who visited a doctor's office or clinic at least once during the survey year, the analysis was restricted to the subpopulation with valid satisfaction ratings.

Table~\ref{tab:sub_satisfaction_not_include} presents the decomposition results for this subpopulation (n = 13,365 for MEPS 2009, and n = 11,546 for MEPS 2016) using the original four mediators (SES, Insurance access, Health Behaviors, Health Status). The results are partly consistent with the primary analysis on the full sample (Table~\ref{tab:PSEs}), though magnitudes vary slightly due to the exclusion of individuals with no healthcare visits (who generally have lower expenditures).

Longitudinal evidence suggests that self-perceived overall health and functional ability influence subsequent levels of satisfaction with medical care; however, there is insufficient evidence of a causal pathway from satisfaction to later health outcomes \citep{hall1993causal}. Thus, Table~\ref{tab:sub_satisfation_include} extends the analysis by including healthcare satisfaction as a fifth mediator ($M_5$), positioned downstream of health status. $\rho_{R \rightarrow M_5 \leadsto Y}$ captures the disparity attributed to  healthcare satisfaction in healthcare experiences. We found that this factor contributed significantly to overall disparities in most comparisons, with the exception of Black-Asian comparisons in both 2009 and 2016, as well as White-Hispanic and Asian-Hispanic comparisons in 2016. Including healthcare satisfaction as a mediator also attenuated the outcome-attributed  disparities ($\rho_{R \rightarrow Y}$) compared to the four-mediator model in Table~\ref{tab:sub_satisfaction_not_include}, suggesting that a portion of the unexplained disparity in the primary analysis may be driven by differential healthcare experiences and satisfaction. Racial and ethnic minorities often face poorer access to healthcare due to financial, social, and cultural barriers and tend to report lower levels of satisfaction \citep{chen2019association, turpin2021differences, ku2023association}. In addition, racial or ethnic concordance between patients and providers is positively associated with healthcare satisfaction, which may influence subsequent healthcare utilization \citep{takeshita2020association, ku2023association}. These results indicate the importance to improve physician workforce diversity and to promote culturally responsive care, particularly in racially or ethnically discordant patient-provider interactions.

\begin{table}[t]
\caption{Disparity components across racial group comparisons, subpopulation with the records of healthcare satisfaction, reported on the scaled geometric mean ratios.}
 % or \footnotesize for slightly larger font
% \setstretch{1}
\footnotesize
\renewcommand{\arraystretch}{0.9}
\setlength{\tabcolsep}{12.5pt} 
\centering
\begin{tabular}{lcccccc}
\toprule
& \multicolumn{3}{c}{\textbf{MEPS data in year 2009}} & \multicolumn{3}{c}{\textbf{MEPS data in year 2016}} \\
\cmidrule(lr){2-4} \cmidrule(lr){5-7}
Disparity & Value & 95\% CI & p value & Value & 95\% CI & p value \\ 
\midrule\addlinespace[2.5pt]
\multicolumn{7}{l}{Whites vs Blacks*} \\[2.5pt]
\midrule\addlinespace[2.5pt]
$\rho_{R \rightarrow M_1 \leadsto Y}$ & 1.031 & 0.993 --- 1.070 & 0.111 & 1.063 & 1.008 --- 1.118 & {\bf 0.026} \\
$\rho_{R \rightarrow M_2 \leadsto Y}$ & 1.020 & 1.000 --- 1.041 & {\bf 0.047} & 1.001 & 0.984 --- 1.018 & 0.920 \\
$\rho_{R \rightarrow M_3 \leadsto Y}$ & 0.982 & 0.943 --- 1.021 & 0.365 & 1.008 & 0.976 --- 1.040 & 0.618 \\
$\rho_{R \rightarrow M_4 \leadsto Y}$ & 1.011 & 0.943 --- 1.079 & 0.754 & 1.040 & 0.977 --- 1.103 & 0.219 \\
$\rho_{R \rightarrow Y}$ & 1.615 & 1.462 --- 1.768 & {\bf $<$0.001} & 1.638 & 1.481 --- 1.794 & {\bf $<$0.001} \\
$\rho_\text{total}$ & 1.663 & 1.477 --- 1.850 & {\bf $<$0.001} & 1.742 & 1.535 --- 1.948 & {\bf $<$0.001} \\
\midrule\addlinespace[2.5pt]
\multicolumn{7}{l}{Whites vs Asians*} \\[2.5pt]
\midrule\addlinespace[2.5pt]
$\rho_{R \rightarrow M_1 \leadsto Y}$ & 0.996 & 0.900 --- 1.091 & 0.932 & 1.110 & 0.962 --- 1.258 & 0.145 \\
$\rho_{R \rightarrow M_2 \leadsto Y}$ & 1.017 & 0.981 --- 1.052 & 0.358 & 1.010 & 0.972 --- 1.049 & 0.599 \\
$\rho_{R \rightarrow M_3 \leadsto Y}$ & 1.046 & 0.970 --- 1.122 & 0.238 & 1.046 & 1.018 --- 1.074 & {\bf 0.001} \\
$\rho_{R \rightarrow M_4 \leadsto Y}$ & 1.243 & 1.116 --- 1.369 & {\bf $<$0.001} & 1.339 & 1.201 --- 1.476 & {\bf $<$0.001} \\
$\rho_{R \rightarrow Y}$ & 2.016 & 1.736 --- 2.297 & {\bf $<$0.001} & 1.865 & 1.583 --- 2.146 & {\bf $<$0.001} \\
$\rho_\text{total}$ & 2.305 & 1.936 --- 2.674 & {\bf $<$0.001} & 2.312 & 1.907 --- 2.716 & {\bf $<$0.001} \\
\midrule\addlinespace[2.5pt]
\multicolumn{7}{l}{Whites vs Hispanics*} \\[2.5pt]
\midrule\addlinespace[2.5pt]
$\rho_{R \rightarrow M_1 \leadsto Y}$ & 1.085 & 1.036 --- 1.134 & {\bf 0.001} & 1.107 & 1.039 --- 1.174 & {\bf 0.002} \\
$\rho_{R \rightarrow M_2 \leadsto Y}$ & 1.140 & 1.104 --- 1.176 & {\bf $<$0.001} & 1.052 & 1.030 --- 1.074 & {\bf $<$0.001} \\
$\rho_{R \rightarrow M_3 \leadsto Y}$ & 1.013 & 0.972 --- 1.054 & 0.545 & 1.001 & 0.972 --- 1.029 & 0.971 \\
$\rho_{R \rightarrow M_4 \leadsto Y}$ & 1.176 & 1.119 --- 1.234 & {\bf $<$0.001} & 1.212 & 1.148 --- 1.275 & {\bf $<$0.001} \\
$\rho_{R \rightarrow Y}$ & 1.757 & 1.606 --- 1.908 & {\bf $<$0.001} & 1.533 & 1.402 --- 1.664 & {\bf $<$0.001} \\
$\rho_\text{total}$ & 2.246 & 2.027 --- 2.465 & {\bf $<$0.001} & 1.888 & 1.689 --- 2.088 & {\bf $<$0.001} \\
\midrule\addlinespace[2.5pt]
\multicolumn{7}{l}{Blacks vs Asians*} \\[2.5pt]
\midrule\addlinespace[2.5pt]
$\rho_{R \rightarrow M_1 \leadsto Y}$ & 0.924 & 0.850 --- 0.997 & {\bf 0.041} & 1.042 & 0.929 --- 1.156 & 0.465 \\
$\rho_{R \rightarrow M_2 \leadsto Y}$ & 1.021 & 0.959 --- 1.083 & 0.501 & 0.981 & 0.949 --- 1.013 & 0.242 \\
$\rho_{R \rightarrow M_3 \leadsto Y}$ & 0.935 & 0.890 --- 0.979 & {\bf 0.004} & 1.006 & 0.974 --- 1.038 & 0.706 \\
$\rho_{R \rightarrow M_4 \leadsto Y}$ & 1.169 & 1.018 --- 1.321 & {\bf 0.028} & 1.263 & 1.104 --- 1.422 & {\bf 0.001} \\
$\rho_{R \rightarrow Y}$ & 1.024 & 0.859 --- 1.188 & 0.778 & 1.132 & 0.935 --- 1.330 & 0.189 \\
$\rho_\text{total}$ & 1.277 & 1.020 --- 1.534 & {\bf 0.035} & 1.361 & 1.081 --- 1.640 & {\bf 0.011} \\
\midrule\addlinespace[2.5pt]
\multicolumn{7}{l}{Blacks vs Hispanics*} \\[2.5pt]
\midrule\addlinespace[2.5pt]
$\rho_{R \rightarrow M_1 \leadsto Y}$ & 1.095 & 1.042 --- 1.147 & {\bf $<$0.001} & 1.069 & 1.018 --- 1.120 & {\bf 0.008} \\
$\rho_{R \rightarrow M_2 \leadsto Y}$ & 1.186 & 1.140 --- 1.231 & {\bf $<$0.001} & 1.066 & 1.036 --- 1.095 & {\bf $<$0.001} \\
$\rho_{R \rightarrow M_3 \leadsto Y}$ & 1.004 & 0.971 --- 1.037 & 0.812 & 1.001 & 0.976 --- 1.026 & 0.916 \\
$\rho_{R \rightarrow M_4 \leadsto Y}$ & 1.120 & 1.055 --- 1.186 & {\bf $<$0.001} & 1.024 & 0.971 --- 1.076 & 0.375 \\
$\rho_{R \rightarrow Y}$ & 0.928 & 0.840 --- 1.017 & 0.113 & 0.826 & 0.750 --- 0.901 & {\bf $<$0.001} \\
$\rho_\text{total}$ & 1.380 & 1.170 --- 1.589 & {\bf $<$0.001} & 1.034 & 0.884 --- 1.185 & 0.655 \\
\midrule\addlinespace[2.5pt]
\multicolumn{7}{l}{Asians vs Hispanics*} \\[2.5pt]
\midrule\addlinespace[2.5pt]
$\rho_{R \rightarrow M_1 \leadsto Y}$ & 1.067 & 0.975 --- 1.160 & 0.155 & 1.105 & 1.037 --- 1.173 & {\bf 0.003} \\
$\rho_{R \rightarrow M_2 \leadsto Y}$ & 1.144 & 1.107 --- 1.182 & {\bf $<$0.001} & 1.042 & 1.017 --- 1.066 & {\bf 0.001} \\
$\rho_{R \rightarrow M_3 \leadsto Y}$ & 0.997 & 0.980 --- 1.013 & 0.715 & 1.000 & 0.985 --- 1.016 & 0.974 \\
$\rho_{R \rightarrow M_4 \leadsto Y}$ & 0.855 & 0.792 --- 0.918 & {\bf $<$0.001} & 0.859 & 0.805 --- 0.913 & {\bf $<$0.001} \\
$\rho_{R \rightarrow Y}$ & 0.862 & 0.785 --- 0.940 & {\bf 0.001} & 0.953 & 0.874 --- 1.031 & 0.238 \\
$\rho_\text{total}$ & 0.983 & 0.798 --- 1.169 & 0.861 & 0.921 & 0.760 --- 1.082 & 0.337 \\
\bottomrule
\end{tabular}
\begin{tablenotes}
\item \textsuperscript{*} Reference group; $M_1$: SES, $M_2$: Insurance access, $M_3$: Health behaviors, $M_4$: Health status.
\end{tablenotes}
\label{tab:sub_satisfaction_not_include}
\end{table}

\begin{table}[t]
\caption{Disparity components across racial group comparisons, subpopulation with the records of healthcare satisfaction, including healthcare satisfaction as mediator, reported on the scaled geometric mean ratios.}
 % or \footnotesize for slightly larger font
% \setstretch{1}
\footnotesize
\renewcommand{\arraystretch}{0.83}
\setlength{\tabcolsep}{12.5pt} 
\centering
\begin{tabular}{lcccccc}
\toprule
& \multicolumn{3}{c}{\textbf{MEPS data in year 2009}} & \multicolumn{3}{c}{\textbf{MEPS data in year 2016}} \\
\cmidrule(lr){2-4} \cmidrule(lr){5-7}
Disparity & Value & 95\% CI & p value & Value & 95\% CI & p value \\ 
\midrule\addlinespace[2.5pt]
\multicolumn{7}{l}{Whites vs Blacks*} \\[2.5pt]
\midrule\addlinespace[2.5pt]
$\rho_{R \rightarrow M_1 \leadsto Y}$ & 1.031 & 0.993 --- 1.070 & 0.111 & 1.063 & 1.008 --- 1.118 & {\bf 0.026} \\
$\rho_{R \rightarrow M_2 \leadsto Y}$ & 1.020 & 1.000 --- 1.041 & {\bf 0.047} & 1.001 & 0.984 --- 1.018 & 0.920 \\
$\rho_{R \rightarrow M_3 \leadsto Y}$ & 0.982 & 0.943 --- 1.021 & 0.365 & 1.008 & 0.976 --- 1.040 & 0.618 \\
$\rho_{R \rightarrow M_4 \leadsto Y}$ & 1.011 & 0.943 --- 1.079 & 0.754 & 1.040 & 0.977 --- 1.103 & 0.219 \\
$\rho_{R \rightarrow M_5 \leadsto Y}$ & 1.052 & 1.028 --- 1.075 & {\bf $<$0.001} & 1.030 & 1.011 --- 1.048 & {\bf 0.001} \\
$\rho_{R \rightarrow Y}$ & 1.591 & 1.444 --- 1.738 & {\bf $<$0.001} & 1.622 & 1.471 --- 1.772 & {\bf $<$0.001} \\
$\rho_\text{total}$ & 1.663 & 1.477 --- 1.850 & {\bf $<$0.001} & 1.742 & 1.535 --- 1.948 & {\bf $<$0.001} \\
\midrule\addlinespace[2.5pt]
\multicolumn{7}{l}{Whites vs Asians*} \\[2.5pt]
\midrule\addlinespace[2.5pt]
$\rho_{R \rightarrow M_1 \leadsto Y}$ & 0.996 & 0.900 --- 1.091 & 0.932 & 1.110 & 0.962 --- 1.258 & 0.145 \\
$\rho_{R \rightarrow M_2 \leadsto Y}$ & 1.017 & 0.981 --- 1.052 & 0.358 & 1.010 & 0.972 --- 1.049 & 0.599 \\
$\rho_{R \rightarrow M_3 \leadsto Y}$ & 1.046 & 0.970 --- 1.122 & 0.238 & 1.046 & 1.018 --- 1.074 & {\bf 0.001} \\
$\rho_{R \rightarrow M_4 \leadsto Y}$ & 1.243 & 1.116 --- 1.369 & {\bf $<$0.001} & 1.339 & 1.201 --- 1.476 & {\bf $<$0.001} \\
$\rho_{R \rightarrow M_5 \leadsto Y}$ & 1.050 & 1.003 --- 1.098 & {\bf 0.038} & 1.108 & 1.072 --- 1.143 & {\bf $<$0.001} \\
$\rho_{R \rightarrow Y}$ & 1.969 & 1.701 --- 2.237 & {\bf $<$0.001} & 1.842 & 1.569 --- 2.115 & {\bf $<$0.001} \\
$\rho_\text{total}$ & 2.305 & 1.936 --- 2.674 & {\bf $<$0.001} & 2.312 & 1.907 --- 2.716 & {\bf $<$0.001} \\
\midrule\addlinespace[2.5pt]
\multicolumn{7}{l}{Whites vs Hispanics*} \\[2.5pt]
\midrule\addlinespace[2.5pt]
$\rho_{R \rightarrow M_1 \leadsto Y}$ & 1.085 & 1.036 --- 1.134 & {\bf 0.001} & 1.107 & 1.039 --- 1.174 & {\bf 0.002} \\
$\rho_{R \rightarrow M_2 \leadsto Y}$ & 1.140 & 1.104 --- 1.176 & {\bf $<$0.001} & 1.052 & 1.030 --- 1.074 & {\bf $<$0.001} \\
$\rho_{R \rightarrow M_3 \leadsto Y}$ & 1.013 & 0.972 --- 1.054 & 0.545 & 1.001 & 0.972 --- 1.029 & 0.971 \\
$\rho_{R \rightarrow M_4 \leadsto Y}$ & 1.176 & 1.119 --- 1.234 & {\bf $<$0.001} & 1.212 & 1.148 --- 1.275 & {\bf $<$0.001} \\
$\rho_{R \rightarrow M_5 \leadsto Y}$ & 1.020 & 1.006 --- 1.034 & {\bf 0.005} & 1.012 & 0.991 --- 1.033 & 0.262 \\
$\rho_{R \rightarrow Y}$ & 1.745 & 1.598 --- 1.892 & {\bf $<$0.001} & 1.531 & 1.401 --- 1.660 & {\bf $<$0.001} \\
$\rho_\text{total}$ & 2.246 & 2.027 --- 2.465 & {\bf $<$0.001} & 1.888 & 1.689 --- 2.088 & {\bf $<$0.001} \\
\midrule\addlinespace[2.5pt]
\multicolumn{7}{l}{Blacks vs Asians*} \\[2.5pt]
\midrule\addlinespace[2.5pt]
$\rho_{R \rightarrow M_1 \leadsto Y}$ & 0.924 & 0.850 --- 0.997 & {\bf 0.041} & 1.042 & 0.929 --- 1.156 & 0.465 \\
$\rho_{R \rightarrow M_2 \leadsto Y}$ & 1.021 & 0.959 --- 1.083 & 0.501 & 0.981 & 0.949 --- 1.013 & 0.242 \\
$\rho_{R \rightarrow M_3 \leadsto Y}$ & 0.935 & 0.890 --- 0.979 & {\bf 0.004} & 1.006 & 0.974 --- 1.038 & 0.706 \\
$\rho_{R \rightarrow M_4 \leadsto Y}$ & 1.169 & 1.018 --- 1.321 & {\bf 0.028} & 1.263 & 1.104 --- 1.422 & {\bf 0.001} \\
$\rho_{R \rightarrow M_5 \leadsto Y}$ & 0.976 & 0.936 --- 1.017 & 0.257 & 1.027 & 0.970 --- 1.085 & 0.352 \\
$\rho_{R \rightarrow Y}$ & 1.017 & 0.864 --- 1.170 & 0.824 & 1.124 & 0.942 --- 1.306 & 0.183 \\
$\rho_\text{total}$ & 1.277 & 1.020 --- 1.534 & {\bf 0.035} & 1.361 & 1.081 --- 1.640 & {\bf 0.011} \\
\midrule\addlinespace[2.5pt]
\multicolumn{7}{l}{Blacks vs Hispanics*} \\[2.5pt]
\midrule\addlinespace[2.5pt]
$\rho_{R \rightarrow M_1 \leadsto Y}$ & 1.095 & 1.042 --- 1.147 & {\bf $<$0.001} & 1.069 & 1.018 --- 1.120 & {\bf 0.008} \\
$\rho_{R \rightarrow M_2 \leadsto Y}$ & 1.186 & 1.140 --- 1.231 & {\bf $<$0.001} & 1.066 & 1.036 --- 1.095 & {\bf $<$0.001} \\
$\rho_{R \rightarrow M_3 \leadsto Y}$ & 1.004 & 0.971 --- 1.037 & 0.812 & 1.001 & 0.976 --- 1.026 & 0.916 \\
$\rho_{R \rightarrow M_4 \leadsto Y}$ & 1.120 & 1.055 --- 1.186 & {\bf $<$0.001} & 1.024 & 0.971 --- 1.076 & 0.375 \\
$\rho_{R \rightarrow M_5 \leadsto Y}$ & 0.966 & 0.945 --- 0.986 & {\bf 0.001} & 0.959 & 0.941 --- 0.977 & {\bf $<$0.001} \\
$\rho_{R \rightarrow Y}$ & 0.952 & 0.870 --- 1.035 & 0.256 & 0.852 & 0.776 --- 0.928 & {\bf $<$0.001} \\
$\rho_\text{total}$ & 1.380 & 1.170 --- 1.589 & {\bf $<$0.001} & 1.034 & 0.884 --- 1.185 & 0.655 \\
\midrule\addlinespace[2.5pt]
\multicolumn{7}{l}{Asians vs Hispanics*} \\[2.5pt]
\midrule\addlinespace[2.5pt]
$\rho_{R \rightarrow M_1 \leadsto Y}$ & 1.067 & 0.975 --- 1.160 & 0.155 & 1.105 & 1.037 --- 1.173 & {\bf 0.003} \\
$\rho_{R \rightarrow M_2 \leadsto Y}$ & 1.144 & 1.107 --- 1.182 & {\bf $<$0.001} & 1.042 & 1.017 --- 1.066 & {\bf 0.001} \\
$\rho_{R \rightarrow M_3 \leadsto Y}$ & 0.997 & 0.980 --- 1.013 & 0.715 & 1.000 & 0.985 --- 1.016 & 0.974 \\
$\rho_{R \rightarrow M_4 \leadsto Y}$ & 0.855 & 0.792 --- 0.918 & {\bf $<$0.001} & 0.859 & 0.805 --- 0.913 & {\bf $<$0.001} \\
$\rho_{R \rightarrow M_5 \leadsto Y}$ & 0.981 & 0.962 --- 1.000 & {\bf 0.049} & 0.999 & 0.982 --- 1.016 & 0.895 \\
$\rho_{R \rightarrow Y}$ & 0.891 & 0.809 --- 0.972 & {\bf 0.008} & 0.937 & 0.858 --- 1.017 & 0.123 \\
$\rho_\text{total}$ & 0.983 & 0.798 --- 1.169 & 0.861 & 0.921 & 0.760 --- 1.082 & 0.337 \\
\bottomrule
\end{tabular}
\begin{tablenotes}
\item \textsuperscript{*} Reference group; $M_1$: SES, $M_2$: Insurance access, $M_3$: Health behaviors, $M_4$: Health status, $M_5$: healthcare satisfaction.
\end{tablenotes}
\label{tab:sub_satisfation_include}
\end{table}

\subsubsection{Mediator ordering} 
\label{app:order}

A wide range of factors contribute to healthcare expenditures. As discussed in the Introduction and in our interpretation of the empirical results, SES, insurance access, health behaviors, and health status are key factors influencing healthcare expenditures. This motivates a more careful examination of the causal ordering among these mediators.

We begin with SES  because it is a relatively upstream determinant of subsequent social and health-related processes \citep{barkley_factors_2008,beydoun_racial_2016,leapman2022mediators,croce_lower_2023}. Income, education, and employment shape access to employer-sponsored or private insurance, eligibility for public coverage, and continuity of insurance over time \citep{o2003employers,litaker2005context,durden2006access,sohn2017racial}. For example, a study using longitudinal data from the Survey of Income and Program Participation found that spells of unemployment place privately insured individuals at substantially increased risk of losing or changing their health insurance coverage \citep{palumbohealth}. At the same time, very low income may qualify individuals for Medicaid or SCHIP, but only when state eligibility criteria and family or immigration status requirements are met \citep{bunch2025health}. Insurance access may then influence health behaviors by improving access to preventive care, counseling, and treatment services \citep{brown2018impact}. A longitudinal cohort study based on Oregon Medicaid enrollment data found that newly insured individuals had higher odds of smoking cessation over 24 months of follow-up than those who remained uninsured \citep{bailey2016effect}. Health behaviors are strong predictors of health status, where smoking and physical inactivity increase the risks of cardiovascular disease, cancer, disability, and all-cause mortality over time \citep{cairney2019physical, zhu2024combined, hu2024impact, wang2025systematic}. For example, smoking cessation is also associated with reduced depression and anxiety as well as improved quality of life \citep{taylor2014change}. Finally, physical multimorbidity would indicate a significantly increased likelihood of catastrophic health expenditure \citep{zhao2020physical}. Taken together, these considerations motivate our primary ordering,  $\text{SES} (M_1) \rightarrow \text{Insurance Access} (M_2) \rightarrow \text{Health Behaviors} (M_3) \rightarrow \text{Health Status}(M_4)$. We note that feedback relationships are likely present in practice (e.g., poor health affecting employment and income), but such dynamic processes cannot be represented within the acyclic framework required for this decomposition.

However, we also acknowledge that alternative mediator orderings may be plausible; for instance, health status may affect health behaviors. On the one hand, some studies suggest that individuals become more physically active after receiving a chronic disease diagnosis, possibly because the diagnosis serves as a signal to adopt a healthier lifestyle \citep{neutel2008changes, yuan2011family}. On the other hand, some studies suggest that patients remain inactive or even become less physically active due to physical barriers \citep{ramage2012adopting, newsom2012health, newsom2012healthGerontology}. To assess the sensitivity of our decomposition to these ordering assumptions, we conducted an additional analysis using an alternative mediator ordering: $\text{SES}\ (M_1) \rightarrow \text{Insurance Access}\ (M_2) \rightarrow \text{Health Status}\ (M_3) \rightarrow \text{Health Behaviors}\ (M_4)$. As shown in Table~\ref{tab:PSEs_order}, although the point estimates change slightly in magnitude, the overall conclusions regarding the main drivers of disparity remain robust. The significance and direction of the disparities attributed to health status are unchanged relative to Table~\ref{tab:PSEs}. The only notable difference is that the disparity attributed to health behaviors in the Black versus Asian comparison in 2016 changes from non-significant (0.983, $p=0.493$) to significant (0.950, $p=0.018$). Overall, these findings suggest that our conclusions are robust to plausible alternative mediator orderings.

\begin{center}
\begin{table}[t]
\caption{Disparity components across racial group comparisons under an alternative mediator ordering, reported on the scaled geometric mean ratios. }
 % or \footnotesize for slightly larger font
%\renewcommand{\arraystretch}{0.55} % Row spacing (default is 1)
\footnotesize
\renewcommand{\arraystretch}{0.9}
\setlength{\tabcolsep}{12.5pt} 
\centering
\begin{tabular}{lcccccc}
\toprule
& \multicolumn{3}{c}{\textbf{MEPS data in year 2009}} & \multicolumn{3}{c}{\textbf{MEPS data in year 2016}} \\
\cmidrule(lr){2-4} \cmidrule(lr){5-7}
Disparity & Value & 95\% CI & p-value & Value & 95\% CI & p-value \\ 
\midrule\addlinespace[2.5pt]
\multicolumn{7}{l}{Whites vs Blacks*} \\[2.5pt]
\midrule\addlinespace[2.5pt]
$\rho_{R \rightarrow M_1 \leadsto Y}$ & 1.045 & 1.006 --- 1.085 & {\bf 0.024} & 1.066 & 1.015 --- 1.117 & {\bf 0.011} \\
$\rho_{R \rightarrow M_2 \leadsto Y}$ & 1.011 & 0.981 --- 1.041 & 0.490 & 0.995 & 0.971 --- 1.020 & 0.710 \\
$\rho_{R \rightarrow M_3 \leadsto Y}$ & 1.012 & 0.948 --- 1.075 & 0.721 & 1.057 & 0.993 --- 1.121 & 0.082 \\
$\rho_{R \rightarrow M_4 \leadsto Y}$ & 1.008 & 0.987 --- 1.029 & 0.462 & 1.010 & 0.991 --- 1.029 & 0.284 \\
$\rho_{R \rightarrow Y}$ & 1.772 & 1.614 --- 1.930 & {\bf $<$0.001} & 1.768 & 1.607 --- 1.929 & {\bf $<$0.001} \\
$\rho_\text{total}$ & 1.901 & 1.680 --- 2.122 & {\bf $<$0.001} & 2.084 & 1.840 --- 2.329 & {\bf $<$0.001} \\
\midrule\addlinespace[2.5pt]
\multicolumn{7}{l}{Whites vs Asians*} \\[2.5pt]
\midrule\addlinespace[2.5pt]
$\rho_{R \rightarrow M_1 \leadsto Y}$ & 1.010 & 0.898 --- 1.122 & 0.861 & 1.035 & 0.946 --- 1.124 & 0.441 \\
$\rho_{R \rightarrow M_2 \leadsto Y}$ & 1.085 & 1.017 --- 1.154 & {\bf 0.015} & 1.003 & 0.959 --- 1.047 & 0.891 \\
$\rho_{R \rightarrow M_3 \leadsto Y}$ & 1.328 & 1.176 --- 1.480 & {\bf $<$0.001} & 1.366 & 1.241 --- 1.491 & {\bf $<$0.001} \\
$\rho_{R \rightarrow M_4 \leadsto Y}$ & 1.025 & 0.961 --- 1.090 & 0.440 & 0.996 & 0.964 --- 1.028 & 0.800 \\
$\rho_{R \rightarrow Y}$ & 2.316 & 1.999 --- 2.632 & {\bf $<$0.001} & 2.028 & 1.770 --- 2.287 & {\bf $<$0.001} \\
$\rho_\text{total}$ & 2.893 & 2.408 --- 3.378 & {\bf $<$0.001} & 2.561 & 2.184 --- 2.937 & {\bf $<$0.001} \\
\midrule\addlinespace[2.5pt]
\multicolumn{7}{l}{Whites vs Hispanics*} \\[2.5pt]
\midrule\addlinespace[2.5pt]
$\rho_{R \rightarrow M_1 \leadsto Y}$ & 1.207 & 1.145 --- 1.269 & {\bf $<$0.001} & 1.280 & 1.206 --- 1.353 & {\bf $<$0.001} \\
$\rho_{R \rightarrow M_2 \leadsto Y}$ & 1.265 & 1.220 --- 1.310 & {\bf $<$0.001} & 1.245 & 1.204 --- 1.285 & {\bf $<$0.001} \\
$\rho_{R \rightarrow M_3 \leadsto Y}$ & 1.291 & 1.219 --- 1.362 & {\bf $<$0.001} & 1.393 & 1.312 --- 1.474 & {\bf $<$0.001} \\
$\rho_{R \rightarrow M_4 \leadsto Y}$ & 1.009 & 0.977 --- 1.042 & 0.573 & 1.002 & 0.974 --- 1.030 & 0.880 \\
$\rho_{R \rightarrow Y}$ & 2.071 & 1.901 --- 2.240 & {\bf $<$0.001} & 1.830 & 1.676 --- 1.984 & {\bf $<$0.001} \\
$\rho_\text{total}$ & 3.705 & 3.318 --- 4.093 & {\bf $<$0.001} & 3.371 & 3.005 --- 3.737 & {\bf $<$0.001} \\
\midrule\addlinespace[2.5pt]
\multicolumn{7}{l}{Blacks vs Asians*} \\[2.5pt]
\midrule\addlinespace[2.5pt]
$\rho_{R \rightarrow M_1 \leadsto Y}$ & 1.064 & 0.940 --- 1.189 & 0.311 & 1.015 & 0.867 --- 1.163 & 0.846 \\
$\rho_{R \rightarrow M_2 \leadsto Y}$ & 1.079 & 0.980 --- 1.179 & 0.118 & 0.963 & 0.878 --- 1.048 & 0.392 \\
$\rho_{R \rightarrow M_3 \leadsto Y}$ & 1.306 & 1.101 --- 1.510 & {\bf 0.003} & 1.316 & 1.119 --- 1.514 & {\bf 0.002} \\
$\rho_{R \rightarrow M_4 \leadsto Y}$ & 0.968 & 0.925 --- 1.011 & 0.149 & 0.950 & 0.909 --- 0.992 & {\bf 0.018} \\
$\rho_{R \rightarrow Y}$ & 0.991 & 0.811 --- 1.171 & 0.922 & 0.921 & 0.777 --- 1.065 & 0.281 \\
$\rho_\text{total}$ & 1.466 & 1.115 --- 1.817 & {\bf 0.009} & 1.210 & 0.913 --- 1.506 & 0.165 \\
\midrule\addlinespace[2.5pt]
\multicolumn{7}{l}{Blacks vs Hispanics*} \\[2.5pt]
\midrule\addlinespace[2.5pt]
$\rho_{R \rightarrow M_1 \leadsto Y}$ & 1.210 & 1.150 --- 1.270 & {\bf $<$0.001} & 1.202 & 1.144 --- 1.260 & {\bf $<$0.001} \\
$\rho_{R \rightarrow M_2 \leadsto Y}$ & 1.432 & 1.371 --- 1.492 & {\bf $<$0.001} & 1.333 & 1.279 --- 1.387 & {\bf $<$0.001} \\
$\rho_{R \rightarrow M_3 \leadsto Y}$ & 1.218 & 1.149 --- 1.288 & {\bf $<$0.001} & 1.157 & 1.095 --- 1.219 & {\bf $<$0.001} \\
$\rho_{R \rightarrow M_4 \leadsto Y}$ & 0.977 & 0.953 --- 1.002 & 0.068 & 0.989 & 0.964 --- 1.013 & 0.367 \\
$\rho_{R \rightarrow Y}$ & 1.031 & 0.952 --- 1.109 & 0.442 & 0.896 & 0.822 --- 0.969 & {\bf 0.005} \\
$\rho_\text{total}$ & 1.973 & 1.670 --- 2.276 & {\bf $<$0.001} & 1.624 & 1.391 --- 1.858 & {\bf $<$0.001} \\
\midrule\addlinespace[2.5pt]
\multicolumn{7}{l}{Asians vs Hispanics*} \\[2.5pt]
\midrule\addlinespace[2.5pt]
$\rho_{R \rightarrow M_1 \leadsto Y}$ & 1.230 & 1.138 --- 1.322 & {\bf $<$0.001} & 1.306 & 1.219 --- 1.393 & {\bf $<$0.001} \\
$\rho_{R \rightarrow M_2 \leadsto Y}$ & 1.255 & 1.205 --- 1.305 & {\bf $<$0.001} & 1.244 & 1.199 --- 1.289 & {\bf $<$0.001} \\
$\rho_{R \rightarrow M_3 \leadsto Y}$ & 0.852 & 0.799 --- 0.904 & {\bf $<$0.001} & 0.839 & 0.786 --- 0.893 & {\bf $<$0.001} \\
$\rho_{R \rightarrow M_4 \leadsto Y}$ & 1.022 & 1.007 --- 1.038 & {\bf 0.004} & 1.009 & 0.989 --- 1.030 & 0.358 \\
$\rho_{R \rightarrow Y}$ & 1.017 & 0.941 --- 1.092 & 0.666 & 1.097 & 1.013 --- 1.180 & {\bf 0.023} \\
$\rho_\text{total}$ & 1.471 & 1.193 --- 1.749 & {\bf 0.001} & 1.410 & 1.139 --- 1.680 & {\bf 0.003} \\
\bottomrule
\end{tabular}
\begin{tablenotes}
\item  \textsuperscript{*}Reference group; $M_1$: SES, $M_2$: Insurance, $M_3$: Health status, $M_4$: Health behaviors.
\end{tablenotes}
\label{tab:PSEs_order}
\end{table}
\end{center}

% ########################################
\clearpage
\section{Simulations}
\label{app:simulation}
% ########################################

%  ---  ---  ---  ---  ---  ---  ---  ---  ---  ---  ---  ---  --- - 
\subsection{Finite sample performance and theoretical guarantees}
\label{app:simu_meps}
%  ---  ---  ---  ---  ---  ---  ---  ---  ---  ---  ---  ---  --- - 

In the first set of simulations, we generated data designed to closely resemble MEPS data. We included three covariates: two continuous and one binary. Mediators $M_1$, $M_3$, and $M_4$ are each two-dimensional; $M_1$ and $M_4$ each include one continuous and one binary variable, while $M_3$ consists of two binary variables. The binary components of each bivariate mediator are generated through corresponding latent variables $M^*$, allowing for internal correlation. In addition, $M_2$ is generated as a unidimensional binary variable. The outcome $Y$ follows a zero-inflated, right-skewed distribution: a binomial model determines whether $Y=0$, and a lognormal model generates positive values of $Y$. The data-generating process is detailed as follows: 
{\small
\begin{align*}
   &  X_1, X_2 \overset{iid}{\sim} \text{Uniform}(0,2),  \quad X_3 \sim \text{Bernoulli}(0.5),  \\
   & R \sim \text{Bernoulli}\big(\text{expit}(
   V_R \mt{ 1 & X_1^{0.5} & X_1^{0.5} X_2^{1.5} X_3 &X_2^2& X_2/(1+X_1+X_3)}^T)  ), \\ \\
% M1 ----------
   & M_1 = \begin{bmatrix}  M_{11} & M_{12} \end{bmatrix}  , 
     M_{12} \sim \text{Bernoulli}(\text{expit}(M^*_{12})), \\
    & \hspace{1cm}
    \begin{bmatrix}
        M_{11} \\
        M^*_{12} 
    \end{bmatrix} 
    \sim
    \mathcal{N} \Bigg( 
    \begin{bmatrix} 
    V_{M_{11}} (1 & R &X_1 X_2 & X_2^{0.5} X_3 & R X_3)^T \\
    V_{M_{12}} (1 & R & X_1^2 &  X_2 &        X_3) ^T 
    \end{bmatrix} ,
    \begin{bmatrix}
        1  & 0.5 \\
        0.5  & 1 
    \end{bmatrix} \Bigg),\\
% M2 ----------
   & M_2 \sim \text{Bernoulli}(\text{expit}(
   V_{M_2}
   \begin{bmatrix}
       1 & R & R X_3 & R M_{11} & M_{12}X_2 & X_1 & M_{11}/(1+X_2) 
   \end{bmatrix}^T
   )), \\
% M3 ----------
   & M_3 = \begin{bmatrix}  M_{31} &   M_{32} \end{bmatrix} , 
    M_{31} \sim \text{Bernoulli}(\text{expit}(M^*_{31})),
    M_{32} \sim \text{Bernoulli}(\text{expit}(M^*_{32})), \\
    & \hspace{1cm}
    \begin{bmatrix}
        M^*_{31} \\
        M^*_{32} 
    \end{bmatrix} 
     \sim
    \mathcal{N} \Bigg( 
    \begin{bmatrix} 
    V_{M_{31}} (1 & R & R M_{11} &    M_{12} & R M_2 & X_1 & X_2 & R X_3)^T \\
    V_{M_{32}} (1 & R & M_{11} &    M_{12} & R M_2 & X_1^{0.5} & X_2 & X_3   ) ^T 
    \end{bmatrix} ,
    \begin{bmatrix}
        1  & 0.5 \\
        0.5  & 1  
    \end{bmatrix} \Bigg),\\
 % M4 ----------
   & M_4 = \begin{bmatrix}  M_{41} &   M_{42} \end{bmatrix},
    M_{42} \sim \text{Bernoulli}(\text{expit}(M^*_{42})),\\
   & \hspace{1cm} 
   \begin{bmatrix}
        M_{41} \\
        M^*_{42} 
    \end{bmatrix} 
     \sim
    \mathcal{N} \Bigg( 
    \begin{bmatrix} 
    V_{M_{41}} (1 & R & M_{11} & M_{12} &  M_2 & M_{31} M_{32} & R X_1 & X_2 &   X_2 X_3)^T \\
    V_{M_{42}} (1 & R & M_{11} & M_{12} &  M_2 & M_{31} M_{32} & X_1 &    X_2 &   X_3) ^T 
    \end{bmatrix} ,
    \begin{bmatrix}
         1  & 0.5 \\
        0.5  & 1 
    \end{bmatrix} \Bigg), \\ \\
% Y ----------
   & Y^* = 
   V_Y{\scriptsize \begin{bmatrix} 
   1 \hspace{0.2cm} R \hspace{0.2cm} M_{11}X_1^{0.5} \hspace{0.2cm} M_{12} X_2^2 \hspace{0.2cm} M_2 X_1^3 X_2^{0.5} \hspace{0.2cm} M_{31} \exp(X_1^{0.1})\hspace{0.2cm} R M_{32} \hspace{0.2cm} M_{41} \hspace{0.2cm} M_{42} \hspace{0.2cm} M_{41} X_1\hspace{0.2cm} R M_2 X_2 \hspace{0.2cm} \text{cos}(X_1 X_2) \hspace{0.2cm} X_3 \hspace{0.2cm} (X_1+X_2)^{0.5} 
   \end{bmatrix}^T }, \\
  & \mathbb{I}(Y>0)  \sim   \text{Bernoulli}(\text{expit}(Y^*)), \\
 & Y \mid Y>0  \sim \text{LogNormal}(\text{log}\mu = 0.4Y^*, \text{log}sd = 0). \numberthis
\end{align*}
}
where  
\begin{align*}
V_R  &= [-0.34, 0.38, -0.24, 0.31, -0.44],\\
V_{M_{11}} &= [-0.09, 0.56, 0.26, 0.23, -0.28], \\
V_{M_{12}}  &= [-0.43, 0.44, 0.17, 0.33,  -0.33], \\
V_{M_{2}}& = [- 0.15, 0.80, 0.36, 0.16, 0.48, -0.23, 0.39], \\
V_{M_{31}} &= [-0.23, 0.61, 0.23, 0.35, 0.48, -0.24, 0.24, 0.34], \\
V_{M_{32}} &= [-0.46, 0.57, 0.33, 0.21, 0.23,  0.13, -0.16, -0.12], \\
V_{M_{41}} &= [-0.50, 0.31, 0.48, 0.17, 0.40, 0.18,  0.37, 0.39, -0.38], \\ 
V_{M_{42}} &= [-0.47, 0.45, 0.31, 0.43, 0.14, 0.39, 0.44, -0.36, -0.49], \\
V_{Y} &= [0.61, 0.57, 0.53, 0.45, 0.81, 0.87, 0.92, 0.23, 0.37, 0.69, 0.95, -0.47, 0.14, -0.64]. 
\end{align*}

Due to the complexity of the data-generating process, closed-form expressions for the nuisance functions required in the EIF are intractable. Therefore, we rely on numerical approximations to estimate the EIF variance. 

\begin{table}[t]
\caption{Comparative performance of the one-step corrected plug-in estimator using super learner versus GLM for nuisance estimation. }
\footnotesize
 % or \footnotesize for slightly larger font
\renewcommand{\arraystretch}{1.2} % Row spacing (default is 1)
\setlength{\tabcolsep}{8pt}    
\begin{tabular}{rcccccccccc}
\toprule
 & \multicolumn{2}{c}{Bias} & \multicolumn{2}{c}{SD} & \multicolumn{2}{c}{MSE} & \multicolumn{2}{c}{Coverage Rate} & \multicolumn{2}{c}{CI width} \\ 
\cmidrule(lr){2-3} \cmidrule(lr){4-5} \cmidrule(lr){6-7} \cmidrule(lr){8-9} \cmidrule(lr){10-11}
n & SL & GLM & SL & GLM & SL & GLM &  SL & GLM &  SL & GLM \\ 
\midrule 
$\rho^+_{R \rightarrow M_1 \leadsto Y}$  & & & & & & & &  & \\
1000 & 0.002 & 0.005 & 0.046 & 0.047 & 0.002 & 0.002 & 0.924 & 0.960 & 0.157 & 0.194 \\ 
2000 & 0.001 & 0.005 & 0.033 & 0.032 & 0.001 & 0.001 & 0.924 & 0.969 & 0.115 & 0.137 \\ 
4000 & 0.000 & 0.004 & 0.023 & 0.021 & 0.001 & 0.000 & 0.929 & 0.975 & 0.084 & 0.096 \\ 
8000 & 0.001 & 0.005 & 0.017 & 0.016 & 0.000 & 0.000 & 0.929 & 0.964 & 0.060 & 0.068 \\ 
\midrule 
$\rho^+_{R \rightarrow M_2 \leadsto Y}$   & & & & & & & &   & \\
1000 & -0.003 & 0.002 & 0.039 & 0.037 & 0.002 & 0.001 & 0.887 & 0.968 & 0.127 & 0.158 \\ 
2000 & 0.000 & 0.004 & 0.028 & 0.026 & 0.001 & 0.001 & 0.890 & 0.974 & 0.095 & 0.112 \\ 
4000 & -0.001 & 0.002 & 0.021 & 0.018 & 0.000 & 0.000 & 0.892 & 0.968 & 0.070 & 0.079 \\ 
8000 & 0.000 & 0.003 & 0.015 & 0.013 & 0.000 & 0.000 & 0.911 & 0.964 & 0.051 & 0.056 \\ 
\midrule
$\rho^+_{R \rightarrow M_3 \leadsto Y}$  & & & & & & & &   & \\
1000 & -0.001 & 0.009 & 0.039 & 0.047 & 0.002 & 0.002 & 0.878 & 0.965 & 0.120 & 0.196 \\ 
2000 & -0.001 & 0.010 & 0.026 & 0.032 & 0.001 & 0.001 & 0.902 & 0.963 & 0.089 & 0.138 \\ 
4000 & 0.000 & 0.010 & 0.019 & 0.024 & 0.000 & 0.001 & 0.908 & 0.946 & 0.064 & 0.097 \\ 
8000 & 0.000 & 0.010 & 0.013 & 0.016 & 0.000 & 0.000 & 0.920 & 0.936 & 0.046 & 0.069 \\ 
\midrule
$\rho^+_{R \rightarrow M_4 \leadsto Y}$  & & & & & & & &   &   \\
1000 & 0.000 & 0.028 & 0.037 & 0.059 & 0.001 & 0.004 & 0.857 & 0.944 & 0.110 & 0.227 \\ 
2000 & -0.001 & 0.029 & 0.026 & 0.038 & 0.001 & 0.002 & 0.884 & 0.935 & 0.083 & 0.160 \\ 
4000 & 0.000 & 0.030 & 0.018 & 0.029 & 0.000 & 0.002 & 0.914 & 0.853 & 0.062 & 0.114 \\ 
8000 & 0.000 & 0.028 & 0.013 & 0.019 & 0.000 & 0.001 & 0.908 & 0.752 & 0.045 & 0.080 \\ 
\midrule
\multicolumn{1}{l}{$\rho^+_{R  \rightarrow Y}$}  & & & & & & & &   & & \\
1000 & -0.003 & 0.000 & 0.040 & 0.088 & 0.002 & 0.008 & 0.906 & 0.947 & 0.132 & 0.348 \\ 
2000 & -0.005 & -0.012 & 0.029 & 0.059 & 0.001 & 0.004 & 0.906 & 0.953 & 0.097 & 0.243 \\ 
4000 & 0.000 & -0.009 & 0.020 & 0.043 & 0.000 & 0.002 & 0.921 & 0.942 & 0.071 & 0.171 \\ 
8000 & 0.000 & -0.011 & 0.015 & 0.030 & 0.000 & 0.001 & 0.910 & 0.938 & 0.051 & 0.120 \\ 
\bottomrule
\end{tabular}
\begin{tablenotes}
\item  The numbers are rounded to 3 digits.
\end{tablenotes}
\label{tab:simu_meps}
\end{table}

%  ---  ---  ---  ---  ---  ---  ---  ---  ---  ---  ---  ---  --- -
\clearpage

\subsection{Robustness to model misspecification}
\label{app:simu_misspecific}
%  ---  ---  ---  ---  ---  ---  ---  ---  ---  ---  ---  ---  --- - 

The simulation data---including variables $(X_1, X_2, X_3, X_4, R, M_1, M_2, M_3, M_4, Y)$---are generated to evaluate the robustness of one-step estimators for counterfactual means under the model misspecification scenarios described in Corollary~\ref{cor:robustness}, and outlined in Table~\ref{tab:misspec_combined}, using the following data-generating models: 
\begin{align*}
X_1, X_2, X_3, X_4  & \overset{iid}{\sim} \text{Uniform}(0,1), \\
R  & \sim \text{Bernoulli}(\text{expit}(V_R \mt{1 & X}^T)), \\
M_1 & \sim \mathcal{N}(V_{M_1}\mt{1 & X & R}^T, 1), \\
M_2 & \sim \mathcal{N}(V_{M_2} \mt{1 & X & R & M_1}^T , 1), \\
M_3 & \sim \mathcal{N}(V_{M_3}\mt{1 & X & R & M_1 & M_2}^T, 1), \\
M_4 & \sim \mathcal{N}(V_{M_4}\mt{1 & X & R & M_1 & M_2 & M_3}^T, 1), \\
Y & \sim \mathcal{N}(V_Y \mt{1 & X & R & M_1 & M_2 & M_3 & M_4}^T, 1) , \numberthis
\end{align*}
where
\begin{align*}
& V_R  = [-0.10, 1.00, 0.20, -0.40,  0.80],\\
& V_{M_1} =[-0.13, 0.23, -0.18, 0.15, -0.16, 0.13],\\
& V_{M_2} =[-0.11, -0.06, 0.20, 0.25, 0.02, -0.12, 0.16],\\
& V_{M_3} =[-0.24, -0.08, -0.15, 0.03, 0.14, 0.06, -0.14, 0.09],\\
& V_{M_4} =[-0.13, -0.09, -0.04, 0.10, -0.25, -0.05, -0.08, 0.19, -0.20],\\
& V_{Y} =[0.43, 0.29, 0.28, -0.26, -0.38, 0.18, 0.39, -0.22, -0.13, 0.28] . 
\end{align*}

The proposed one-step estimators of $\gamma^+_{R \rightarrow Y}$ and $\gamma^+_{R \rightarrow M_k \leadsto Y}$ are constructed using estimates of the nuisance functions $Q = \{\pi, \{g_k, \mu_k, \mathscr{B}_k, \mathscr{C}_{\mathscr{B}_k} : \forall k\}, \mathscr{C}_{\mu_4}\}$. These nuisance functions can be consistently estimated via GLMs based on linear combinations of the predictors, as follows: 
{\small
\begin{align*}
\pi(X) = \text{expit}(\theta_0 \mt{1 & X }^T), \quad & 
g_k(\overline{M}_k, X) = \text{expit}(\theta_k \mt{1 & X & \overline{M}_k}^T) , \\
\mu_k(\overline{M}_k, R, X) =  \alpha_k \mt{1 & X & R & \overline{M}_k}^T, \quad &
\mathscr{B}_k(\overline{M}_{k-1}, R, X) = \delta_{k-1} \mt{1 & X & R & \overline{M}_{k-1}}^T,\\
\mathscr{C}_{\mathscr{B}_k}(R, X) =  \nu_{\mathscr{B}_k} \mt{1 & X & R}^T, \quad &
\mathscr{C}_{\mu_4}(R, X) =\nu_{\mu_4} \mt{1 & X & R}^T . \numberthis
\end{align*} }

To evaluate the impact of model misspecification, a set of transformed covariates is generated from the true covariates $X$ as $X^{\text{false}} = (X_1^2,\ e^{X_2},\ X_3^{0.3},\ (X_4 + X_3^{0.3}) / (e^{X_2} + X_1^2))$. These transformed covariates are then used to construct misspecified versions of the nuisance functions, denoted $Q^{\text{false}}$, using GLMs: 
{\small
\begin{align*}
\pi(X^{\text{false}})  = \text{expit}(\theta^*_0 \mt{1 & X^{\text{false}} }^T), \quad & 
g_k(\overline{M}_k, X^\text{false}) = \text{expit}(\theta^*_k \mt{1 & X^{\text{false}} & \overline{M}_k}^T) , \\
\mu_k(\overline{M}_k, R, X^\text{false}) =  \alpha^*_k \mt{1 & X^{\text{false}} & R & \overline{M}_k}^T, \quad &
\mathscr{B}_k(\overline{M}_{k-1}, R, X^\text{false}) = \delta^*_{k-1} \mt{1 & X^{\text{false}} & R & \overline{M}_{k-1}}^T,\\
\mathscr{C}_{\mathscr{B}_k}(R, X^\text{false}) =  \nu^*_{\mathscr{B}_k} \mt{1 & X^{\text{false}} & R}^T, \quad &
\mathscr{C}_{\mu_4}(R, X^\text{false}) =\nu^*_{\mu_4} \mt{1 & X^{\text{false}} & R}^T . \numberthis
\end{align*}}

\begin{figure}[!h]
    \centering
    \includegraphics[width=15cm]{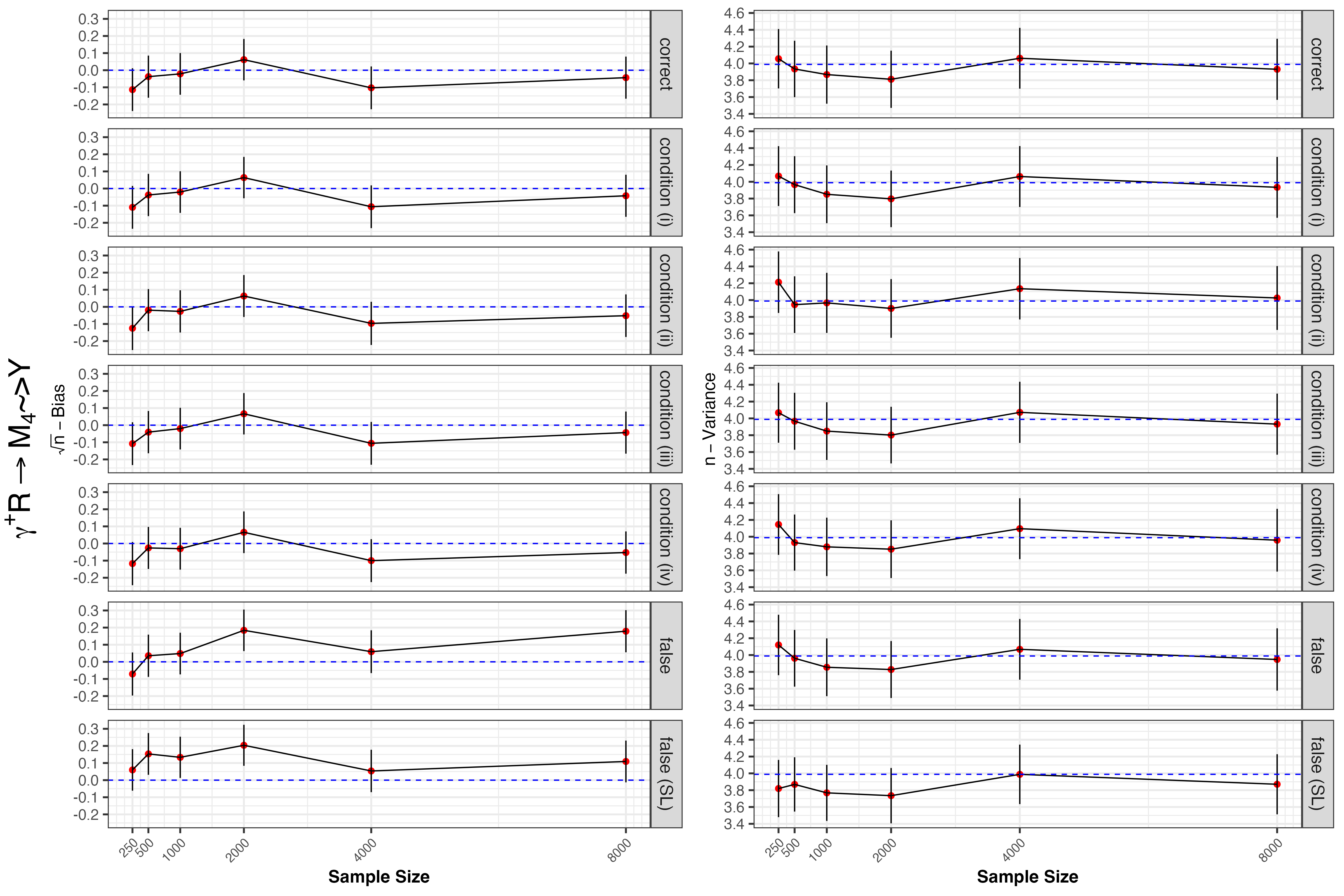}
    \caption{Simulation results demonstrating $\sqrt{n}$-consistency of one-step estimators under various nuisance misspecification scenarios. ``False'' refers to estimators using fully misspecified GLM-based nuisance functions ($Q^{\text{false}}$), while ``False (SL)'' refers to those using misspecified nuisance functions estimated via super learner ($Q^{\text{SL}}$).  }
    \label{fig:misspecific}
\end{figure}

The one-step estimators under each condition are derived by combining estimated nuisance functions from both $Q$ and $Q^{\text{false}}$. We also consider a scenario in which all nuisance functions are misspecified, serving as a baseline for comparison. Additionally, the variables $(X^{\text{false}}, R, \overline{M}_4, Y)$ are used to estimate nuisance functions $Q^{\text{SL}}$ via super learner, yielding the corresponding estimators. The results are presented in Figure~\ref{fig:misspecific}. 

\clearpage
\begin{figure}[h]
    \centering
    \includegraphics[width=15cm]{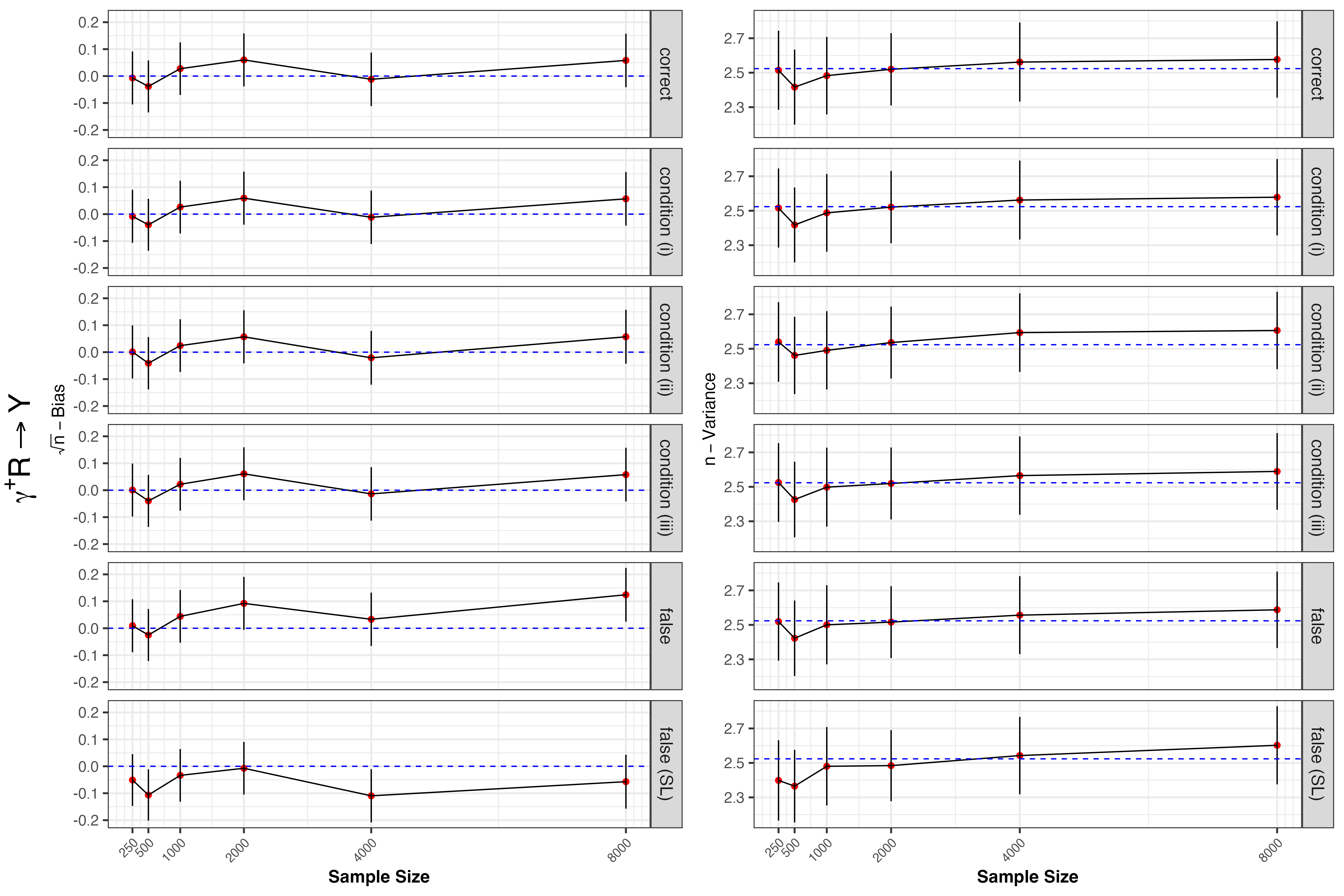}
\end{figure}
\begin{figure}[!h]
    \centering
    \includegraphics[width=15cm]{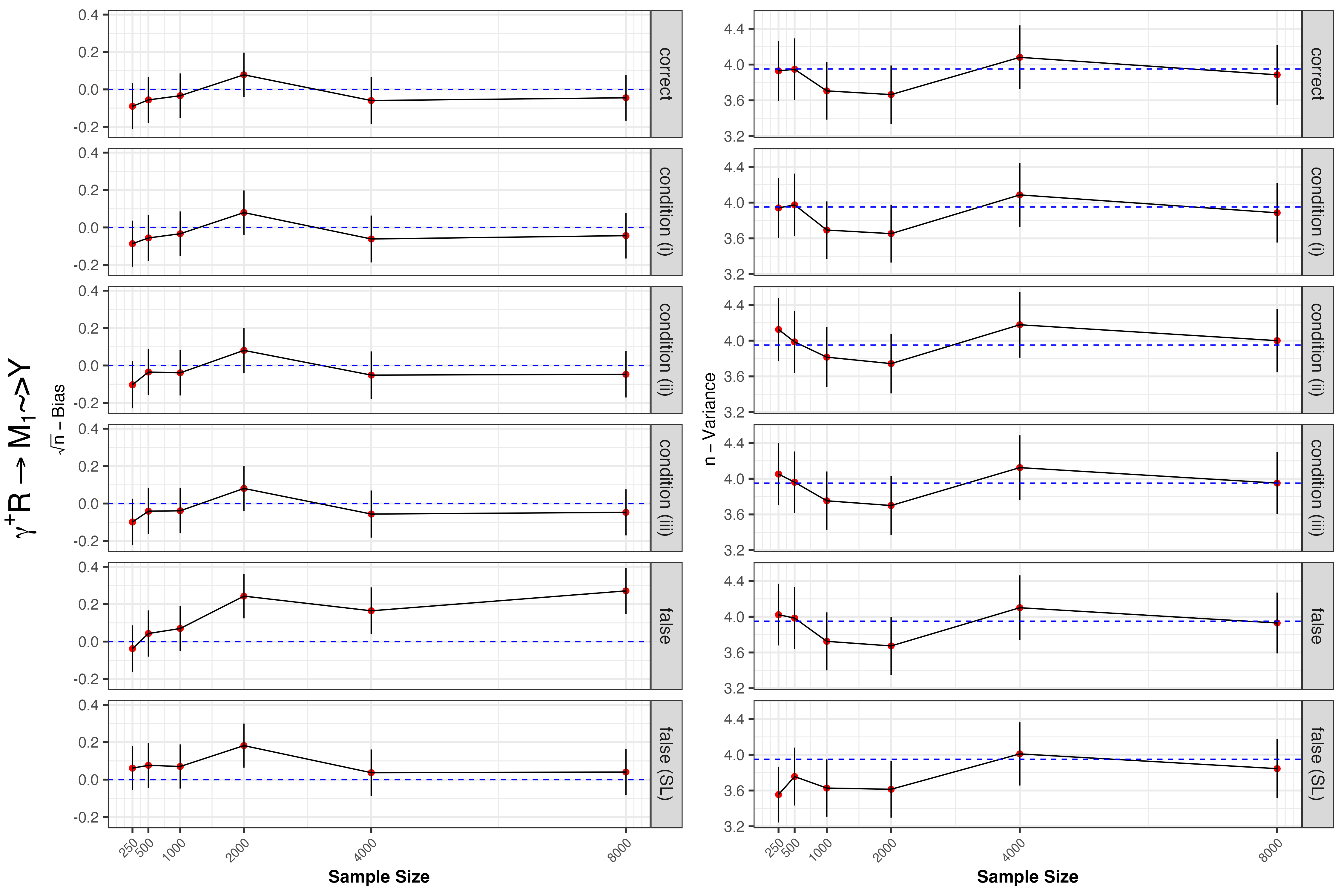}
\end{figure} 
\newpage
\begin{figure}[!h]
    %\hspace{-1.5cm}
    \centering
    \includegraphics[width=15cm]{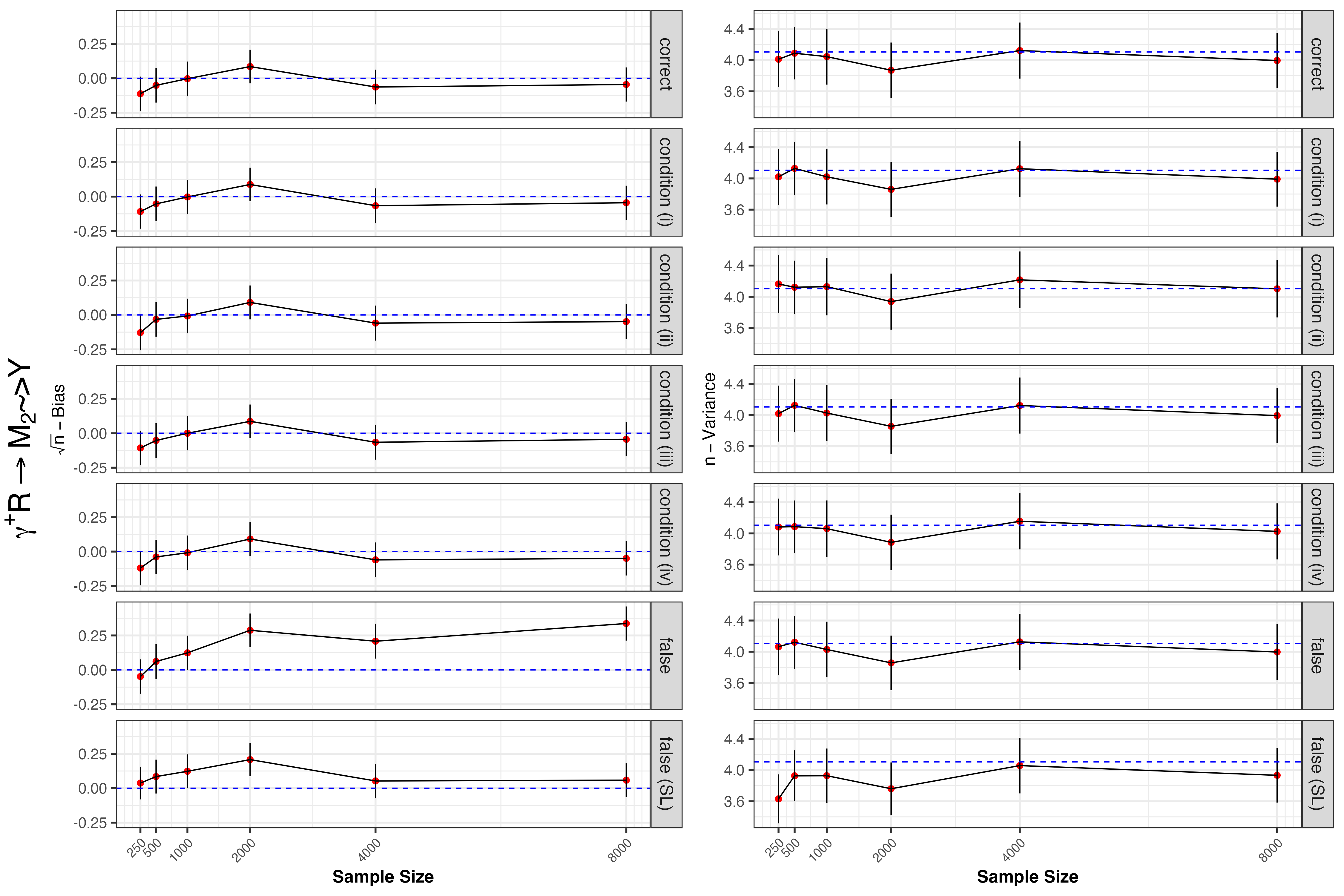}
\end{figure} 
\begin{figure}[!h]
    %\hspace{-1.5cm}
    \centering
    \includegraphics[width=15cm]{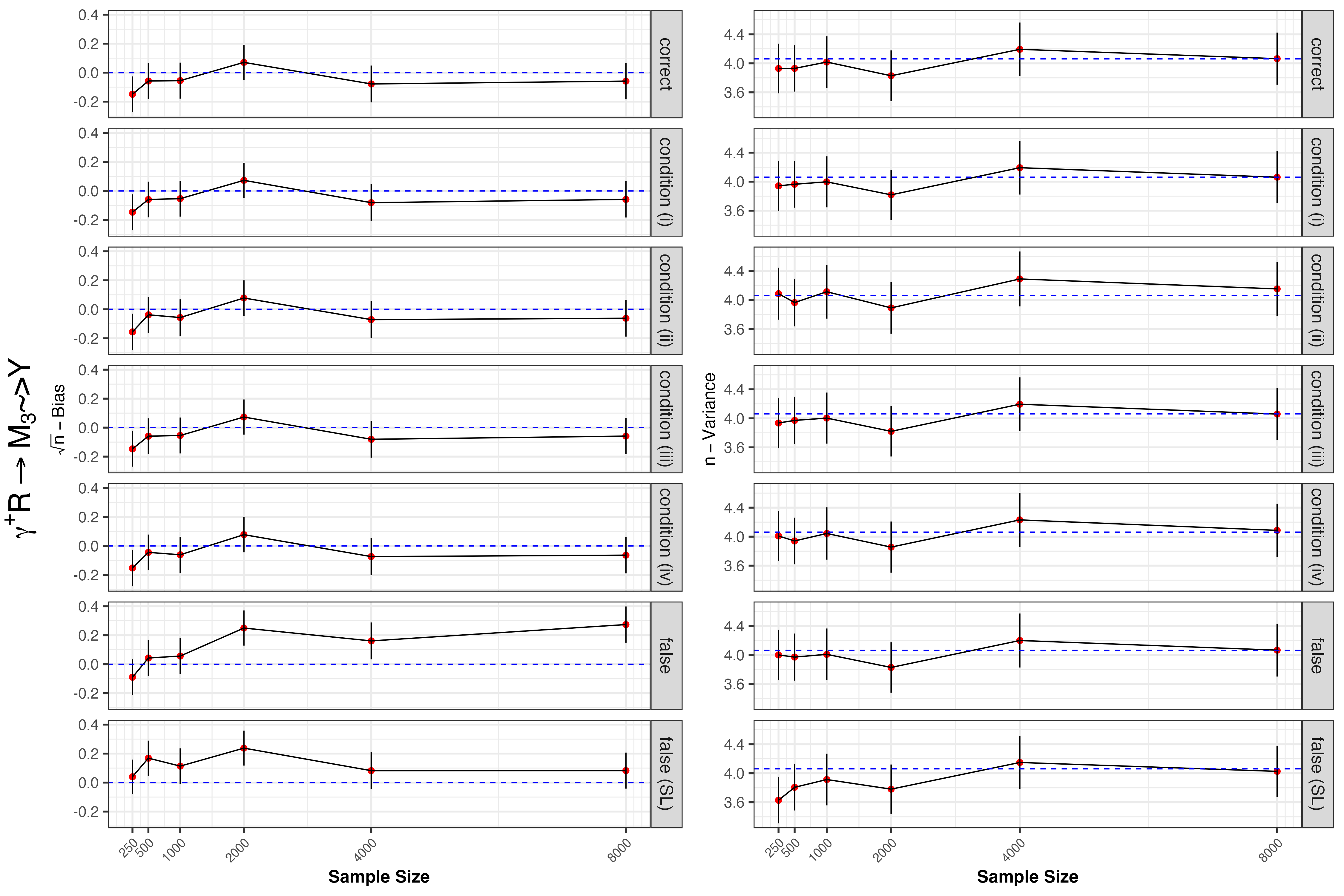}
\end{figure} 
\newpage

% *******************
% reference for supp
% *******************
% \putbib[Fairness]  % Only shows references cited in appendix
\putbib[Fairness]            % insert the bib from the bibunit

\end{bibunit}

\end{document}